\documentclass[PAPER, atlasdraft=false, texlive=2023, UKenglish, cernpreprint, texmf, orcidlogo]{atlasdoc}
 
\usepackage{atlaspackage}
\usepackage{array} 
\usepackage{atlasbiblatex}
 
\usepackage{atlasphysics}
 
\graphicspath{{logos/}{figures/}}
 
\usepackage{ANA-HIGG-2020-25-PAPER-defs}

\AtlasTitle{Integrated and differential fiducial cross-section measurements for the vector boson fusion production of the Higgs boson in the $H{\rightarrow\,}WW^{\ast}{\rightarrow\,}e\nu\mu\nu$ decay channel at 13 $\text{TeV}$ with the ATLAS detector}
 
\AtlasVersion{4.0}
\AtlasAbstract{
The vector-boson production cross-section for the Higgs boson decay in the $H{\rightarrow\,}WW^{\ast}{\rightarrow\,}e\nu\mu\nu$ channel is measured as a function of kinematic observables sensitive to the Higgs boson production and decay properties as well as integrated in a fiducial phase space. The analysis is performed using the proton--proton collision data collected by the ATLAS detector in Run 2 of the LHC at $\sqrt{s}= 13$ $\text{TeV}$ center-of-mass energy corresponding to an integrated luminosity of 139 fb$^{-1}$. The different flavor final state is studied by selecting an electron and a muon originating from a pair of $W$ bosons and compatible with the Higgs boson decay. The data are corrected for the effects of detector inefficiency and resolution, and the measurements are compared with different state-of-the-art theoretical predictions. The differential cross-sections are used to constrain anomalous interactions described by dimension-six operators in an Effective Field Theory.
}
 
\author{The ATLAS Collaboration}
 
\AtlasRefCode{HIGG-2020-25}
\PreprintIdNumber{CERN-EP-2023-025}

\AtlasJournal{Phys.\ Rev.\ D}
\AtlasJournalRef{Phys. Rev. D 108, 072003}
\AtlasDOI{10.1103/PhysRevD.108.072003}

\hypersetup{pdftitle={ATLAS document},pdfauthor={The ATLAS Collaboration}}
 
\begin{document}
 
\maketitle
 
\tableofcontents

\section{Introduction}
\label{sec:intro}

\newcommand{\AtlasCoordFootnote}{
ATLAS uses a right-handed coordinate system with its origin at the nominal interaction point (IP)
in the center of the detector and the \(z\)-axis along the beam pipe.
The \(x\)-axis points from the IP to the center of the LHC ring,
and the \(y\)-axis points upwards.
Cylindrical coordinates \((r,\phi)\) are used in the transverse plane,
\(\phi\) being the azimuthal angle around the \(z\)-axis.
The pseudorapidity is defined in terms of the polar angle \(\theta\) as \(\eta = -\ln \tan(\theta/2)\).
Angular distance is measured in units of \(\Delta R \equiv \sqrt{(\Delta\eta)^{2} + (\Delta\phi)^{2}}\).}

A particle consistent with the Standard Model (SM) Higgs boson predictions and a mass of approximately 125 \GeV was discovered by the ATLAS and CMS Collaborations~\cite{HIGG-2012-27,CMS-HIG-12-028} in 2012  using proton–proton (\pp) collision data produced by the Large Hadron Collider (LHC) at CERN\@. Its properties were studied in 2011 and 2012 data sets at 7 and 8 \TeV\ center-of-mass energies ($\sqrt{s}$), referred to as Run 1, and with $\sqrt{s} = 13$ \TeV\ data collected between 2015 and 2018, referred to as Run 2.
 
In the vector-boson-fusion (VBF) processes, quarks from the colliding protons radiate weak vector bosons that fuse to form the Higgs boson. The VBF process is the second most frequent Higgs boson production mechanism at the LHC, following the gluon--gluon fusion (ggF) process~\cite{CAHN1984196} where the interacting gluons produce a Higgs boson predominantly through a top-quark loop. The VBF signature is characterized by the presence of jets from each of the interacting quarks with a large rapidity gap between them and a large invariant mass of the dijet system. Radiative hadronic activity between the two jets is suppressed because of the absence of color connection between the two quarks. 
The measured cross-section of the VBF production process directly probes the Higgs boson couplings to \W and \Z bosons.
The VBF production of the Higgs boson was measured by the ATLAS and CMS experiments in several decay channels and the combined results at 13 \TeV\ center-of-mass energy are presented in Refs.~\cite{HIGG-2021-23,CMS-HIG-22-001} with the LHC Run 2 data set.
 
The \hww decay channel has the second-largest branching fraction and allowed for the most precise Higgs boson cross-section measurements in Run 1~\cite{HIGG-2015-07}.
The VBF production of the \hwwlnln channel  was previously studied by the CMS~\cite{CMS-HIG-16-042,CMS-HIG-12-036} and the ATLAS~\cite{HIGG-2016-07,HIGG-2013-13,HIGG-2017-13} Collaborations using a partial Run 2 data set corresponding to an integrated luminosity of approximately 36~\fb\, and at lower center-of-mass energies of 7~\TeV\ and 8~\TeV\ in Run 1. The observation of the VBF \hwwlnln process was reported by the ATLAS Collaboration using the full Run 2 data set at 13 \TeV\ center-of-mass energy, corresponding to 139 $\ifb$~\cite{HIGG-2021-20}. 
The CMS Collaboration has also reported the measurement of the VBF \hwwlnln process and its differential cross-section in the simplified template cross-section framework, using the full Run 2 data set of 138 $\ifb$~\cite{CMS-HIG-20-013}.
Figure~\ref{fig:FeynmanDiagram} shows an example of a tree-level Feynman diagram for the Higgs boson production via VBF in the \hww decay channel.
\begin{figure}[tb!]
\centering
\includegraphics[width=0.40\linewidth]{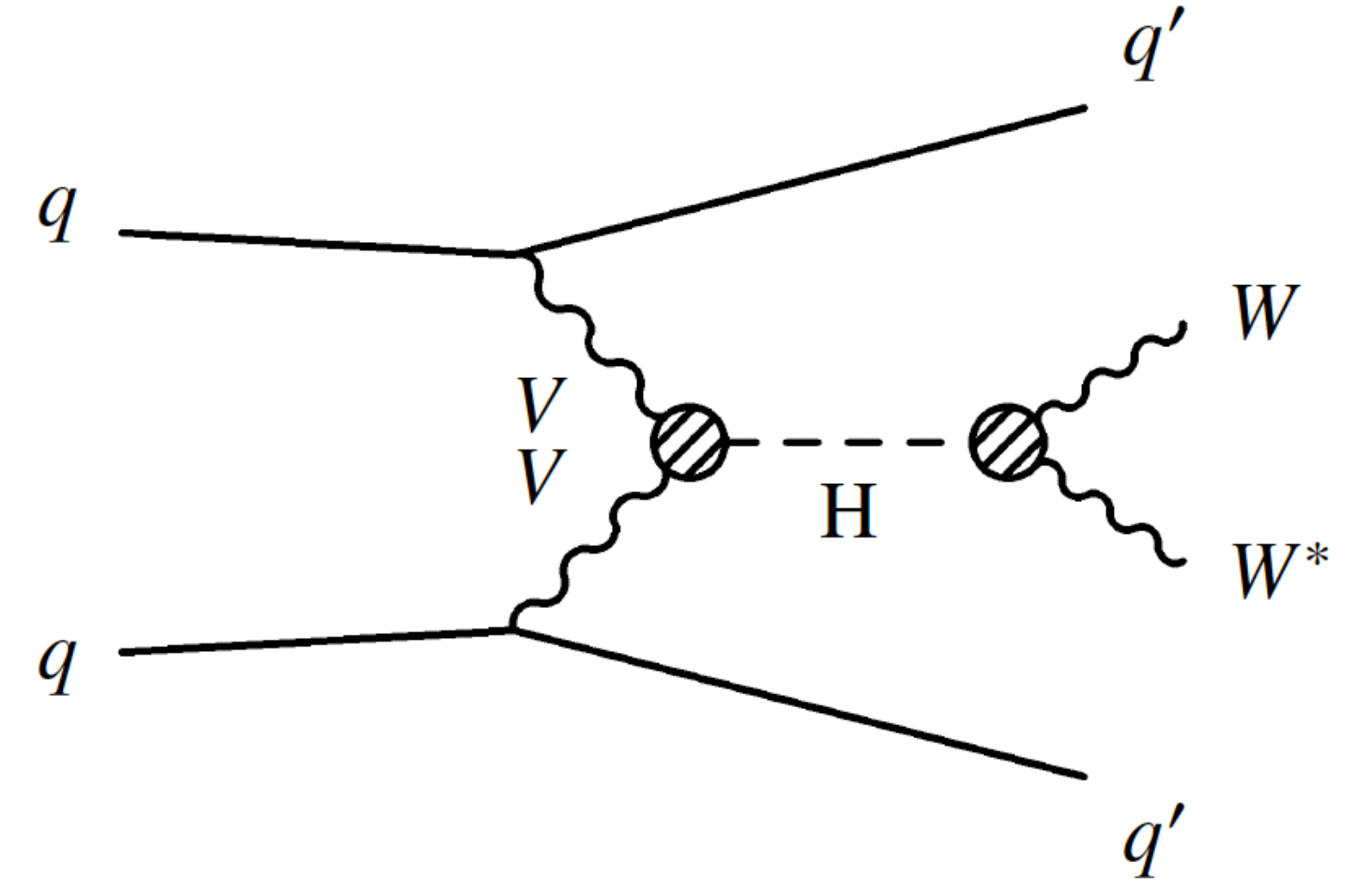}
\caption{Diagram for the Higgs boson production in the VBF mode with \V representing a \W or \Z vector boson. The Higgs boson coupling vertices are marked with circles.  }
\label{fig:FeynmanDiagram}
\end{figure}

This paper presents the  measurement of fiducial cross-sections for the VBF \hwwlnln channel. The full ATLAS Run 2 data set, consisting of proton--proton collision data at $\sqrt{s}= 13$ \TeV\ center-of-mass energy taken between 2015 and 2018, is used. The total integrated luminosity, after imposing data quality requirements, is 139 \ifb.
 
Several SM processes result in final states with two electrically charged leptons, neutrinos, and jets. Therefore, stringent selection requirements are applied to enhance the data sample with signal events and suppress the contamination from background processes. Selection criteria were studied and optimized relative to previous ATLAS VBF \hwwlnln studies~\cite{HIGG-2021-20} for an accurate measurement of the production cross-section in a fiducial phase space as well as differentially. Such selections aim to maximize the statistical significance of the VBF \hwwlnln signal process and minimize the impact of the uncertainties associated with the background estimations and signal modeling.
The $\hwwlnln$ event candidates are identified by selecting one electron and one muon with opposite charges and close in azimuthal angle. The backgrounds from low-energy resonances and $Z/\gamma^*$+jets events are further suppressed by imposing kinematic requirements on the two charged leptons, while the background from top-quark production processes is reduced by imposing a veto on events with $b$-quark jets. The VBF production mode is selected by requiring at least two energetic jets and is distinguished from other Higgs boson production modes (most notably ggF) by applying further kinematic requirements.
Those requirements include event vetoes based on the lepton and the jet kinematics as well as additional requirements on the invariant mass and rapidity separation of the two leading jets, i.e., the jets with the greatest transverse momentum (\pT).

The signal contribution is extracted from a simultaneous binned likelihood fit of multivariate analysis discriminants to data in several kinematic regions.
The contamination of the major background processes in the data sample is estimated from data in signal-enhanced regions (signal regions) and in background-enhanced regions (control regions).
The measured signal cross-section is corrected (unfolded) for detector inefficiency and resolution and is measured in a fiducial region close to the experimental event selection thus minimizing the model dependence in the extrapolation from the detector-level signal region to the particle-level fiducial region. The unfolding to particle level is implemented using detector response corrections directly in the likelihood fit.
 
The differential fiducial cross-section is measured as a function of variables sensitive to the properties of the Higgs boson, such as spin, parity, interactions, production and decay, as well as those of the associated jets.
Statistical correlations among pairs of these differential cross-sections are calculated to allow future re-interpretations of the experimental results.
The cross-section measurements are compared with fixed-order calculations at leading-order (LO) and next-to-leading-order (NLO) in quantum-chromodynamics (QCD) and electroweak (EW) corrections as well as with parton-shower simulations.
 
The differential fiducial cross-section measurements are also interpreted in an Effective Field Theory (EFT) formalism to set limits on anomalous couplings that may affect the interaction vertices in the VBF production of the Higgs boson and in its decay to the \lnln final state.

The paper is organized as follows. An overview of the ATLAS detector is given in Section~\ref{sec:detector}, while the data and simulated signal and background samples are described in Section~\ref{sec:dataset_mc}. The selection of the Higgs boson candidate events is detailed in Section~\ref{sec:obj_evnt_reco}, while Section~\ref{sec:fid_phasespace_obs} outlines the fiducial phase-space definitions for the integrated and the differential cross-section measurements as well as the observables that are unfolded to particle-level in the differential cross-section measurements. Section~\ref{sec:bkg} describes the background estimate techniques. The statistical analysis used to extract the cross-sections and the unfolding strategy together with the estimate of systematic uncertainties are described in Section~\ref{sec:method}.
Experimental results and their comparisons to SM predictions are presented in Section~\ref{sec:result}, while their interpretation to constrain anomalous interactions are discussed in Section~\ref{sec:eft}. Concluding remarks are given in Section~\ref{sec:conclusion}.


\section{ATLAS detector}
\label{sec:detector}

The ATLAS detector~\cite{PERF-2007-01} at the LHC covers nearly the entire solid angle around the collision point. 
It consists of an inner tracking detector surrounded by a thin superconducting solenoid, electromagnetic and hadron calorimeters,
and a muon spectrometer incorporating three large superconducting air-core toroidal magnets.
 
The inner-detector system (ID) is immersed in a \qty{2}{\tesla} axial magnetic field
and provides charged-particle tracking in the range of \(|\eta| < 2.5\)\footnote{\AtlasCoordFootnote}.
The high-granularity silicon pixel detector covers the vertex region and typically provides four measurements per track,
the first hit normally being in the insertable B-layer (IBL) installed before Run~2~\cite{ATLAS-TDR-19,PIX-2018-001}.
It is followed by the silicon microstrip tracker (SCT), which usually provides eight measurements per track.
These silicon detectors are complemented by the transition radiation tracker (TRT),
which enables radially extended track reconstruction up to \(|\eta| = 2.0\).
The TRT also provides electron identification information
based on the fraction of hits (typically 30 in total) above a higher energy-deposit threshold corresponding to transition radiation.
 
The calorimeter system covers the pseudorapidity range \(|\eta| < 4.9\). Within the region \(|\eta|< 3.2\), electromagnetic calorimetry is provided by barrel and
endcap high-granularity lead/liquid-argon (LAr) calorimeters,
with an additional thin LAr presampler covering \(|\eta| < 1.8\)
to correct for energy loss in material upstream of the calorimeters.
Hadron calorimetry is provided by the steel/scintillator-tile calorimeter,
segmented into three barrel structures within \(|\eta| < 1.7\), and two copper/LAr hadron endcap calorimeters.
The solid angle coverage is completed with forward copper/LAr and tungsten/LAr calorimeter modules
optimized for electromagnetic and hadronic energy measurements respectively.
 
The muon spectrometer (MS) comprises separate trigger and
high-precision tracking chambers measuring the deflection of muons in a magnetic field generated by the superconducting air-core toroidal magnets.
The field integral of the toroids ranges between \num{2.0} and \qty{6.0}{\tesla\metre}
across most of the detector.
Three layers of precision chambers, each consisting of layers of monitored drift tubes, cover the region \(|\eta| < 2.7\),
complemented by cathode-strip chambers in the forward region, where the background is highest.
The muon trigger system covers the range \(|\eta| < 2.4\) with resistive-plate chambers in the barrel, and thin-gap chambers in the endcap regions.
 
Events are selected by the first-level trigger system implemented in custom hardware,
followed by selections made by algorithms implemented in software in the high-level trigger~\cite{TRIG-2016-01}.
The first-level trigger accepts events from the \qty{40}{\MHz} bunch crossings at a rate below \qty{100}{\kHz},
which the high-level trigger further reduces to record events to disk at about \qty{1}{\kHz}.
 
An extensive software suite~\cite{ATL-SOFT-PUB-2021-001} is used in data simulation, in the reconstruction
and analysis of real and simulated data, in detector operations, and in the trigger and data acquisition
systems of the experiment.


\section{Data samples and Monte Carlo event simulation}
\label{sec:dataset_mc}

The analysis was performed on \pp collision data collected by the ATLAS experiment at a center-of-mass energy of $\sqrt{s}$ = 13 \TeV. The data were recorded between 2015 and 2018, corresponding to an integrated luminosity of 139 \ifb. The data were subjected to quality requirements~\cite{DAPR-2018-01}, including the removal of events recorded when relevant detector components were not operating correctly.
 
Monte Carlo (MC) event generators were used to simulate the signal and background events produced in the \pp collisions. These samples were used to optimize the selection and identification of the signal process, evaluate systematic uncertainties, and correct the data for detector inefficiency and resolution. The MC simulations were optimized to match the varying experimental conditions in the data collected in the 2015-2018 period. Predictions at parton level using fixed-order calculations are also made at different orders in QCD and EW higher-order corrections for the  VBF \hwwlnln signal process, and, together with the MC simulations at particle level for the signal process, are compared with the unfolded experimental cross-section measurement, as reported in Section~\ref{sec:result}.

Higgs boson production and decay into a pair of $W$ bosons were simulated for each of the four main production modes: ggF, VBF, and associated $WH$/$ZH$ productions. The Higgs boson production in association with a heavy quark pair (\ttH, \bbH) was found to give a negligible contribution. All Higgs boson samples were generated with a Higgs boson mass of 125 \GeV. In the following, the details of the production of VBF \hwwlnln signal and background samples from other Higgs boson production modes and SM background processes are given.

The signal and background events were processed with the \GeantFour~\cite{Agostinelli:2002hh} simulation of the ATLAS detector~\cite{SOFT-2010-01} and reconstructed using the same algorithms as used for the data.
The simulated energy deposits in the calorimeters and the simulated momenta are corrected using dedicated procedures detailed in Refs.~\cite{ATLAS:2022jjr,PERF-2017-01}.
Differences in lepton trigger, reconstruction and isolation efficiencies between detector-level simulation and data are corrected on an event-by-event basis using \pT- and $\eta$- dependent scale factors for each lepton~\cite{PERF-2015-10,PERF-2017-01}. The effect of multiple \pp interactions (pile-up) in the same or nearby bunch crossings is accounted for using inelastic \pp interactions generated by \PYTHIAV{8.186}~\cite{Sjostrand:2007gs} using the A3 tune~\cite{ATL-PHYS-PUB-2016-017} and the
\nnpdftwo parton distribution function (PDF) set~\cite{Ball:2012cx}. These inelastic \pp interactions were added to the signal and background samples that were processed with the ATLAS detector simulation and were weighted such that the distribution of the average number of reconstructed \pp interactions in simulation matches the one observed in the data.

\subsection{Signal predictions}
\label{sec:dataset_mc_signal}
 
The predictions for the signal VBF \hwwlnln processes, including contributions from $\W\rightarrow\tau\nu$ decays, were simulated using calculations that implement different approximations in QCD and EW higher-order corrections as well as different modeling of parton showering (PS), hadronization, and underlying event (UE). A summary of the simulations for the signal process is presented in Table~\ref{tab:MCSignalSamples} and further details of each generator are given below.

The default MC sample for signal events was generated with \PowhegvTwo~\cite{Nason:2004rx, Frixione:2007vw, Alioli:2010xd, Nason:2009ai}, interfaced with \PYTHIAV{8.230}~\cite{Sjostrand:2014zea}~to model the PS, hadronization, and UE\@. The dipole-recoil option was enabled in \PYTHIAV{8} to simulate the recoil of radiation in the VBF process and reduce hard and central radiation~\cite{ATL-PHYS-PUB-2019-007}. The PDF set \nnpdfnlo was used in the matrix-element (ME) calculations and the AZNLO tune~\cite{STDM-2012-23} of \PYTHIAV{8} parameters was used in the MC generation. The \EvtGen~v1.6.0 program~\cite{Lange:2001uf} was used for the properties of the bottom and charm hadron decays.
The \POWHEGva\ prediction is accurate to NLO in QCD corrections and tuned to match calculations with effects due to finite heavy-quark masses and soft-gluon resummations up to next-to-next-to-leading-logarithm (NNLL).
The MC prediction was normalized to a cross-section calculated up to the full NLO in QCD and EW couplings~\cite{Ciccolini:2007jr,Ciccolini:2007ec} with an additional approximate next-to-next-to-leading order (NNLO) QCD correction~\cite{Bolzoni:2010xr}.
The sample normalization accounts for the decay branching ratios calculated with \HDECAY~\cite{Djouadi:1997yw, Spira:1997dg, Djouadi:2006bz} and \PROPHECYF~\cite{Bredenstein:2006ha, Bredenstein:2006rh, Bredenstein:2006nk}.
Systematic variations were generated as event weights to account for uncertainties associated with PDFs (using the 30 variations of the \texttt{PDF4LHC15} PDF set~\cite{Butterworth:2015oua}), $\alpha_s$ (using 0.117 and 0.119 as variations), and higher-order effects (varying renormalization $\mu_R$ and factorization $\mu_F$ scales independently by factors of 0.5 and 2 with $0.5 \le \mu_F/ \mu_R \le 2$ as a constraint on their ratio).
This sample is referred to as \POWHEGva+\PYTHIAV{8}\ production.
 
A second MC signal sample was generated using the same events generated with \POWHEG, but interfaced to \HERWIGV{7.1.3}~\cite{Bahr:2008pv,Bellm:2015jjp} as an alternative model of PS, hadronization, and UE\@. The UE is modeled with \JIMMY~\cite{Butterworth:1996zw}, using the H7UE set of tuned parameters~\cite{Bellm:2015jjp} based on the \texttt{MMHT2014LO} PDF set~\cite{Harland-Lang:2014zoa}.  This sample is referred to as \POWHEGva+\HERWIGV{7} production.
 
A third MC signal sample was produced with the \MADGRAPHFive~\cite{Alwall:2014hca} generator interfaced to \HERWIGV{7.1.6} to assess effects related to the matching between the ME and the PS\@. The \MADGRAPHFive\, generator implements the CKKW scheme~\cite{Catani:2001cc} to remove overlaps between the ME and the PS and calculates the matrix element of the full process, including the Higgs boson decay by \MadSpin~\cite{Artoisenet:2012st}. This calculation is accurate to NLO in QCD corrections
and utilizes the \texttt{NNPDF30NLO}~\cite{Ball:2014uwa} PDF set. The \EvtGen~v1.7.0 program was used to simulate the bottom and charm hadron decays.
The H7UE set of tuned \HERWIGV{7} parameters and the \texttt{MMHT2014LO} PDF set  were used for the UE\@.
This sample is referred to as \MGFiveHw\, production.
 
Another set of three distinct predictions for the signal process were produced at LO and NLO accuracies in the strong and electroweak couplings using the \vbfnlo\, generator version 2.7.1~\cite{Arnold:2011wj}. In contrast to the other MC samples, the events generated by \vbfnlo\, were produced only at particle or parton levels. The momentum transfer of the exchanged gauge boson was used as the dynamic scale of the process, the Fermi constant $G_F$ was used as the EW scheme, and different NLO PDF sets were used in the event generation, such as \texttt{CT14}~\cite{Dulat:2015mca}, \texttt{MMHT14}~\cite{Harland-Lang:2014zoa}, and \nnpdfnlo~\cite{Ball:2014uwa}. The individual QCD and EW NLO correction factors relative to the LO cross-section ($k$-factors) are 0.81 and 0.99, respectively, in the considered phase space while their simultaneous application amounts to an overall 0.79 correction as it includes the combined QCD and EW contributions.
The ME calculation by \vbfnlo\, at LO was also interfaced with \PYTHIAV{8.244} to simulate PS, hadronization, and UE, using the A14 set of tuned parameters~\cite{ATL-PHYS-PUB-2014-021}.
In this event generation, the properties of the bottom and charm hadron decays were simulated using the \EvtGen~v1.7.0 program.
The parton-level fixed-order predictions calculated with \vbfnlo\, at LO and NLO in QCD and EW couplings are referred to as \vbfnlo$@$LO and \vbfnlo$@$NLO, respectively, while the \vbfnlo\, prediction at LO showered with \PYTHIAV{8} is referred to as \vbfnlo$@$LO+\PYTHIAV{8} production.
Finally, a correction was calculated for the \vbfnlo$@$NLO and  \vbfnlo$@$LO predictions to account for the missing $\tau$-lepton decays in this generator. Such a correction was applied multiplicatively in each bin of \vbfnlo$@$NLO kinematic distributions as the ratio of \vbfnlo$@$LO+\PYTHIAV{8} productions with and without $\tau$-lepton decays. This correction is on average 13\% and ranges between 5\% and 25\% in the selected phase space.
Three sources of uncertainties were accounted for in all \vbfnlo\, predictions, i.e., PDF, $\alpha_s$, and QCD scales, using the same procedures as in the \POWHEGva+\PYTHIAV{8}\ production.

\begin{table}[t!]
\fontsize{8.2}{10}\selectfont 
\centering
\caption{Summary of generators used for simulating the signal VBF \hwwlnln processes. The details and the corresponding references are provided in the body of the text. }
\label{tab:MCSignalSamples}
\begin{tabular}{c|ccccc}
\hline\hline
\noalign{\vspace{0.05cm}}
Simulation        &  Generator  &   ME          &   PDF    &   Shower           & UE \& PS \\
Name              &             &  Accuracy     &          & \& Hadronization   &  Parameter Set \\
\noalign{\vspace{0.05cm}}
\hline\hline
\noalign{\vspace{0.05cm}}
\POWHEGva+\PYTHIAV{8}   &    \PowhegvTwo           &   NLO QCD \& EW       &    \nnpdfnlo   &  \PYTHIAV{8.230}          &   AZNLO      \\
&                          &   + approx. NNLO QCD  &                 &  +\EvtGen~v1.6.0          &              \\
\POWHEGva+\HERWIGV{7}   &    \PowhegvTwo           &   NLO QCD \& EW       &    \nnpdfnlo   &  \HERWIGV{7.1.3}     &   H7UE      \\
&                          &   + approx. NNLO QCD  &                 &  +\EvtGen~v1.6.0 &             \\
\MGFiveHw               &    \MADGRAPHFive         &   NLO QCD, LO EW &    \texttt{NNPDF30NLO} &   \HERWIGV{7.1.6}     &   H7UE     \\
&                          &                       &                    &  \EvtGen~v1.7.0 &   \\
\vbfnlo$@$LO            &    \vbfnlo\, 2.7.1   &   LO QCD \& EW        &    \nnpdfnlo                       &  -                &   -     \\
&                     &                       &    \texttt{CT14}, \texttt{MMHT14}  &                   &              \\
\vbfnlo$@$NLO           &    \vbfnlo\, 2.7.1   &   NLO QCD \& EW       &    \nnpdfnlo                       &  -                &   -      \\
&                     &                       &    \texttt{CT14}, \texttt{MMHT14}  &                   &              \\
\vbfnlo$@$LO+\PYTHIAV{8}  &    \vbfnlo\, 2.7.1   &   LO QCD \& EW        &    \nnpdfnlo                       &  \PYTHIAV{8.244}  &   A14     \\
&                     &                       &    \texttt{CT14}, \texttt{MMHT14}  & +\EvtGen~v1.7.0   &  \\
\hline
\hline
\end{tabular}
\end{table}

\subsection{Background Monte Carlo samples}
\label{sec:dataset_mc_bkg}
 
The Higgs boson production and decay into a pair of $W$ bosons were simulated for the ggF and associated $WH$/$ZH$ production modes (collectively referred to as $VH$).
 
The simulation of the ggF Higgs boson production used the \Powheg\, method~\cite{Nason:2004rx, Frixione:2007vw, Alioli:2010xd} for merging the NLO Higgs boson + jet cross-section with the PS and the \MiNLO method~\cite{Hamilton:2012np, Campbell:2012am, Hamilton:2012rf} to simultaneously achieve NLO accuracy for the inclusive Higgs boson production. The PS and non-perturbative effects were simulated using \PYTHIAV{8.212}. In a second step, a reweighting procedure (\NNLOPS~\cite{Hamilton:2013fea, Hamilton:2015nsa}), exploiting the Higgs boson rapidity distribution, was applied using \HNNLO program~\cite{Catani:2007vq,Grazzini:2008tf} to achieve NNLO accuracy in the strong coupling constant.
The cross-section was normalized to next-to-next-to-next-to-leading order (N$^3$LO) in the strong coupling, and included NLO EW corrections~\cite{deFlorianUpdated:2016spz, Anastasiou:2016cez, Anastasiou:2015ema, Dulat:2018rbf, Harlander:2009mq, Harlander:2009bw, Harlander:2009my, Pak:2009dg, Actis:2008ug, Actis:2008ts, Bonetti:2018ukf}.
To assess the impact of the PS model and its matching to the ME, an independent sample of ggF events was simulated with \Powheg\ interfaced to \HERWIGV{7}.
 
Higgs boson production in association with a vector boson, $VH$, was simulated using \PowhegvTwo\ \MiNLO interfaced with \PYTHIAV{8} for PS and non-perturbative effects.
The prediction is accurate to NLO for the production of $VH$ plus one jet.
The MC prediction was normalized to cross-sections calculated at NNLO in the strong coupling with NLO electroweak corrections~\cite{Ciccolini:2003jy, Brein:2003wg, Brein:2011vx, Denner:2014cla, Brein:2012ne}.
 
The ggF Higgs boson samples were generated with \nnpdfnnlo PDFs in the ME while the $VH$ samples with \nnpdfnlo PDFs. Both sets of processes used the AZNLO tune of \PYTHIAV{8} parameters and the \EvtGen~v1.2.1 program~\cite{Lange:2001uf} to simulate the properties of the bottom and charm hadron decays. The normalizations account for the decay branching ratios calculated with \HDECAY\, and \PROPHECYF.

The main sources of other SM backgrounds include events from the production of dibosons, top-quark, $Z/\gamma^*$+jets, \Wjets, and multijets.
 
The QCD-induced $WW$ processes were simulated with the \SHERPAV{2.2.2}~\cite{Bothmann:2019yzt} generator and \nnpdfnnlo set of PDFs. These events originating from the scattering of two quarks or a quark and a gluon were simulated in the fully leptonic final states using matrix elements at NLO accuracy in QCD for up to one additional parton and at LO accuracy for up to three additional parton emissions.
To assess the impact of the PS model and its matching to the ME, independent samples of $WW$ events were simulated with \Powheg\, interfaced to the \HERWIGV{++} and \PYTHIAV{8} parton showers.
Samples for the loop-induced $gg \to VV$ processes (with $V$ indicating $W$ and $Z$ bosons) were generated with \SHERPAV{2.2.2} using a LO-accurate ME for up to one additional parton emission and \nnpdfnnlo set of PDFs. The $WW$ sample was normalized to the NLO $gg \to VV$ cross-section~\cite{Caola:2015psa} corresponding to a $k$-factor of 2.3.
The electroweak $WW$ production in the final state with two opposite charge leptons and two neutrinos in association with two jets was simulated using the \MADGRAPHFive\, generator with LO-accurate matrix elements and with the \nnpdfnlo PDF set interfaced to \PYTHIAV{8}.
 
Other QCD-induced diboson processes, i.e., $ZZ$ and $WZ/\gamma^*$, were simulated at NLO accuracy in QCD with \Powheg\, generator and the \texttt{CT10NLO} PDF set, interfaced to the \PYTHIAV{8} parton shower, except the events with final states with four charged leptons, and three charged leptons and a neutrino, which were simulated with \SHERPAV{2.2.2} using matrix elements at NLO accuracy in QCD for up to one additional parton and at LO accuracy for up to three additional parton emissions and the \nnpdfnnlo set of PDFs.
 
The production of $V\gamma$ final states was simulated with the \SHERPAV{2.2.8} generator. Matrix elements were calculated at NLO QCD accuracy for up to one additional parton and at LO accuracy for up to three additional parton emissions.
For all diboson samples, the ME calculations were matched and merged with the \SHERPA parton shower based on the 
dipole factorization~\cite{Gleisberg:2008fv,Schumann:2007mg} using the MEPS@NLO prescription~\cite{Hoeche:2011fd,Hoeche:2012yf,Catani:2001cc,Hoeche:2009rj} and the virtual QCD correction were provided by the \OpenLoops\ library~\cite{Cascioli:2013gfa}.
The default set of tuned parameters in \SHERPA was used for hadronization and UE activity, based on the \nnpdfnnlo set of PDFs.
 
The QCD-induced production of $Z/\gamma^*$+jets was simulated with the \SHERPAV{2.2.1} generator using NLO matrix elements for up to two partons and LO matrix elements for up to four partons calculated with the \Comix and \OpenLoops libraries~\cite{Cascioli:2011va,Denner:2016kdg}.
They were matched with the \SHERPA parton shower using the MEPS@NLO prescription and used the default set of tuned parameters in \SHERPA for hadronization and UE activity.
The \nnpdfnnlo set of PDFs was used and the samples were normalized to an NNLO prediction~\cite{Anastasiou:2003ds}. To assess the impact of the implementation of matrix element calculations and the matching to the PS, an alternative $Z/\gamma^*$+jets sample was simulated with \MADGRAPHFive\, interfaced to \PYTHIAV{8}.
Electroweak production of $\ell\ell jj$ final states was also generated with \SHERPAV{2.2.1}, but using LO matrix elements with up to four additional parton emissions.

The production of \ttbar\ events is modeled using the \powhegbox~v2 generator at NLO with the \nnpdfnlo PDF set and the \hdamp\ parameter\footnote{The \hdamp\ parameter is a resummation damping factor and one of the parameters that controls the matching of \Powheg\ matrix elements to the parton shower and thus effectively regulates the high-\pt\ radiation against which the \ttbar\ system recoils.} set to 1.5~\mtop~\cite{ATL-PHYS-PUB-2016-020}.
To correct for a known mismodeling of the leading lepton \pt due to missing higher-order corrections, an NNLO reweighting was applied to the sample~\cite{Czakon:2017wor}.
The events were interfaced to \PYTHIAV{8.230} to model the PS, hadronization, and UE, with parameters set according to the A14 tune.
The decays of bottom and charm hadrons were performed by \EvtGen~v1.6.0.
Samples were normalized using cross-sections calculated at NNLO+NNLL~\cite{Czakon:2013goa}.
To assess the impact of the PS model, an independent sample of \ttbar\ events was simulated with \Powheg\  interfaced to \HERWIGV{7}, while to assess the impact of the matrix element calculations and their matching to the PS, a sample simulated with \MADGRAPHFive\, interfaced to \PYTHIAV{8} was used.
 
The associated production of top quarks with $W$ bosons (mainly $tW$) is modeled using the \powhegbox~v2 generator at NLO in the strong coupling using the five-flavor scheme and the \nnpdfnlo set of PDFs.
A diagram removal scheme~\cite{Frixione:2008yi} was used to remove interference and overlap with \ttbar\ production.
The events were interfaced to \PYTHIAV{8.230} using the A14 parameter tune.
The decays of bottom and charm hadrons were performed by \EvtGen~v1.6.0.
 
The \Wjets and multijet backgrounds are estimated from data as detailed in Section~\ref{sec:bkg}, while MC samples were used to validate the methodology and evaluate uncertainties. For such studies, dedicated samples for \Wjets and $Z/\gamma^*$+jets processes were simulated with \Powheg\, \MiNLO with the \texttt{CT10NLO} PDF set, interfaced to \PYTHIAV{8} with the AZNLO tune of parameters.
As an alternative generator \MADGRAPHFive\, was used with \nnpdfnlo as PDF sets and interfaced to \PYTHIAV{8} tuned with the A14 parameter set.


\section{Event reconstruction and selection}
\label{sec:obj_evnt_reco}

The VBF $\hwwlnln$ event candidates were selected by requiring each event to have exactly one electron and one muon of opposite charge, in addition to at least two jets with specific kinematic configurations that enhance the VBF production mode. Events with a same-flavor lepton pair are not considered because they have a significant background from $Z/\gamma^*$+jets processes. The details of the selection requirements are described below.

\subsection{Event and object reconstruction}
\label{sec:obj_reco}
 
Candidate events were required to have at least one vertex with at least two associated tracks with \pT $>$ 500~\MeV. The vertex with the highest $\Sigma\pT^2$ of the associated tracks was considered to be the primary vertex.
 
Candidate events were recorded using a combination of single-lepton triggers and a dilepton $e$--$\mu$ trigger to maximize the total trigger efficiency~\cite{TRIG-2018-05,TRIG-2018-01}.
The requirement on the transverse momentum threshold for single-electron (single-muon) triggers was 24 (20) \GeV\ in 2015 together with loose isolation requirements. In the remainder of the Run 2 data-taking period, because of the higher instantaneous luminosities, the trigger thresholds of the single-lepton triggers were increased to 26 \GeV\ and more restrictive isolation (and identification for electrons) requirements were applied to both the lepton flavors. Additionally, single-lepton triggers
with higher \pT thresholds but with no isolation or with loosened identification criteria were used to increase the efficiency. The dilepton electron-muon trigger required a $\pT$ threshold of 17 \GeV\ for the electron and 14 \GeV\ for the muon.
 
Electron candidates are reconstructed through association of energy clusters in the electromagnetic calorimeter with well-reconstructed tracks that are extrapolated to the calorimeter~\cite{PERF-2017-01}. Electrons are required to satisfy $|\eta| < 2.47$, excluding the transition region $1.37 < |\eta| < 1.52$ between the barrel and end-caps in the LAr calorimeter.
Muon candidates are reconstructed from a global fit of matched tracks from the inner detector and the muon spectrometer~\cite{PERF-2015-10,MUON-2018-03}, and are required to satisfy $|\eta| < 2.5$.
 
To reject particles misidentified as prompt leptons, several identification requirements as well as isolation and impact parameter criteria~\cite{EGAM-2018-01, PERF-2015-10,MUON-2018-03} are applied.
For electrons, a likelihood-based identification method~\cite{EGAM-2018-01} is employed, which uses several discriminating variables such as shower shapes, track properties, and track-cluster-matching. Electrons must satisfy the quality requirements described in Ref.~\cite{EGAM-2018-01} that vary according to their transverse momenta: the \textit{Tight} working point in the electron \pT range of 15--25 \GeV, which has an efficiency of approximately 70\% for electrons in this energy range, and
the \textit{Medium} working point for electrons with $\pT > 25$ \GeV, which has an efficiency of approximately 85\% for an electron with $\pT \sim 40$ \GeV.
For muons, a cut-based identification method~\cite{MUON-2018-03} is employed and the \textit{Tight} quality criterion, as defined in Ref.~\cite{,MUON-2018-03}, is applied to combined tracks with $\pT > 15$ \GeV\ with an efficiency of $\sim 93\%$.
The impact parameter requirements are $|z_0\sin\theta| < 0.5$ mm and $|d_0|/\sigma_{d_0} < 5\ (3)$ for electrons (muons)\footnote{The parameters $z_0$ and $d_0$ are the longitudinal and transverse impact parameters, respectively, defined in terms of the point of closest approach between the associated track and the primary vertex.}.
Leptons are also required to be isolated using information from inner detector  tracks and energy clusters in the calorimeters in a cone around the lepton. The expected muon (electron) isolation efficiency is at least 89\% ($75$\%) with \pT (\eT) of 20 -- 25 \gev and it reaches 90\% or greater efficiency for  \pT  (\et) greater than 40 \gev, using the working points defined in Refs.~\cite{EGAM-2018-01, PERF-2015-10,MUON-2018-03}.
At least one of the offline reconstructed leptons must be matched to an online lepton candidate that triggered the recording of the event.
In the case where the dilepton trigger is solely responsible for the recording of the event, each lepton must correspond to one of the trigger objects.
This trigger matching scheme also requires the $\pT$ of the offline lepton to be at least 1 \GeV\ above the trigger-level threshold.
 
Jets are reconstructed within the calorimeter acceptance using the anti-$k_t$ jet clustering algorithm using the FastJet code~\cite{Cacciari:2008gp,Fastjet} with a radius parameter of $R = 0.4$ and particle flow objects as input~\cite{PERF-2015-09}, which use tightly reconstructed tracks associated with the primary vertex.
The energy of the jets is corrected for the non-compensating calorimeter response, noise threshold effects, energy loss from inactive material, and pile-up contamination. As part of the jet energy calibration a pile-up correction based on the concept of jet area is applied to the jet candidates~\cite{JETM-2018-05}. A kinematic preselection of $\pt > 20~\GeV$ and $|\eta| <  4.5$ is applied on the jets.
 
The jet-vertex-tagger (JVT) multivariate discriminant selection~\cite{PERF-2014-03} is used to separate hard-scatter jets from pile-up jets within the acceptance of the inner detector of $|\eta| < 2.5$ for jets with $20 < \pt <  60~\GeV$, by utilizing calorimeter and tracking information.
 
Within the inner detector acceptance, jets with $\pT > 20~\GeV$ originating from the fragmentation of $b$-hadrons ($b$-jets) are identified using a deep-learning neural network,
known as DL1r~\cite{FTAG-2018-01,PERF-2012-04}, based on several lower-level taggers, which use relevant quantities such as the associated track impact parameters and information from secondary vertices. The chosen $b$-tagging operating point has an efficiency of 85\% for selecting jets containing $b$-hadrons, as estimated from a sample of simulated \ttbar events and validated with data.
 
An overlap removal procedure is applied to ensure that jets, electrons, and muons are not double counted.
If two electrons share calorimetric energy clusters or a track, the lower-$\pT$ electron is removed. If a muon shares an ID track with an electron, the electron is removed.
For electrons and jets, the jet is removed if $\Delta R$(jet, $e$) $<$ 0.2 and the jet is not tagged as a $b$-jet. For any surviving jets, the electron is removed if $\Delta R$(jet, $e$) $<$ min(0.4, 0.04 + 10 \GeV/$\pT^{e}$).
For muons and jets, the jet is removed if $\Delta R$(jet, $\mu$) $<$ 0.2, the jet is not tagged as a $b$-jet and the jet has less than three associated tracks with $\pT$ $>$ 500 \MeV\ or both the following conditions are met: the $\pT$ ratio of the muon and the jet is larger than 0.5 ($\pT^{\mu} / \pT^{\rm{jet}}>0.5$) and the ratio of the muon $\pT$ to the sum of $\pt$ of tracks with $\pT$ $>$ 500 \MeV\ associated to the jet is larger than 0.7.
For any surviving jets, the muon is removed if $\Delta R$(jet, $\mu$) < min(0.4, 0.04 + 10 \GeV/$\pT^{\mu}$).

The missing transverse momentum $\vmet$ (with magnitude \met) is defined as the negative vector sum of the $\pt$ of all the selected leptons and jets together with reconstructed tracks (referred to as \textit{soft term}) that are not matched to these objects but are consistent with originating from the primary vertex of the $pp$ collision~\cite{PERF-2016-07}.

\subsection{Event selection}
\label{sec:evt_sel}
 
The event selection was optimized to select VBF $\hwwlnln$ event candidates and comprises a preselection, applied commonly in the signal region (SR) and in the control regions (CRs) used to evaluate the backgrounds, defined in Section~\ref{sec:bkg}, followed by a dedicated SR selection.
The event preselection and SR selection requirements based on event kinematics are summarized in Table~\ref{tab:EvtSelection}.
At the preselection level, events are first required to have exactly two\footnote{While the transverse momentum and pseudorapidity requirements remain the same, the isolation criteria for additional leptons that are vetoed is looser than the nominal working point used for the two leading leptons and in the case of additional electrons or muons, a \textit{Medium} identification quality~\cite{MUON-2018-03} is used.}
different-flavor and opposite-sign charged leptons ($e$, $\mu$) to reduce the background from diboson processes. The highest-$\pT$ (leading) lepton is required to have $\pT > 22$ \GeV\ and the subleading lepton $\pT > 15$ \GeV. The invariant mass $\mll$ of the lepton pair is required to be greater than 10 \GeV\ to suppress background contributions from the $Z/\gamma^*$+jet and hadronic resonances.
In addition, at least two jets must be reconstructed in the event with $\pT > 30~\GeV$ and $|\eta|<4.5$. Among them, the two $\pT$-leading jets are tagged as originating from the VBF process.
To reject background from top-quark production, events containing $b$-jets with $\pT > 20~\GeV$ are vetoed.
 
In addition to those selection criteria, further requirements are applied to define the SR\@.
A veto on the background from $Z/\gamma^*\rightarrow \tau\tau$ production is implemented by requiring $\mtt < m_{Z} - 25$ \GeV, where $m_{Z}$ is the $Z$ boson invariant mass (set to 91.1876 \GeV) and $\mtt$ is the invariant mass of the hypothetical $\tau$-lepton pair. This is calculated in the collinear approximation using the direction and magnitude of the measured missing transverse momentum and projecting it along the direction defined by the reconstructed charged lepton system~\cite{Plehn:1999xi}.
To enhance the contribution of events with VBF topology, events are rejected if they contain either additional jets with $\pT > 20~\GeV$ that lie in the rapidity interval spanned by the dijet system (\textit{central jet veto}, or CJV) or either lepton is outside the rapidity interval spanned by the dijet system (\textit{outside lepton veto}, or OLV).
To further suppress the background contamination and enhance the selection of the Higgs boson events in the SR, the invariant mass of the system of the two VBF-tagged jets (\mjj)  is required to be greater than 450 \GeV, the absolute difference of the rapidity of the two tagged jets (|\dyjj|) must be greater than 2.1, and the absolute difference of the azimuthal angles of the two selected leptons (|\dphill|) must be less than 1.4 rad.

\begin{table}[t!]
\centering
\caption{
Kinematic event selection criteria that define the signal region and the fiducial region. The central horizontal double line discriminates between preselection (top) and selection (bottom) requirements for the signal region. In the table $\ell$ stands for $e$ or $\mu$, and the definitions of lepton identification and isolation in the signal region are detailed in Section~\ref{sec:obj_reco}. For the fiducial region the leptons are dressed as detailed in Section~\ref{sec:fid_phasespace_obs}. The overlap removal procedure between leptons and jets in the signal region is detailed in Section~\ref{sec:obj_reco}. }
\label{tab:EvtSelection}
\begin{tabular}{l|c|c}
\hline\hline
Selection Requirements                                   & Signal Region                                & Fiducial Region\\
\hline\hline
Lepton pair flavors                                     &  \multicolumn{2}{c}{$e$-$\mu$} \\
\hline
Lepton pair charge                                       &  \multicolumn{2}{c}{0} \\
\hline
Leading (subleading) lepton $\pT$                        &  \multicolumn{2}{c}{$>$ 22 \GeV\ ($>$ 15 \GeV)} \\
\hline
\multirow{4}{*}{Lepton $\eta^{\ell}$}                     &  \multicolumn{2}{c}{$|\eta^{\mu}| < 2.5$}\\
&  $0 < |\eta^{e}| < 1.37$                    &   \multirow{3}{*}{$|\eta^{e}| < 2.5$}  \\
&        or                        & \\
&  $1.52 < |\eta^{e}| < 2.47$      & \\
\hline
No. of additional leptons                                &  \multicolumn{2}{c}{0} \\
\hline
$\Delta R(\ell,\ell)$                                    &  overlap removal                       &   $>$ 0.1 \\
\hline
$\mll$ 	                                                 &  \multicolumn{2}{c}{$> 10 ~\GeV$} \\
\hline
$\Delta R(\ell, \textrm{jet})$                           &  overlap removal                       &   $>$ 0.4 \\
\hline
No. of jets ($\pT > 30 ~\GeV$, $|\eta| < 4.5$)           &  \multicolumn{2}{c}{$\ge 2$} \\
\hline
No. of $b$-jets ($\pT > 20 ~\GeV$, $|\eta| < 2.5$)       &  \multicolumn{2}{c}{0} \\
\hline
\hline
$\mtt$                                                   & \multicolumn{2}{c}{$< m_Z - 25~\GeV$} \\
\hline
Central jet veto ($\pT > 20 ~\GeV$)                      & \multicolumn{2}{c}{\checkmark} \\
\hline
Outside lepton veto                                      & \multicolumn{2}{c}{\checkmark} \\
\hline
$\mjj$                                                   & \multicolumn{2}{c}{$>$ 450 \GeV} \\
\hline$|\dyjj|$                                                & \multicolumn{2}{c}{$>$ 2.1} \\
$|\dphill|$                                              & \multicolumn{2}{c}{$<$ 1.4 rad} \\
\hline
\hline
\end{tabular}
\end{table}


\section{Fiducial phase space, acceptance and measured observables}
\label{sec:fid_phasespace_obs}

The integrated and differential cross-sections for the VBF \hwwlnln process are evaluated in a fiducial phase space defined by particle-level kinematic criteria for leptons and jets that closely follow those applied in the experimental SR event selection.  The fiducial phase space is defined in Table~\ref{tab:EvtSelection}.
Only stable particles (with a mean lifetime $c\tau > 10$ mm) are considered in the fiducial region. Electrons and muons are required to satisfy this criterion and not to originate from hadrons or hadron decay products.  Their respective momenta after quantum-electrodynamics (QED) final-state radiation are vectorially added to the momenta of photons emitted in a cone of size $\Delta R = 0.1$ around the lepton direction to form \textit{dressed} charged leptons.  Photons that originate from hadron decays are excluded. A minimal separation of $\Delta R = 0.1$ is required between the two charged leptons.
Final-state particles with lifetimes greater than 30 ps are clustered into jets (referred to as particle-level jets) using the same algorithm as in detector-level jets, i.e., with the anti-$k_t$ algorithm with radius parameter $R = 0.4$. The selected charged leptons and neutrinos from \W-boson decays are not included in the jet clustering.
Events containing $b$-tagged jets with $\pT > 20$~\GeV, and $|\eta| < 4.5$ are vetoed. A jet is identified as $b$-tagged if it is ghost-associated~\cite{Cacciari:2008gn} with one or more weakly decaying $b$-hadrons with \pT $>$ 5 \GeV.
Events are vetoed if a selected jet is closer than $\Delta R = 0.4$ to a selected lepton.

The integrated fiducial cross-section is defined as
 
\begin{equation}
\sigma^{\mathrm{fid}} = \frac{N_{\rm data}^{\text{SR}} - N_{\rm bkg}^{\text{SR}}}{C\, \times {\cal L} }
\label{eq:xSectFid}
\end{equation}
 
where the numerator is the number of signal events in the SR, defined as the number of events observed in data ($N_{\rm data}^{\text{SR}}$) after the total number of estimated background events ($N_{\rm bkg}^{\text{SR}}$) is subtracted away, ${\cal L}$ is the integrated luminosity, and $C$ is a factor that accounts for detector inefficiencies.  The factor $C$ is estimated in simulation as the ratio of the number of signal events with one electron and one muon (including those from $\tau$-lepton decays) passing the selection requirements at detector level as listed in Section~\ref{sec:obj_evnt_reco} to those passing the fiducial selection (including $\W\rightarrow\tau\nu$ decays) at particle level.
The factor $C$ has a value of 0.40 with an uncertainty of 2.4\%, including experimental and theoretical sources, as detailed in Section~\ref{sec:syst}. The numerical value of the factor $C$ is driven by the selections on lepton and jet kinematics.
 
The integrated fiducial cross-section can be extrapolated to the fully inclusive VBF Higgs boson production cross-section $\sigma^{\mathrm{tot}}$ as defined in Ref.~\cite{deFlorianUpdated:2016spz}, with no Higgs or \W boson decays and no kinematic requirements. It is obtained by dividing the integrated fiducial cross-section by an acceptance factor $A$ such that $\sigma^{\mathrm{tot}}= \sigma^{\mathrm{fid}} / A$.  Conversely, this formula can be used as a definition of the factor $A$. Thus, the numerator of the factor $A$ is calculated using the integrated fiducial cross-section predicted by \POWHEGva+\PYTHIAV{8}\, scaled such that $\sigma^{\mathrm{tot}}$ matches the value calculated in Ref.~\cite{deFlorianUpdated:2016spz}.
The uncertainty of the integrated fiducial prediction includes the QCD scale, PDF, $\alpha_s$, and generator and parton shower model uncertainties, as detailed in Section~\ref{sec:syst}. The denominator of the factor $A$ is the fully inclusive cross-section, which is calculated together with its uncertainty in Ref.~\cite{deFlorianUpdated:2016spz} in the QCD NNLO and EW NLO approximation. In these approximations, the factor $A$ is $(5.5 \pm 0.3) \times 10^{-4}$.
 
The differential cross-sections are measured in the fiducial phase space as functions of variables sensitive to the Higgs boson production and decay, using an unfolding method detailed in Section~\ref{sec:method_diff}. They include the transverse momentum of the leading (highest-\pT) and sub-leading (second-highest-\pT) jets, and charged leptons (\pTjone, \pTjtwo, \pTlone and \pTltwo, respectively), the dijet and dilepton invariant masses for the two leading jets and charged leptons, respectively, the absolute rapidity intervals spanned by the two leading jets (|\dyjj|) and by the two leading charged leptons (|\dyll|), the azimuthal intervals spanned by the two leading charged leptons (|\dphill|) and by the more forward ($\phi_{\rm{fwd}}$) and more central ($\phi_{\rm{central}}$) jets ($\dphijj= \phi_{\rm{fwd}}-\phi_{\rm{central}}$), the cosine of the polar angle ($\theta^*_{\eta}$) defined by the two leading charged lepton directions relative to the beam direction in a frame where the two charged leptons are back-to-back in the $r-\theta$ plane~\cite{Banfi:2010cf}, the \pT of the dilepton system (\pTll), and the Higgs boson transverse momentum (\pTH) calculated as the magnitude of the vectorial \vpT sum of charged leptons and neutrinos from the \W boson decay. 
The distributions of $\pT^{H}$ and $\pT^{\ell\ell}$ provide tests of perturbative and non-perturbative QCD calculations and are sensitive to the structure of the Higgs boson interactions. The leptonic distributions of $\pTlone$, $\pTltwo$, and $\mll$ are sensitive to properties of the Higgs boson decays and can be sensitive to beyond-SM physics. Three angular variables of the charged leptons are also measured: $|\dyll|$, $|\dphill|$, and  $\theta^*_{\eta}$.
The leptonic variables are also of interest for their sensitivity to the spin and parity of the Higgs boson, as well as to higher-order EW corrections to the Higgs boson decay.
Variables related to jets probe the VBF production mechanisms of the Higgs boson, QCD radiation effects and are sensitive to contributions from beyond-SM physics. The jet variables include $\pTjone$, $\pTjtwo$, $|\dyjj|$, and $\mjj$. The signed angle in the transverse plane of the two jets, $\dphijj$, is also a test of the spin, charge (C), and parity (P) of the Higgs boson.


\section{Background estimate}
\label{sec:bkg}

 
Several processes contribute to the background contamination in the SR\@. These include top-quark pair ($\ttbar$) and single-top-quark ($Wt$) production collectively referred to as top-quark background, non-resonant $WW$ production, other diboson ($WZ$, $ZZ$, $W\gamma$, $W\gamma^*$) production, and $Z/\gamma^*$+jets (mainly $Z/\gamma^*\rightarrow \tau\tau$) events. Background processes with mis-identified charged leptons are denoted in the following as \textit{mis-Id} and comprise $W$+jets, in which one jet is mis-identified as a lepton, and the smaller contribution from multijet events with more than one mis-identified lepton.
Higgs boson production through mechanisms other than VBF is also considered as background.
The contamination from major background processes, such as top-quark, non-resonant $VV$, $Z/\gamma^*$+jets, and mis-Id as well as ggF Higgs boson production, is estimated from the fit to data in the SR or simultaneously in the SR and dedicated CRs, while the remaining (minor) background processes are modeled with MC simulation.
 
The backgrounds from top-quark and non-resonant $VV$ events have very similar properties, therefore their combined yield is obtained from the data while their relative contributions are estimated from MC simulation. The common yield of top-quark plus non-resonant $VV$ background is determined in a background-enriched subset of the SR that provides enough events (45\%, 20\%, and 5\%
contributions for top-quark, non-resonant $WW$, and other diboson processes, respectively). A dedicated multivariate analysis technique provides discrimination relative to other processes, as discussed in Section~\ref{sec:method}.  Alternative methods to estimate the top-quark and non-resonant $VV$ processes contaminations were studied, such as defining independent CRs and accounting for independent normalization factors for these two classes of processes. They yielded consistent results but with larger model dependence owing to the extrapolation from kinematic regions slightly different from the SR\@.
 
The background contamination in the SR from $Z/\gamma^*$+jets events is determined from a dedicated CR ($Z/\gamma^*$+jets CR) that is chosen to be kinematically close to the SR, except for the requirements on \dphill\, and \dyjj\,, which are replaced by requirements on the invariant masses of the reconstructed $\ell$-$\ell$ and $\tau$-$\tau$ systems, i.e., $\mll<80 ~\GeV$ and 66.2 \GeV\ < $\mtt$ < 116.2 \GeV\, to enhance the contribution from $Z/\gamma^*\rightarrow \tau\tau$ events. The $Z/\gamma^*$+jets CR has a purity of about 74\%.

The ggF Higgs boson production yield is extracted using a control region (ggF CR) with the same selection as the SR except that the CJV or the OLV but not both failed.
The purity of the ggF Higgs boson process in the ggF CR is approximately 2\% and is enhanced with a dedicated multivariate analysis technique discussed in Section~\ref{sec:method}. Although the ggF CR has a low purity in ggF Higgs boson production, together with the SR it allows the minimization of the modeling uncertainty associated with this background source, thanks to the complementary information it provides on this and other processes. The choice of such a ggF CR aims at reducing the impact of the modeling uncertainty originating from higher-order corrections in the estimate of the ggF Higgs boson background in the SR, at a small expense of a slightly larger statistical uncertainty in the ggF Higgs boson yield in the ggF CR\@. It was verified that this selection of the ggF CR has only a small extrapolation uncertainty in the ggF production yield into the SR\@.

The yields of top-quark plus non-resonant $VV$, $Z/\gamma^*$+jet, and ggF Higgs boson productions are extracted in a simultaneous fit in the SR, the $Z/\gamma^*$+jets CR, and the ggF CR, as described in Section~\ref{sec:method}.
Modeling uncertainties for the backgrounds together with other sources of uncertainties, as presented in Section~\ref{sec:syst}, are included as nuisance parameters in the fit.
 
The mis-Id background is estimated by using a data-driven technique. A CR enriched in $\W$+jets events ($\W$+jets CR) is defined by applying the same kinematic requirements as those used in the SR, except for the |\dphill| $<$ 1.4 rad requirement.
Moreover, the identification criteria for one of the two lepton candidates are modified such that one lepton is required to fail the nominal identification criteria and satisfy a looser set of requirements, and is referred to as an \textit{anti-identified} lepton. In the $\W$+jets CR, the expected contribution of mis-Id background is about 76\% of the selected events.
After subtracting the expected contribution from processes with two prompt charged leptons, the expected yield from the mis-Id background in the SR is extrapolated from the observed number of events in the $\W$+jets CR by applying the full event selection on the mis-Id estimate~\cite{HIGG-2021-20}.
The extrapolation factor is determined in a sample enriched with $Z/\gamma^*$+jets events, where the $Z/\gamma^*$ boson decays to a same-flavor pair of electrons or muons, and an additional lepton candidate recoils against the $\Z/\gamma^*$ boson.
The extrapolation factor is defined as the ratio of the numbers of events in which the additional lepton candidate is identified and anti-identified, and is measured in bins of lepton $\pT$ and $|\eta|$.
A correction factor, calculated in MC simulations, accounts for the different compositions of sources of mis-identified charged leptons (such as hadrons, non-prompt leptons from heavy-flavor decays and photons) between the $Z/\gamma^*$+jets and the $\W$+jets samples.
The contribution from processes with two mis-identified charged leptons are accounted for in the extrapolation by applying a correction term evaluated in a sample where both the lepton candidates are anti-identified. The contribution originating from two mis-identified charged leptons accounts for about $20\%$ of the total mis-identified lepton yield in the SR.
The same extrapolation factors and corrections are implemented to estimate the mis-Id background contribution in the SR, $Z/\gamma^*$+jets CR, and ggF CR samples.
 
The observed numbers of events in the signal and control regions are shown in Table~\ref{tab:SR-CR_events}.
 
\begin{table}[!ht]
\centering
\caption{Expected and observed (pre-fit) event yields in the SR as well as in the $\Z/\gamma^*$+jets and ggF CRs, for an integrated luminosity of 139 $\ifb$ collected at $\sqrt{s}$ = 13 \TeV. The quoted uncertainties correspond to the statistical uncertainty, together with the experimental and theory modeling systematic uncertainties. The sum of all the contributions may differ from the total value because of rounding.}
\label{tab:SR-CR_events}
\begin{tabular}{
l
r@{\(\,\pm\,\)}l
r@{\(\,\pm\,\)}l
r@{\(\,\pm\,\)}l
}
\hline
\hline
\noalign{\vspace{0.05cm}}
Sample  &    \multicolumn{2}{c}{SR}   &  \multicolumn{2}{c}{$\Z/\gamma^*$+jets CR}   &   \multicolumn{2}{c}{ggF CR}       \\
\hline\hline
\noalign{\vspace{0.05cm}}
Signal (\POWHEGva+\PYTHIAV{8}) &    \multicolumn{2}{c}{$110$}                   &  \multicolumn{2}{c}{$\phantom{0}13$}      &  \multicolumn{2}{c}{$\phantom{000}86$}                \\
ggF Higgs                      &    \multicolumn{2}{c}{$\phantom{0}39$}      &  \multicolumn{2}{c}{$\phantom{00}4$}       &  \multicolumn{2}{c}{$\phantom{00}450$}                 \\
Other Higgs                    &    \multicolumn{2}{c}{$\phantom{00}3$}       &  \multicolumn{2}{c}{$\phantom{0}10$}      &  \multicolumn{2}{c}{$\phantom{000}78$}                \\
Top                            &    \multicolumn{2}{c}{$420$}                   &  \multicolumn{2}{c}{$\phantom{0}41$}      &  \multicolumn{2}{c}{{$11\,000$}}                \\
$Z/\gamma^*$+jets              &    \multicolumn{2}{c}{$\phantom{0}79$}      &  \multicolumn{2}{c}{$320$}                   &  \multicolumn{2}{c}{$\phantom{0}1\,400$}                \\
$VV$                           &    \multicolumn{2}{c}{$280$}                  &  \multicolumn{2}{c}{$\phantom{0}32$}      &  \multicolumn{2}{c}{$\phantom{0}4\,300$}                \\
$V\gamma$                      &    \multicolumn{2}{c}{$\phantom{0}13$}      &  \multicolumn{2}{c}{$\phantom{0}14$}      &  \multicolumn{2}{c}{$\phantom{00}210$}                \\
Mis-Id                         &    \multicolumn{2}{c}{$\phantom{0}47$}      &  \multicolumn{2}{c}{$\phantom{0}12$}     &   \multicolumn{2}{c}{$\phantom{00}810$}             \\
\hline
Total Signal+Background       & $\phantom{0000}1\,000$  &  $120$ &  \multicolumn{2}{c}{$\phantom{00000}450\pm160$} &  $\phantom{000000}18\,800$  &  $2\,600$        \\
\hline
Data   &  \multicolumn{2}{c}{$916$}     & \multicolumn{2}{c}{$406$}      &  \multicolumn{2}{c}{$18\,228$}                  \\
\hline
\hline
\end{tabular}
\end{table}


\section{Cross-section determination}
\label{sec:method}

 
This section details the method employed to measure the integrated and differential fiducial cross-sections. The yields for the Higgs boson signal and the background processes are extracted from fits to dedicated discriminant distributions in data in the SR and two orthogonal CRs, i.e., the $Z/\gamma^*$+jets and the ggF CRs, defined in~\cref{sec:obj_evnt_reco,sec:bkg}.
This procedure allows a simultaneous measurement of the signal-process cross-section and the contributions from the dominant background processes.
 
The integrated and differential cross-sections for the VBF \hwwlnln process are extracted in the fiducial phase space defined in Section~\ref{sec:fid_phasespace_obs} by minimizing the negative logarithm of a binned profile-likelihood~\cite{Cowan:2010js} that includes corrections for detector inefficiency and resolution effects. The shapes of signal and irreducible background processes, i.e., \ttbar and non-resonant $VV$, $Z/\gamma^*$+jets, and ggF Higgs boson, in the discriminant distributions are obtained from the simulated samples described in Section~\ref{sec:dataset_mc}. Their yields are determined in the SR from the fit, while the yield of the reducible mis-Id background is estimated in data from a dedicated control region as described in Section~\ref{sec:bkg}. Its overall normalization and shape are constrained within systematic uncertainties in the fit.
 
The extended likelihood \textit{L} consists of a product of likelihood functions ($\mathcal{P}$s) of the SR yield and those of the two CRs:
\begin{equation}
\begin{split}
\label{eqn:likelihood}
\textit{L} = & \mathcal{P}(N_{\text{data}}^{\text{SR}}| N_{\text{signal}}^{\text{SR}} +  \sum_{k=1}^{n_{\mathrm{bkg}}} \mu_{\text{bkg-}k}  \cdot N_{\text{bkg-}k}^{\mathrm{SR}} + \sum_{j=1}^{m_{\mathrm{bkg}}} N_{\text{bkg-}j}^{\mathrm{SR}}) \cdot \\
& \prod_{n=1}^{2} \mathcal{P}(N_{\text{data}}^{\text{CR-}n}| N_{\text{signal}}^{\text{CR-}n} + \sum_{k=1}^{n_{\mathrm{bkg}}} \mu_{\text{bkg-}k}\cdot N_{\mathrm{bkg-}k}^{\mathrm{CR-}n} + \sum_{j=1}^{m_{\mathrm{bkg}}} N_{\mathrm{bkg-}j}^{\mathrm{CR-}n}) \cdot \prod_{i=1}^{n_{\theta_i}}\mathcal{N}(\tilde{\theta_i}| \theta_i)
\end{split}
\end{equation}
 
where $N_{\text{data}}^{\text{SR}}$ and $N_{\text{data}}^{\text{CR-}n}$ are the numbers of events observed in data in the signal region and control region CR-$n$ (the ggF CR or the $Z/\gamma^*$+jets CR), respectively, and $N_{\text{signal}}^{\text{SR}}$ and $N_{\text{signal}}^{\text{CR-}n}$ are the expected signal yields in the SR and CR-$n$, respectively. The parameter $\mu_{\text{bkg-}k}$ is the normalization of background process $k$ and $N_{\mathrm{bkg-}k}^{\text{SR}}$ ($N_{\mathrm{bkg-}j}^{\text{SR}}$) and $N_{\mathrm{bkg-}k}^{\text{CR-}n}$ ($N_{\mathrm{bkg-}j}^{\text{CR-}n}$) are the expected yields of the background contribution $k$ ($j$) in the SR and CR-$n$, respectively. The parameter $n_{\mathrm{bkg}}$ is the number of background processes whose normalizations are left floating in the likelihood functions and are determined from the fit to data, while $m_{\mathrm{bkg}}$ is the number of background processes whose yields are obtained from MC simulation or from control samples in data not included in the simultaneous fit. The $\mathcal{N}$ term is a normal function that parameterizes the systematic effect of type $i$ as a function of a constrained nuisance parameter $\theta_i$ and the associated estimate of the corresponding effect $\tilde{\theta_i}$, see Section~\ref{sec:syst}.
The parameter $n_{\theta_i}$ represents the number of parameters modeling the systematic uncertainties considered in the analysis. The likelihood functions in the SR and CRs are defined as Poisson probability density functions~\cite{Cranmer:1456844} built from binned histograms of discriminant distributions (also referred to as templates) using MC events for signal and backgrounds, except the mis-Id background, for which a data-based estimate is used.
The signal and background templates are also functions of the nuisance parameters~\cite{Cranmer:1456844} such that systematic uncertainties can affect the discriminator distribution shapes for the signal process and irreducible backgrounds whose yields are extracted from the fit, and the normalization for the backgrounds whose yields are fixed to SM predictions. This methodology accounts for correlations of systematic uncertainties between signal and background estimates and between bins of the measured observables.
 
Table~\ref{tab:SR-CR_events} shows that a method based on event counting would lead to a poor sensitivity to the signal process. Therefore discriminant distributions are used to exploit differences in kinematics between processes. Multivariate discriminants, based on a boosted decision tree (BDT) algorithm~\cite{TMVA}, are used to further enhance the sensitivity to signal events and ggF Higgs boson events in the SR and the ggF CR, respectively.
In the SR, the template corresponds to a BDT score distribution optimized to discriminate between the VBF signal and the top+$VV$ background.
The transverse mass (\mT) distribution is used as the template in the $Z/\gamma^*$+jets CR, while a dedicated BDT score distribution is used in the ggF CR\@. The BDTs are trained and optimized on simulated events in the SR or the ggF CR as detailed in Section~\ref{sec:BDTs}.
 
Templates are smoothed to minimize statistical fluctuations in the MC samples, using the Gaussian distribution kernel density approximation implemented in the RooKeysPdf algorithm in the RooFit package~\cite{Cranmer:2000du}.
 
The normalization of the sum of top-quark and non-resonant $VV$ processes ($\mu_{\text{top}+VV}$) is determined in the signal region by  leveraging the discrimination power of the multivariate discriminants.
The inclusion of the two control regions in the fit allows the normalizations of the $Z/\gamma^*$+jets ($\mu_{\text{\Zjets}}$) and ggF Higgs boson ($\mu_{\text{ggF}}$) backgrounds  to be constrained by data in the SR as well as in their respective CRs\@.
The normalization parameters of all other processes are fixed to their respective theoretical predictions, or to the data-driven estimate for the case of the mis-Id background.
In the ggF CR, the normalization parameters for the top-quark plus $VV$ template
($\mu_{\text{top}+VV, \text{ggF}}$) and the $Z/\gamma^*$+jets template ($\mu_{\text{\Zjets}, \text{ggF}}$) are left floating and fit to data as parameters independent from those used in the other regions for the same processes. This is done to minimize the modeling dependence in the control and signal regions. The normalizations of all other processes are kept fixed as in the case of the $Z/\gamma^*$+jets CR\@.

\subsection{Multivariate discriminants}
\label{sec:BDTs}

 
Several BDTs are utilized in the analysis and are trained independently 
to discriminate between one process type against others. The BDT algorithms implemented in the TMVA package~\cite{TMVA} are used. Their scores are in the [$-1$, 1] range and reflect the compatibility between an event and the process for which the BDT is trained, with the purity increasing as the score value increases. The BDTs were trained on MC samples and tested on statistically independent MC events. The BDT hyperparameters were tuned to maximize the area under the receiver operating characteristic (ROC) curve while avoiding overtraining by ensuring that the Kolmogorov-Smirnov test score between training and test samples was high. For each BDT, several variables were studied to be used as inputs to the BDT training, but only variables significantly improving the discrimination power (i.e., area under the ROC curve) were retained.
For each BDT a minimum set of variables that balances high discrimination and good modeling was chosen (high scores in the Kolmogorov-Smirnov test).  The BDT score bin boundaries and the number of bins were also optimized to balance between high signal significance and small model dependence. The modeling by MC simulations for a representative set of distributions is illustrated in Figures~\ref{fig:Det_VBFbdtInput} and ~\ref{fig:Det_ggFCRbdtInput}.
\begin{figure}[ht]
\centering
\subfloat[]{\includegraphics[width=0.45\linewidth]{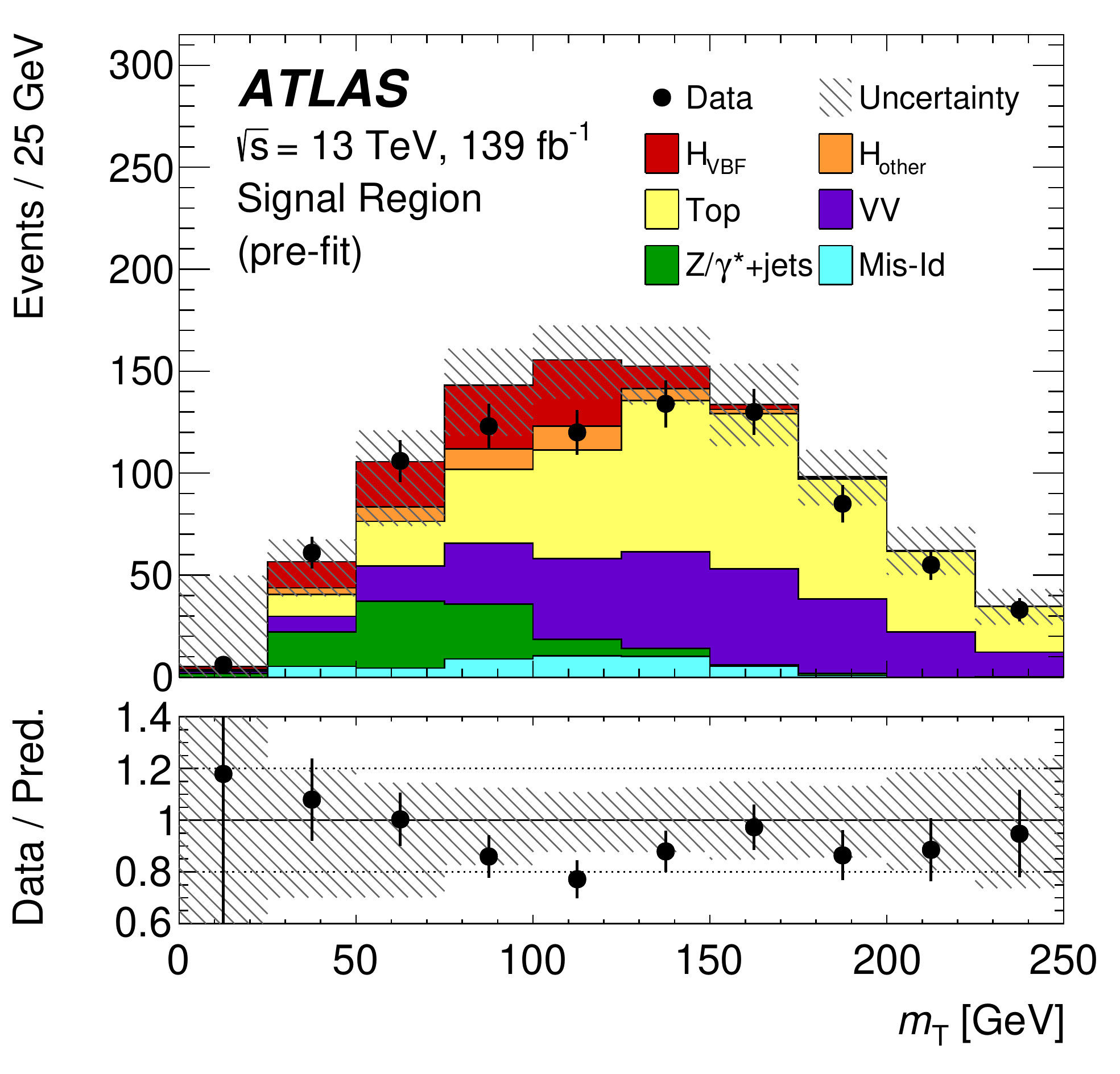}\label{fig:sr_MT}} \hfill
\subfloat[]{\includegraphics[width=0.45\linewidth]{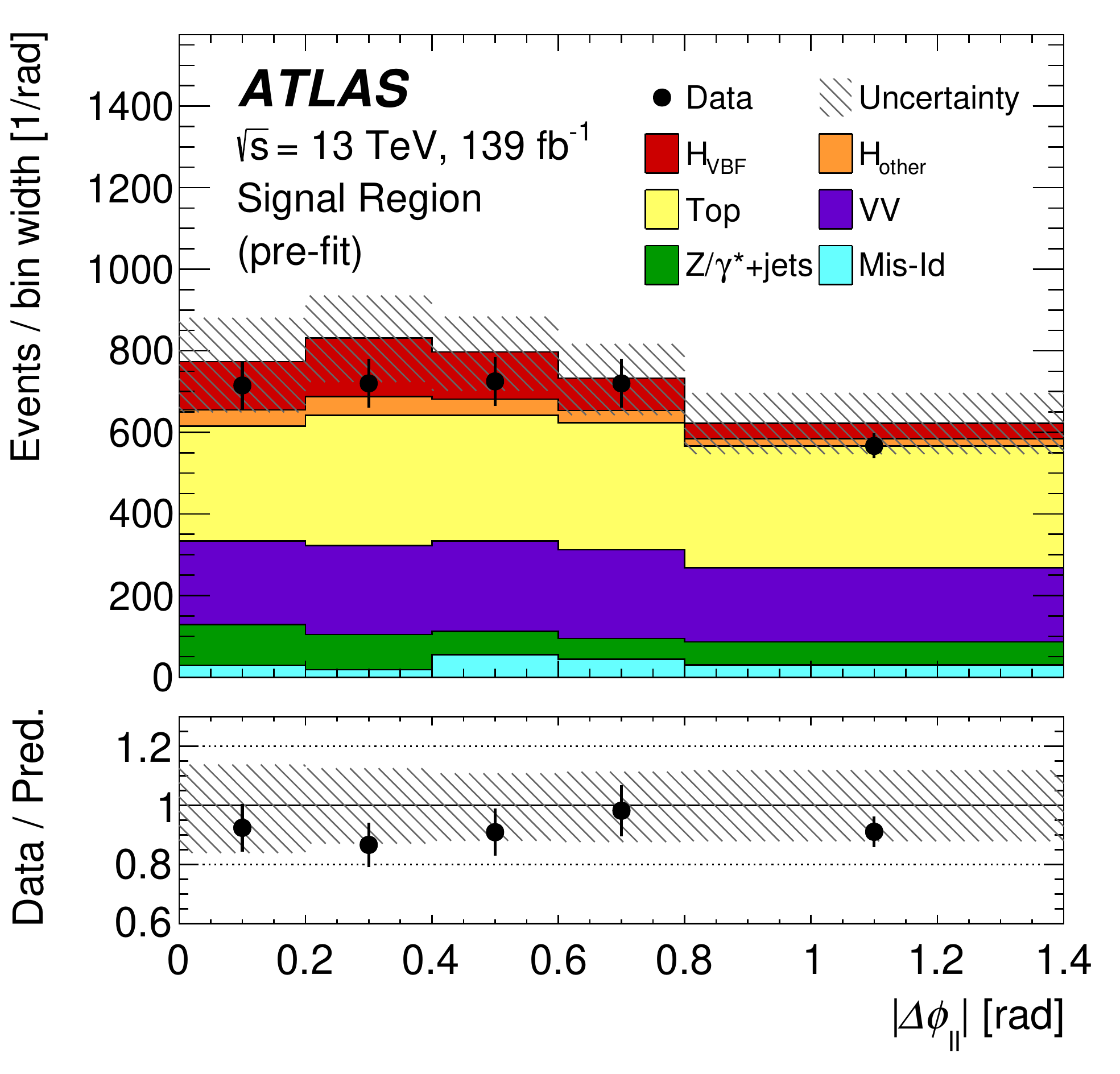}\label{fig:sr_dphill}} \\
\subfloat[]{\includegraphics[width=0.45\linewidth]{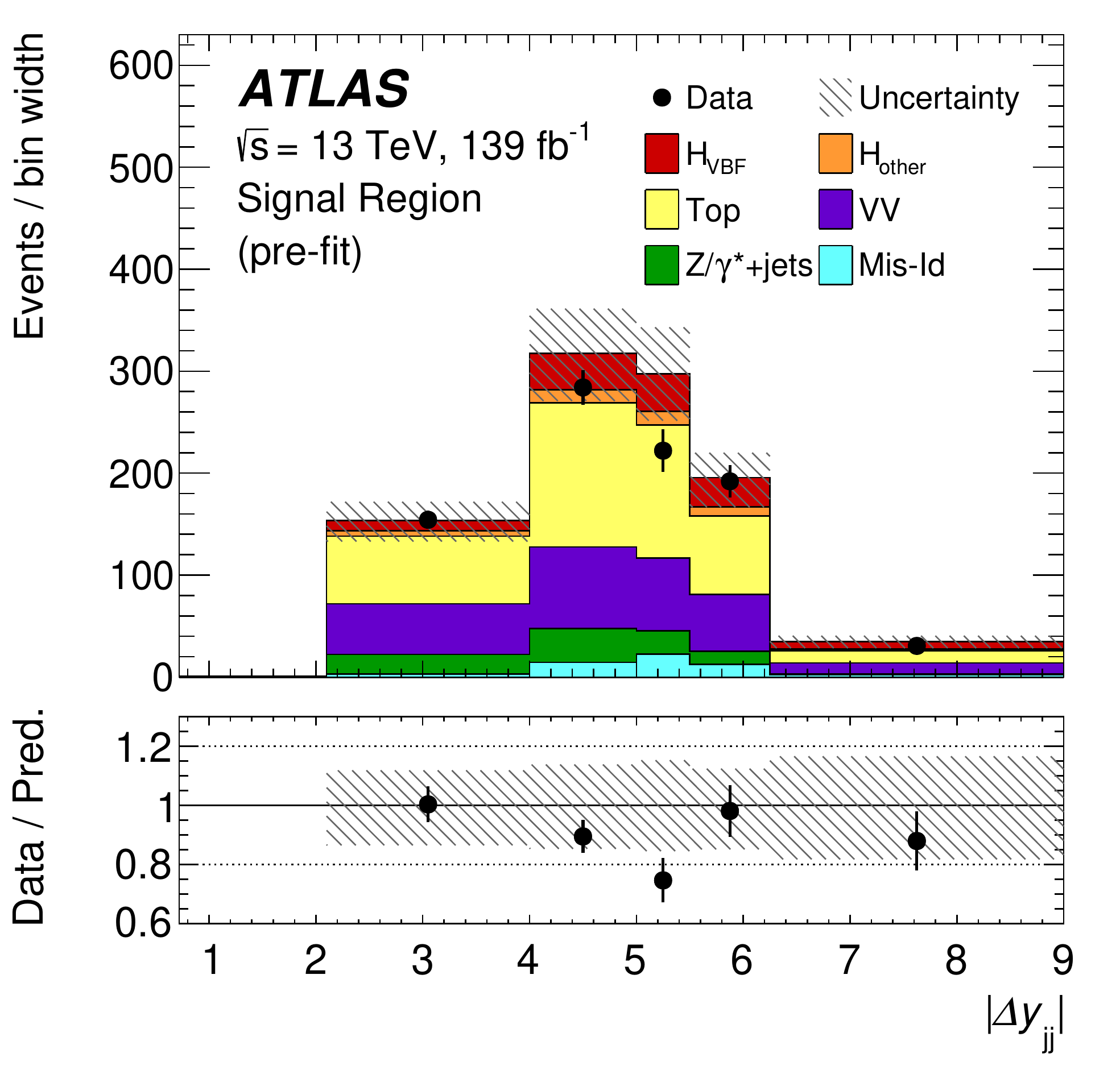}\label{fig:sr_dyjj}} \hfill
\subfloat[]{\includegraphics[width=0.45\linewidth]{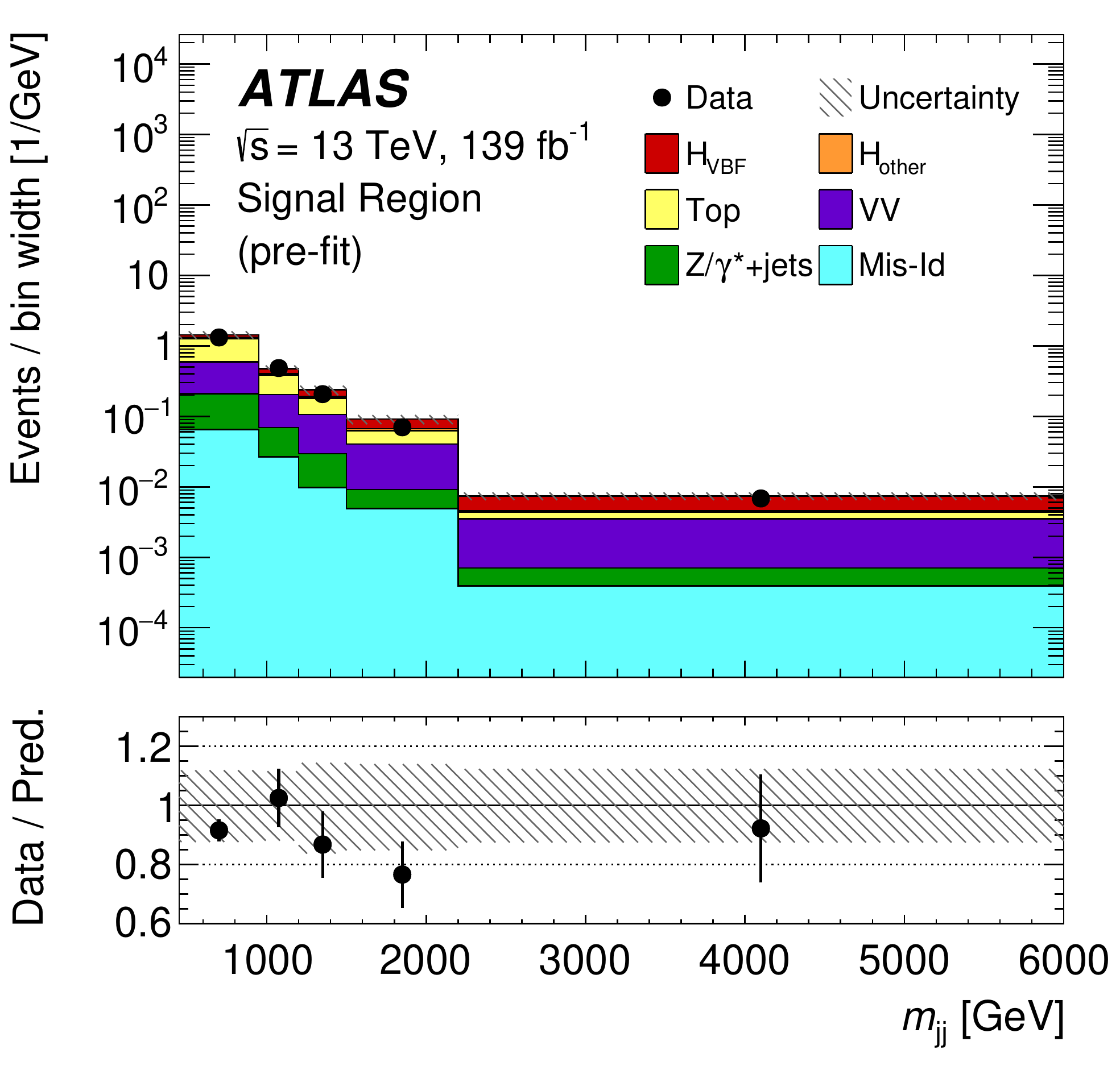}\label{fig:sr_mjj}}
\caption{The observed and expected (pre-fit) distributions of \protect\subref{fig:sr_MT} $\mt$, \protect\subref{fig:sr_dphill} $|\dphill|$, \protect\subref{fig:sr_dyjj} $|\dyjj|$, and \protect\subref{fig:sr_mjj} $\mjj$ in the signal region.
The uncertainty bars on the data points indicate the statistical uncertainties, while the uncertainties in the prediction are shown by the hatched band, which includes experimental and theoretical uncertainties of the signal and the backgrounds, as discussed in Section~\ref{sec:syst}. The $\rm{H_{other}}$ contribution is dominated by ggF production.
The bottom panel shows the ratio of the data to the expected distributions.
}
\label{fig:Det_VBFbdtInput}
\end{figure}
 
\begin{figure}[ht]
\centering
\subfloat[]{\includegraphics[width=0.45\linewidth]{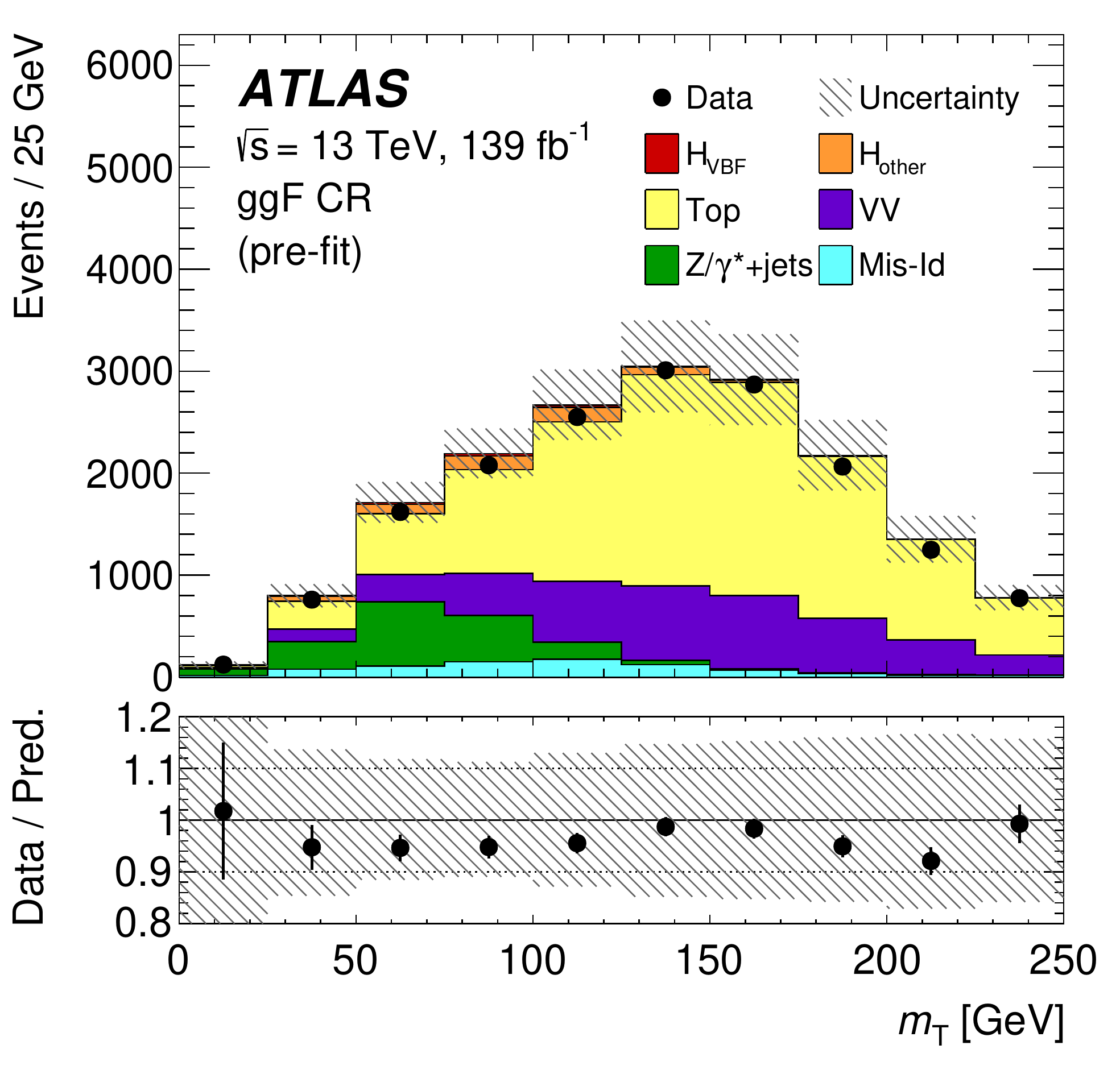}\label{fig:ggfcr_MT}} \hfill
\subfloat[]{\includegraphics[width=0.45\linewidth]{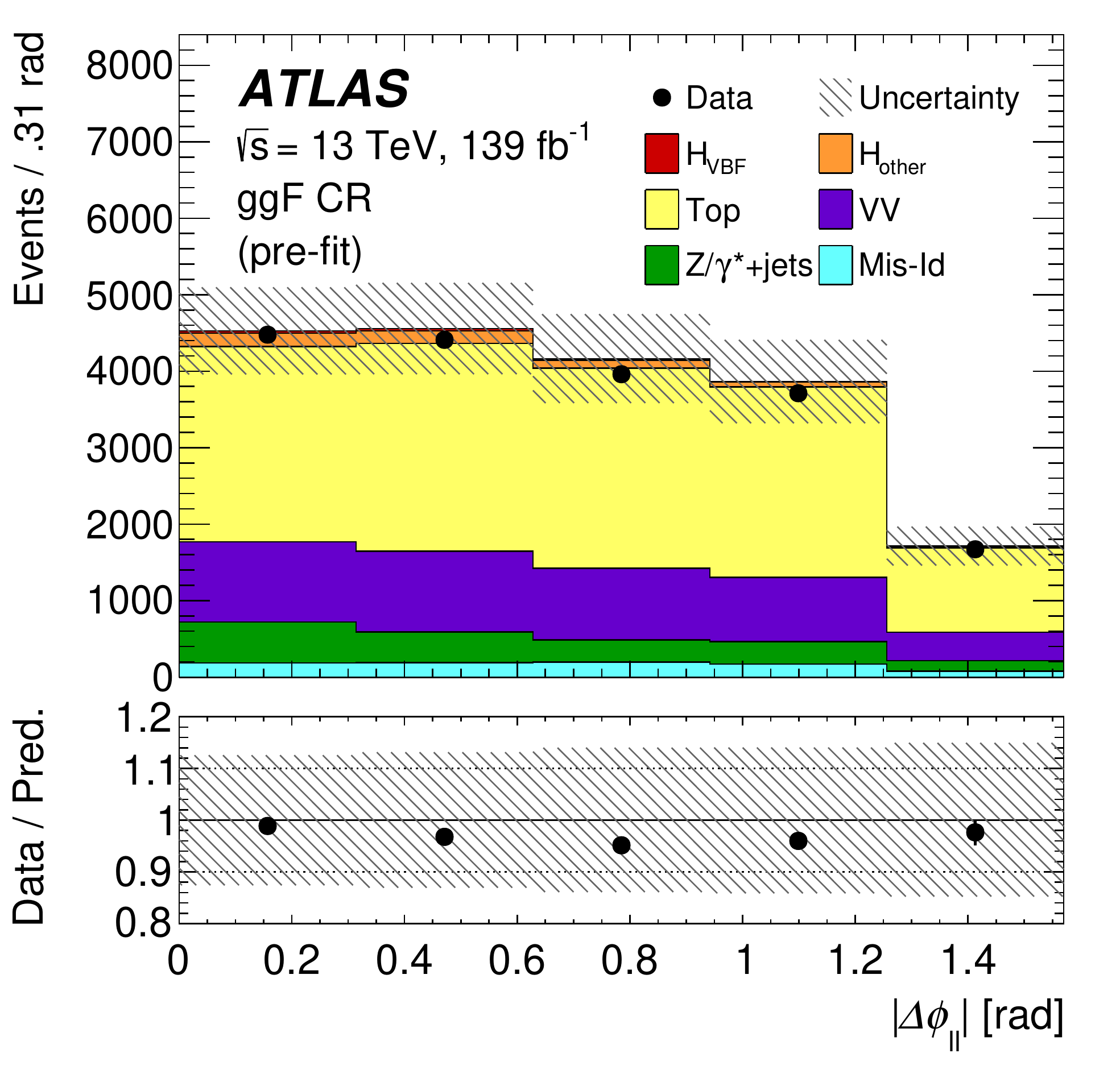}\label{fig:ggfcr_dphill}} \\
\subfloat[]{\includegraphics[width=0.45\linewidth]{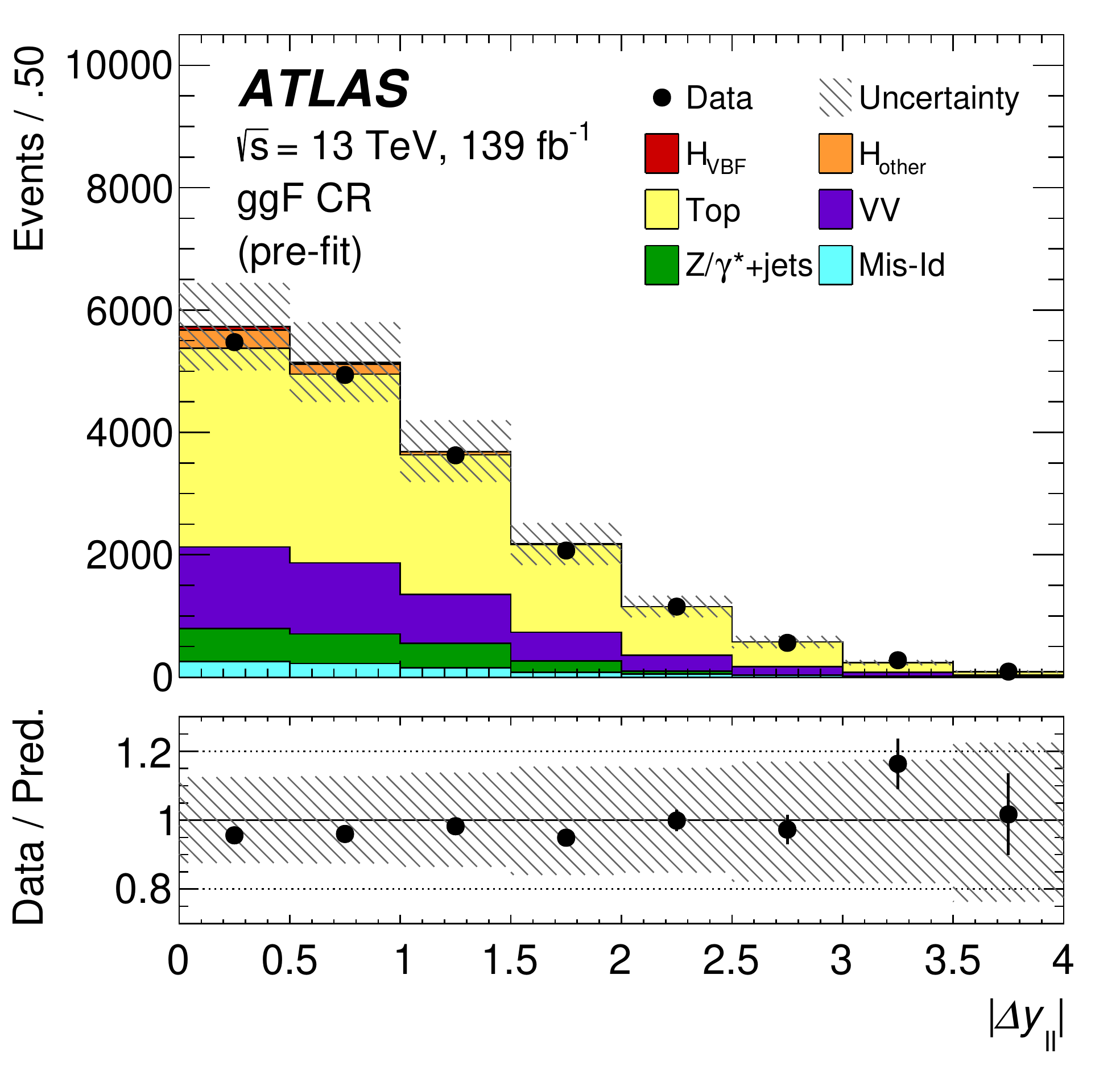}\label{fig:ggfcr_dyll}} \hfill
\subfloat[]{\includegraphics[width=0.45\linewidth]{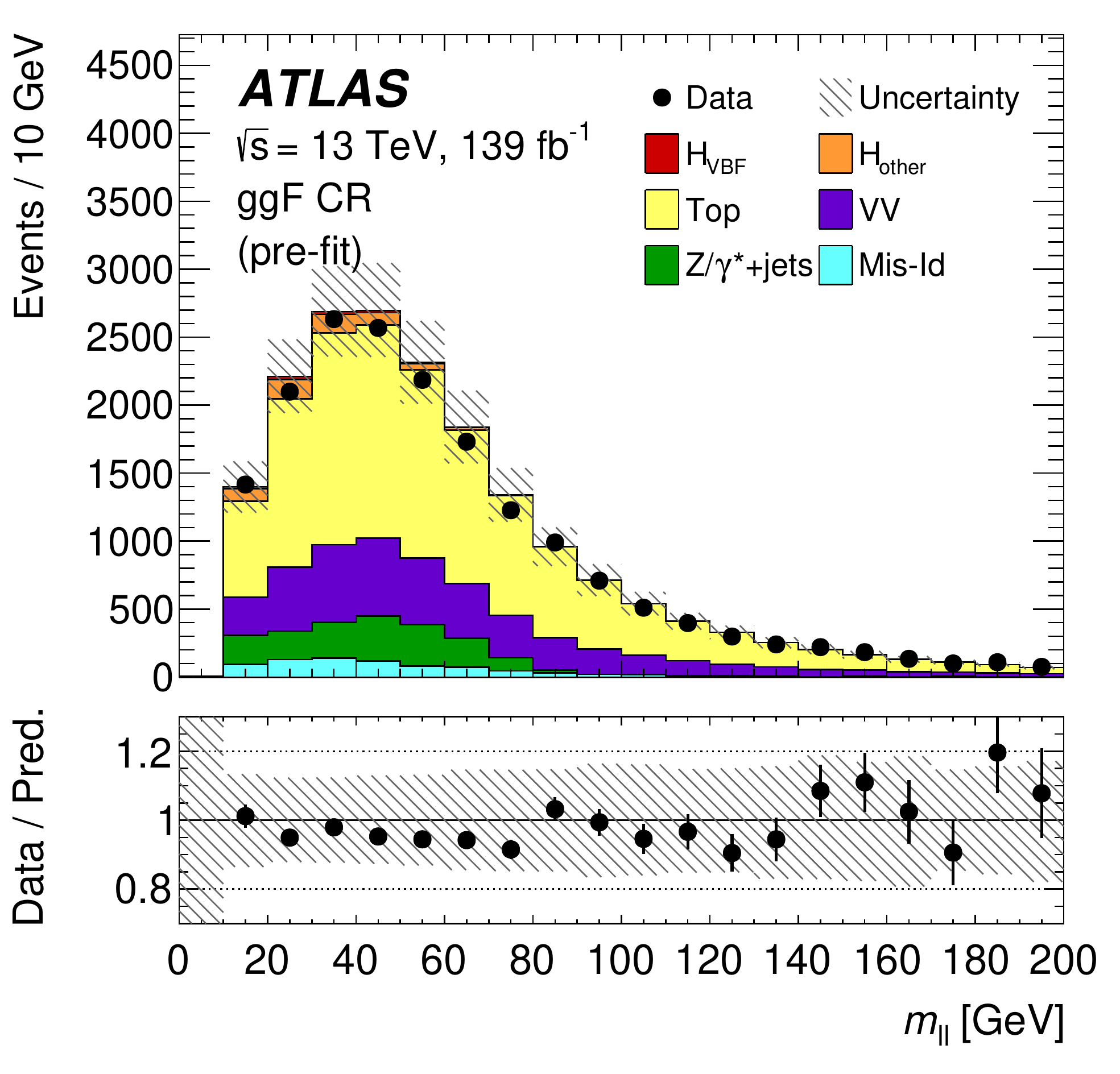}\label{fig:ggfcr_mll}}
\caption{The observed and expected (pre-fit) distributions of \protect\subref{fig:ggfcr_MT} $\mt$, \protect\subref{fig:ggfcr_dphill} $|\dphill|$, \protect\subref{fig:ggfcr_dyll} $|\dyll|$, and \protect\subref{fig:ggfcr_mll} $\mll$ in the ggF CR\@.
The uncertainty bars on the data points indicate the statistical uncertainties, while the uncertainties in the prediction are shown by the hatched band, which includes experimental and theoretical uncertainties of the signal and the backgrounds, as discussed in Section~\ref{sec:syst}. The $\rm{H_{other}}$ contribution is dominated by ggF production.
The bottom panel shows the ratio of the data to the expected distributions.
}
\label{fig:Det_ggFCRbdtInput}
\end{figure}

A bi-dimensional discriminant is formed in the SR by utilizing two distinct BDTs, i.e., $D_{\text{VBF}}$ and $D_{\text{top+VV}}$.
For the measurement of the integrated cross-section, the $D_{\text{VBF}}$ and $D_{\text{top+VV}}$ BDTs are trained in a phase space slightly larger than the SR to increase the event sample, i.e., with the \mjj $>$ 450 \GeV\ cut relaxed to \mjj $>$ 200 \GeV\ and the requirement on |\dphill| $<$ 1.4 rad removed. The bi-dimensional plane is constructed by utilizing the $D_{\text{VBF}}$ score on one axis and the $D_{\text{top+VV}}$ score on the other axis.
 
The following twelve variables are chosen as input to the $D_{\text{VBF}}$, which is trained to discriminate the signal process against the sum of the top-quark and $VV$ processes: $\mll$, $|\dphill|$, $|\dyll|$, the lepton $\eta$-centrality ($\Sigma_{\ell} C_{\ell}$, where $C_{\ell}$ = $|2\eta_{\ell}-\sum_{j}\eta_{j} | / \Delta\eta_{jj}$), which quantifies the positions of the leptons relative to the leading jets in pseudorapidity~\cite{Barger:1994zq}, $\mT$, $\mjj$, $|\dyjj|$, $|\dphijj|$, $\pTjone$, $\pTjtwo$, $\eta^{j_1}$, and $\eta^{j_2}$. No variable dominates the discrimination between signal and background. However, the highest-ranked ones are $\mt$, $|\dphill|$, $|\dyjj|$, and $\mjj$, and their distributions are shown in Figure~\ref{fig:Det_VBFbdtInput}.
 
The $D_{\text{top+VV}}$ BDT is trained to discriminate the sum of top-quark and $VV$ processes against all other processes and uses eight input variables: $|\dyjj|$, $|\dphill|$, $\mt$, leading and sub-leading jet rapidities, $|\dphijj|$, the invariant mass of leading lepton and jet, and $\Sigma_{\ell} C_{\ell}$.

The data contained in the [+0.5, +1.0] range of $D_{\text{VBF}}$ define a sub-set of the signal region that is named ``signal region 1'' (SR1) in the following. For the data falling in the SR1, the $D_{\text{VBF}}$ score is used as a discriminant and along this axis four bins are defined in the [+0.5, +1.0] range.
Similarly, an orthogonal sub-set of the signal region, i.e., ``signal region 2'' (SR2), is defined by the events falling into the [-1.0, +0.5] range of $D_{\text{VBF}}$ score. For events in the SR2, the $D_{\text{top+VV}}$ score is used as a discriminant, and four bins are chosen in the [-1.0, +1.0] range of $D_{\text{top+VV}}$.
 
In SR1 (SR2) the expected VBF signal-to-background ratio is 0.36 (0.015), while the expected percentage of top+$VV$ events relative to all other processes is 45\% (88\%). In the bin with the highest $D_{\text{VBF}}$ score, the expected VBF signal-to-background ratio is greater than 1.3, while in the bin with the highest $D_{\text{top+VV}}$ score the expected purity of top-quark and $VV$ events is approximately 97\%.
 
The BDT used in the ggF CR, named $D_{\text{ggF-CR}}$, is trained to discriminate the ggF Higgs boson production against all other processes in a phase space slightly larger than the ggF CR, i.e., with the requirement on |\dphill| $<$ 1.4 rad removed. It uses $\mt$, $\mll$, $|\dphill|$, $|\dyll|$, $|\dphijj|$, $\pTjone$, $\pTjtwo$, and $\met$ as input variables to the training. The highest-ranked ones are $\mt$, $|\dphill|$, $|\dyll|$, and $\mll$, and their distributions are shown in Figure~\ref{fig:Det_ggFCRbdtInput}. The $D_{\text{ggF-CR}}$ BDT score distribution is split into four bins and in the bin with the highest $D_{\text{ggF-CR}}$ BDT scores the expected purity of ggF Higgs boson events is approximately 7\%. Despite such a small purity,  the ggF CR allows the minimization of the modeling uncertainty associated with the ggF Higgs boson production with two accompanying jets thanks to the accuracy achieved in the $\mu_{\text{top}+VV, \text{ggF}}$ and $\mu_{\text{\Zjets}, \text{ggF}}$ normalization factors in this region.
 
The definitions and training of the $D_{\text{top+VV}}$ and $D_{\text{ggF-CR}}$ BDTs are also used in measurements of differential cross-sections. The $D_{\text{VBF}}$ is re-trained in the SR for each of the differential cross-section measurements with a reduced set of input variables with no significant loss of discrimination thanks to the information provided by the remaining variables. More specifically, the measured variable is excluded to avoid  biasing the differential measurement towards the SM prediction used for the training. For example, for the measurement of the cross-section as a function of $\mjj$, the $\mjj$ variable is excluded from the BDT training.
In addition, variables that are highly correlated to the measured distribution are also excluded from a BDT input variable list. Specifically, $\mll$ is removed from the BDTs used for measurements of the $|\dphill|$ and $|\dyll|$ distributions, conversely $|\dphill|$ and $|\dyll|$ are removed as inputs to the BDTs used for the measurement of the $\mll$ distribution, while $\mjj$, $|\dyjj|$, $\pTjone$, and $\pTjtwo$ are removed as inputs to the BDTs used for the measurements of $|\dyjj|$, $\mjj$, $\pTjtwo$, and $\pTjone$, respectively.
 
Similarly to the case of the integrated cross-section, four bins are used for the $D_{\text{VBF}}$ BDT scores, while two bins are used for the $D_{\text{top+VV}}$ BDT scores in each bin of a kinematic distribution that is measured, except the $\dphijj$ distribution in which four bins are used for the $D_{\text{top+VV}}$ BDT scores. The same classification scheme used for the integrated cross-section is used for a kinematic distribution $i$: all events with $D_{\text{VBF},i}$ $>$ 0.5  are projected onto the $D_{\text{VBF},i}$ score axis in the SR1$_{i}$, while those with $D_{\text{VBF},i}$ $<$0.5 are projected onto the $D_{\text{top+VV}}$ score axis in the SR2$_{i}$.
The signal-to-background ratio in the $D_{\text{VBF}}$ bins varies as a function of the bin of the measured variable and in the BDT bins with the highest score it is in the approximate range of $0.4 - 3.0$.


\subsection{Integrated fiducial cross-section measurement}
\label{sec:method_fiducial}

Figure~\ref{fig:Det_DVBF-DTopVV} shows the data and pre-fit templates for the $D_{\text{VBF}}$ and $D_{\text{top+VV}}$ discriminant distributions in the SR1 and SR2, while Figure~\ref{fig:Det_BDT_CR} shows the \mT distribution in the $Z/\gamma^*$+jets CR and the $D_{\text{ggF-CR}}$ discriminant distribution in the ggF CR, for the extraction of the integrated fiducial cross-section.
\begin{figure}[tb!]
\centering
\subfloat[]{\includegraphics[width=0.45\linewidth]{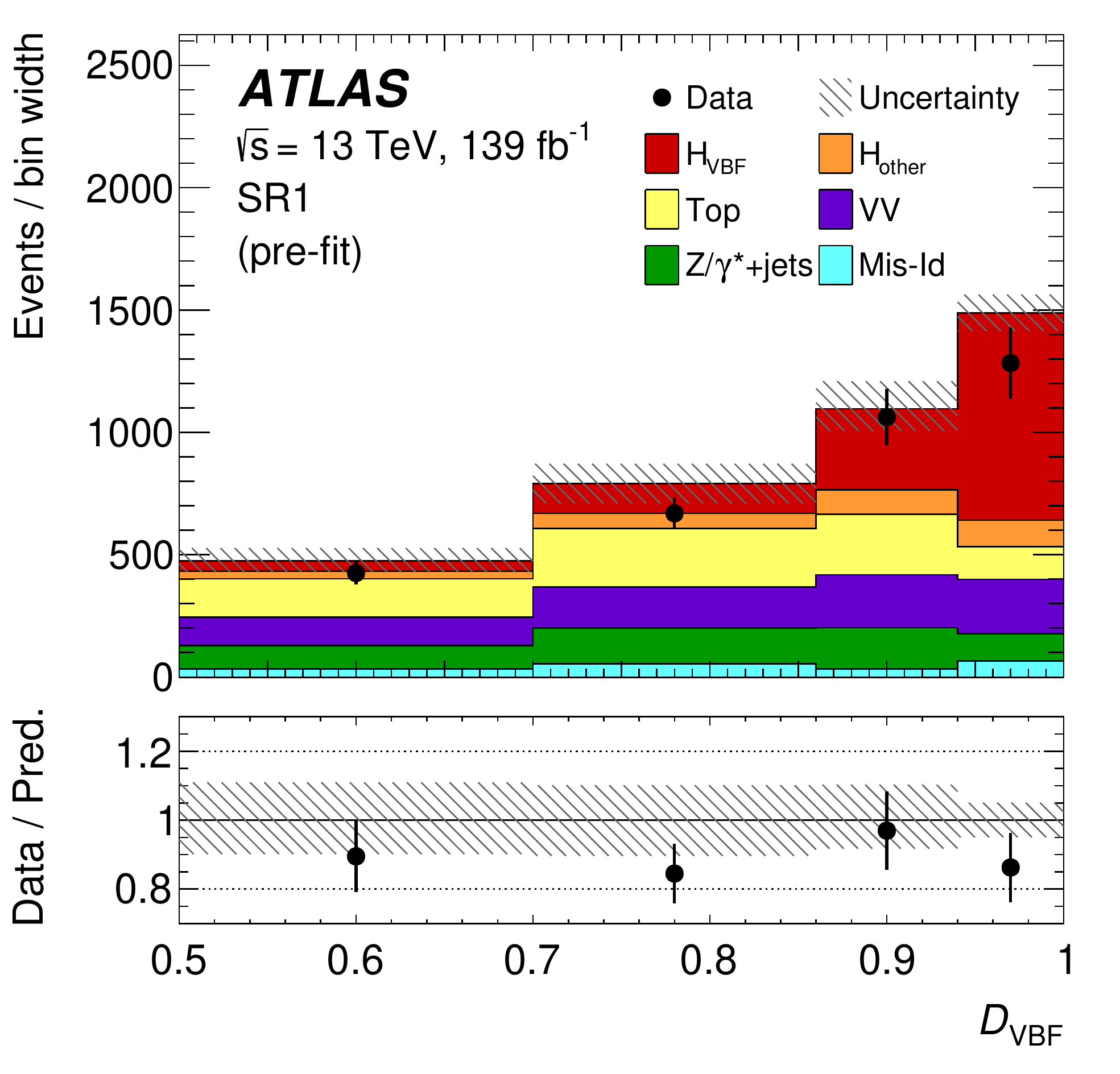}\label{fig:sr1_bdt_vbf}} \hfill
\subfloat[]{\includegraphics[width=0.45\linewidth]{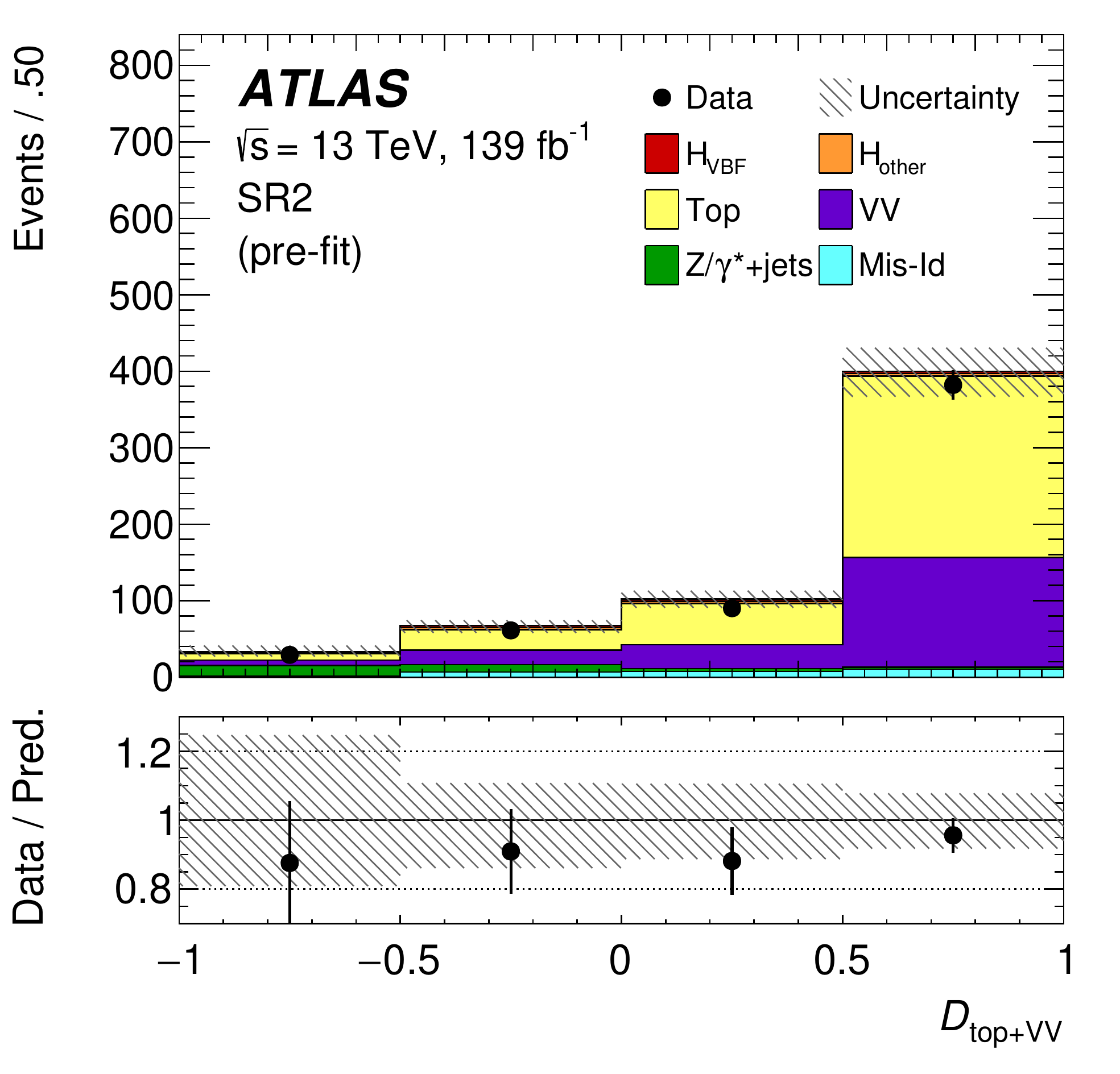}\label{fig:sr2_bdt_topwwall}}
\caption{The observed and expected (pre-fit) distributions of the \protect\subref{fig:sr1_bdt_vbf} $D_{\text{VBF}}$ and \protect\subref{fig:sr2_bdt_topwwall} $D_{\text{top+VV}}$ discriminants in SR1 and SR2, respectively.  The uncertainty bars on the data points indicate the statistical uncertainties, while the uncertainties in the prediction are shown by the hatched band, which includes experimental and theoretical uncertainties of the signal and the backgrounds, as discussed in Section~\ref{sec:syst}. The $\rm{H_{other}}$ contribution is dominated by ggF production.
The bottom panels show the ratios of the data to the expected distributions.
}
\label{fig:Det_DVBF-DTopVV}
\end{figure}
\begin{figure}[tb!]
\centering
\subfloat[]{\includegraphics[width=0.45\linewidth]{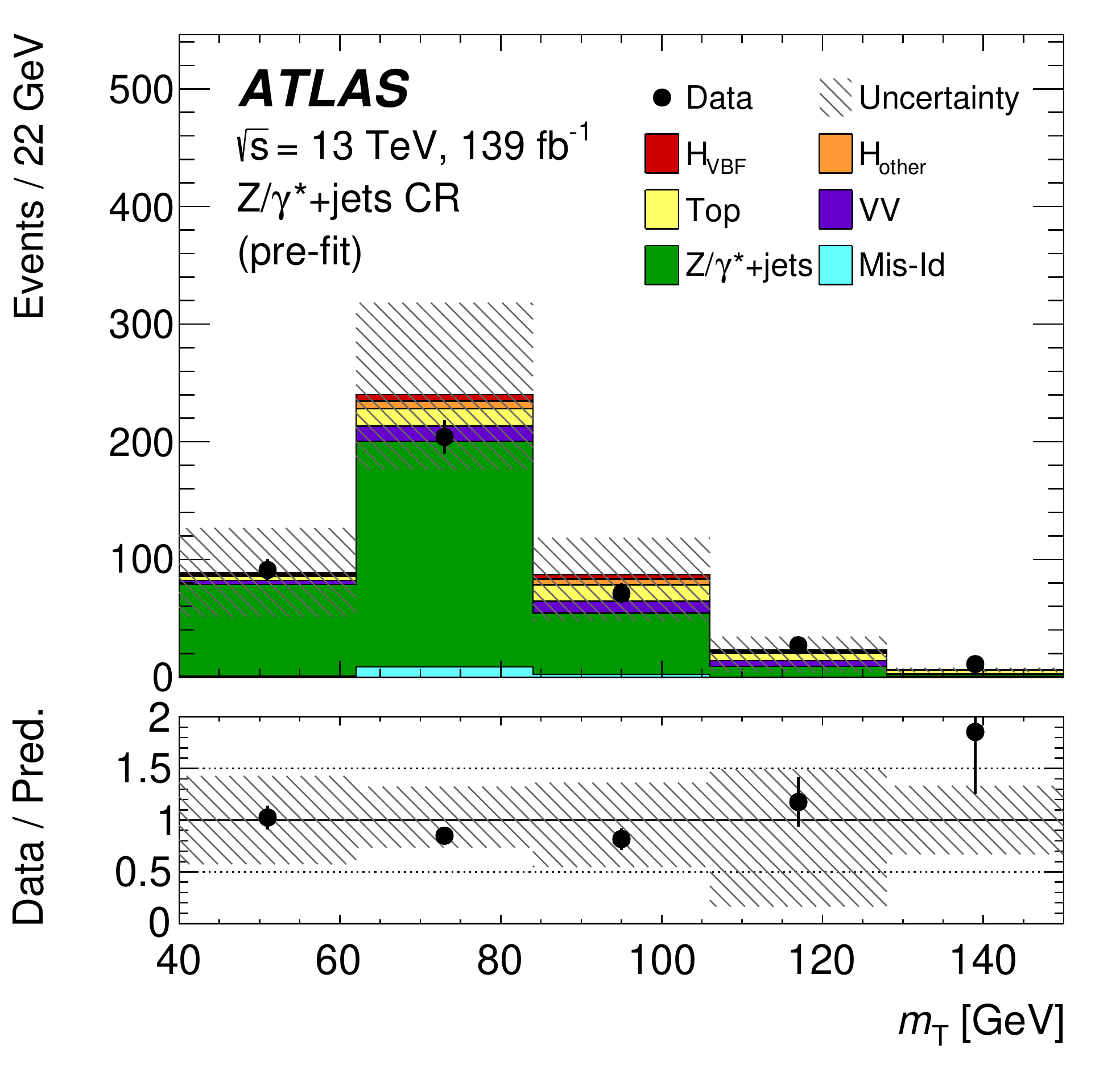}\label{fig:dycr_mt}} \hfill
\subfloat[]{\includegraphics[width=0.45\linewidth]{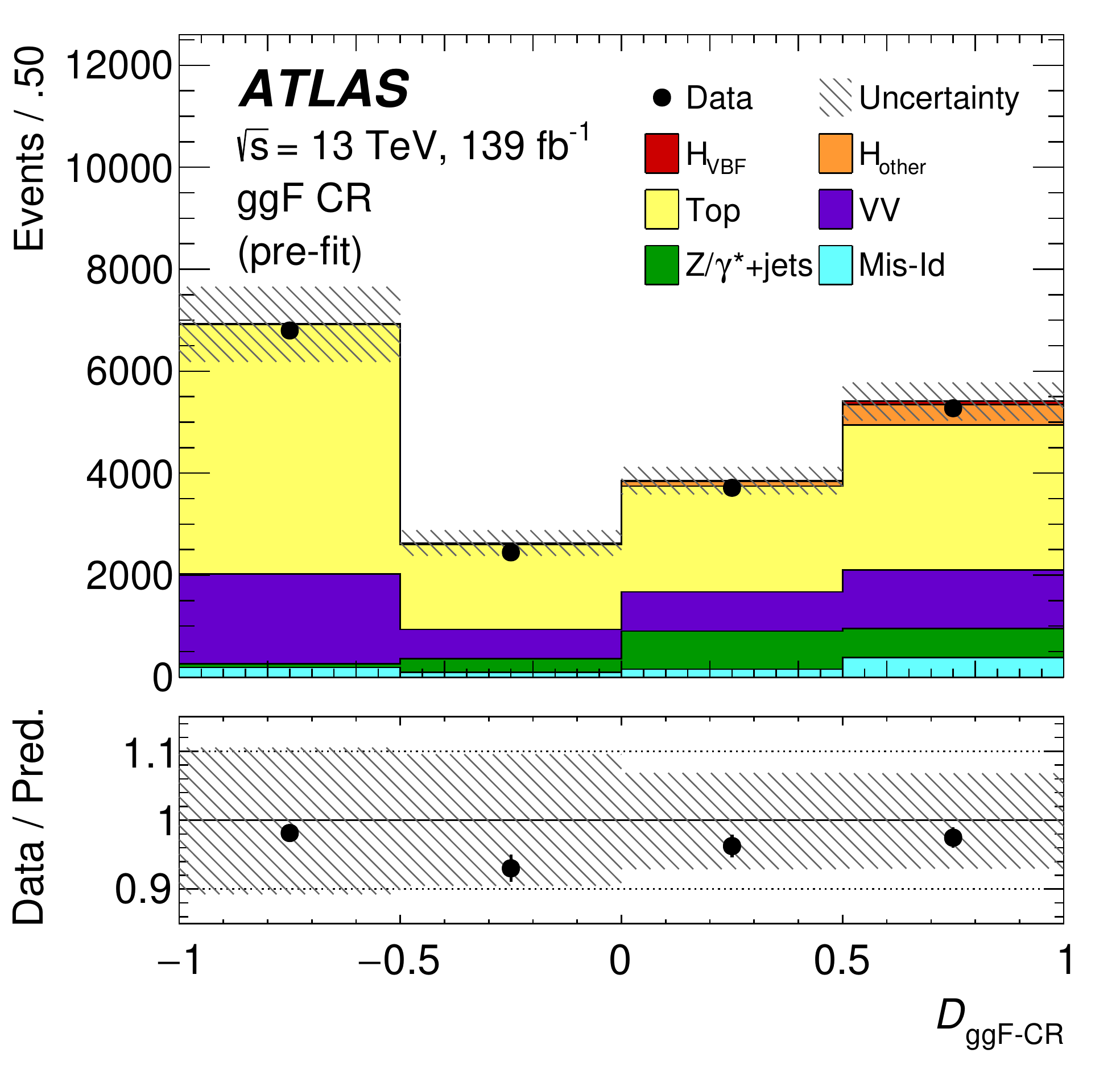}\label{fig:ggfcr1_bdt_ggfcr1}}
\caption{The observed and expected (pre-fit) distributions of \protect\subref{fig:dycr_mt} $\mT$ in the $\Zjets$ CR and \protect\subref{fig:ggfcr1_bdt_ggfcr1} the $D_{\text{ggF-CR}}$ discriminant in ggF CR.
The uncertainty bars on the data points indicate the statistical uncertainties, while the uncertainties in the prediction are shown by the hatched band, which includes experimental and theoretical uncertainties of the signal and the backgrounds, as discussed in Section~\ref{sec:syst}. The $\rm{H_{other}}$ contribution is dominated by ggF production. The bottom panels show the ratios of the data to the expected distributions.
}
\label{fig:Det_BDT_CR}
\end{figure}
 
The signal yield $N_{\text{signal}}^{\text{SR}}$ in Equation~(\ref{eqn:likelihood}) is expressed as:
 
\begin{eqnarray}
\label{eqn:likelihood_fid}
N_{\text{signal}}^{\text{SR}} =    \sigma^{\mathrm{fid}}  \cdot \mathcal{T}(D_{\text{VBF}}, D_{\text{top+VV}}) \cdot{\cal L} \cdot C
\end{eqnarray}
 
where  $\sigma^{\mathrm{fid}}$ is the measured cross-section, as defined in Eq.~(\ref{eq:xSectFid}), and is determined from the fit to data.
The $\mathcal{T}$ term is the signal probability density function (template) formulated in each bin of the two-dimensional $D_{\text{VBF}} - D_{\text{top+VV}}$ discriminant distribution. The integrated luminosity and detector efficiency are indicated with $\cal L$ and $C$, respectively. The technique was validated with ensemble tests based on MC simulations and by integrating the differential cross-section measurements.


\subsection{Differential fiducial cross-section measurements}
\label{sec:method_diff}

\begin{figure}[tb!]
\centering
\includegraphics[width=0.65\linewidth]{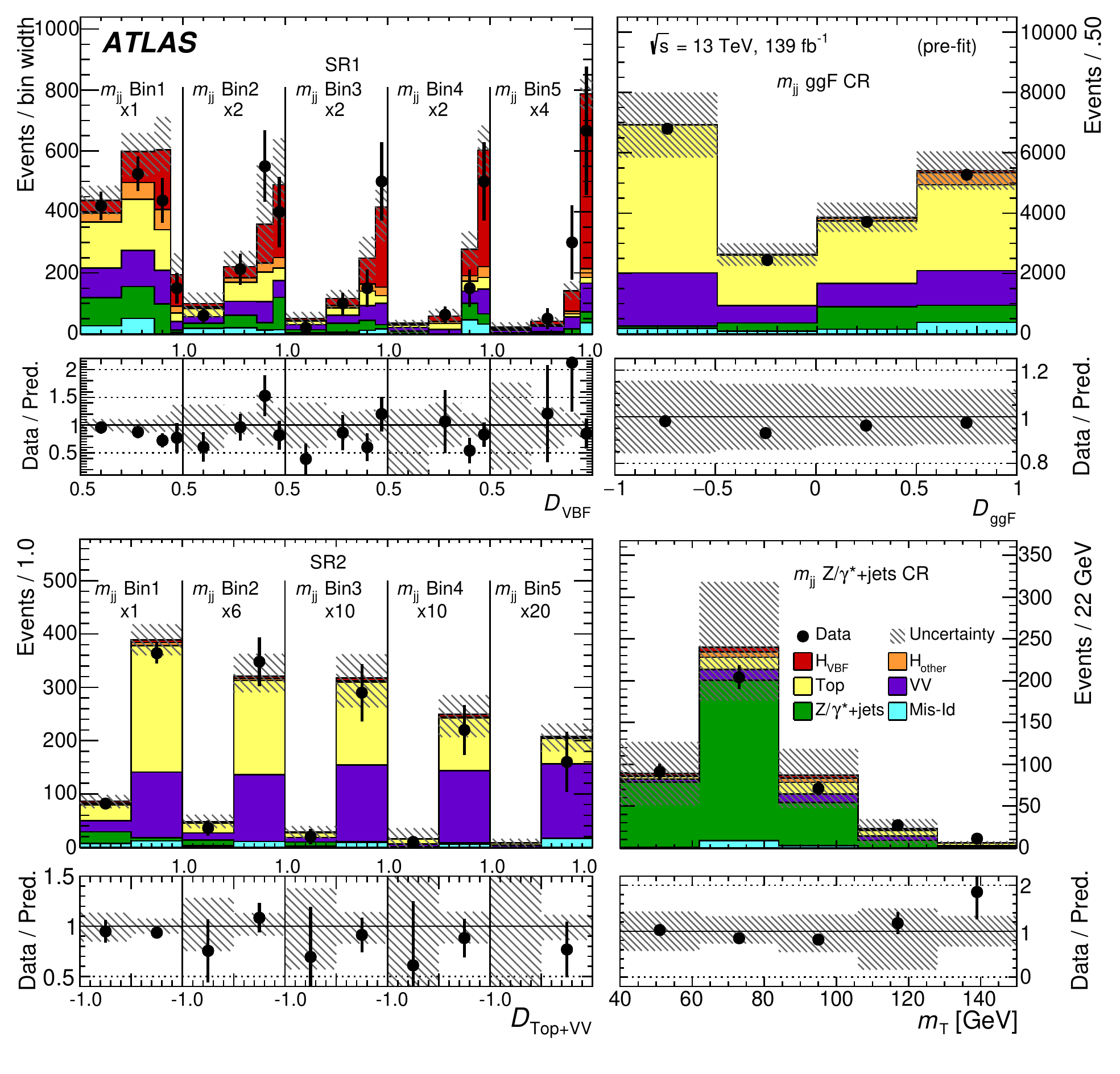}
\caption{The pre-fit distributions of the discriminants in data and in the MC templates at the detector-level for the measurement of the \mjj differential cross-section. The distributions of $D_{\text{VBF}}$  and $D_{\text{top+VV}}$ discriminants in SR1 (top left) and SR2 (bottom left), respectively, are shown for each bin of the observable in the ranges [0.5, 1.0] for $D_{\text{VBF}}$ and [-1.0, 1.0] for $D_{\text{top+VV}}$, while \mT and $D_{\text{ggF-CR}}$  are shown in the $Z/\gamma^*$+jets CR (bottom right) and in the ggF CR (top right), respectively. For visibility, the event yield in each bin of a kinematic distribution is multiplied by the factor specified in the legend.  }
\label{fig:prefit_bdt_mjj}
\end{figure}
\begin{figure}[tb!]
\centering
\includegraphics[width=0.65\linewidth]{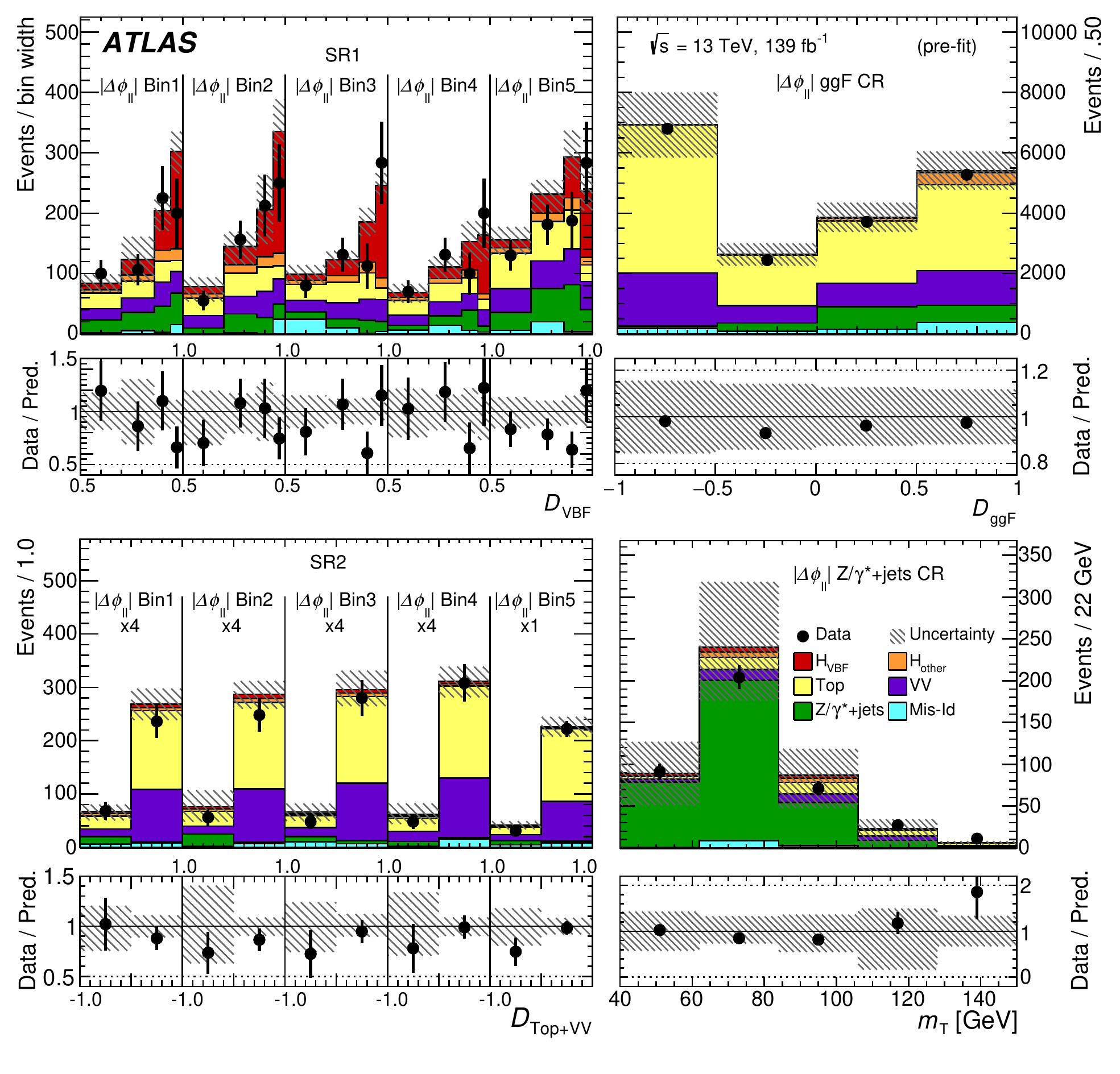}
\caption{The pre-fit distributions of the discriminants in data and in the MC templates at the detector-level for the measurement of the |\dphill| differential cross-section. The distributions of $D_{\text{VBF}}$  and $D_{\text{top+VV}}$ discriminants in SR1 (top left) and SR2 (bottom left), respectively, are shown for each bin of the observable in the ranges [0.5, 1.0] for $D_{\text{VBF}}$ and [-1.0, 1.0] for $D_{\text{top+VV}}$, while  \mT  and $D_{\text{ggF-CR}}$ are shown in the $Z/\gamma^*$+jets CR (bottom right) and in the ggF CR (top right), respectively. For visibility, the event yield in each bin of a kinematic distribution is multiplied by the factor specified in the legend.  }
\label{fig:prefit_bdt_dphill}
\end{figure}
For the measurement of differential fiducial cross-sections, the signal and the background yields are determined from the fit in each bin of the measured variable, following a procedure similar to the one described in Section~\ref{sec:method_fiducial}.
Differently from the integrated fiducial cross-section, the two-dimensional $D_{\text{VBF}} - D_{\text{top+VV}}$ discriminant is evaluated in each bin of the measured variable and dedicated free floating normalization parameters are introduced as a function of the measured variable.  Such a technique was developed to reduce the model dependence in the estimate of the dominant top-quark and $VV$ background sources in the differential cross-section measurements.
The distributions of the discriminants in data, compared with the MC templates at the detector-level are shown in Figures~\ref{fig:prefit_bdt_mjj} and~\ref{fig:prefit_bdt_dphill} for the case of \mjj and |\dphill| differential cross-section measurements, respectively.
 
An unfolding procedure is applied in the likelihood formalism to correct for migrations between bins of the variable distribution at the detector-level and applies fiducial as well as reconstruction efficiency corrections.
A built-in regularization technique based on the method developed in Ref.~\cite{HIGG-2018-29} was studied. This technique implements a constraint on the second derivative of the differential spectrum and is fully incorporated in the likelihood minimization step, taking into account the full correlation information of the degrees of freedom involved in the fit and induced by the regularization parameter.
The expected number of signal events $N_{\text{signal,}i}^{\text{SR}}$ is a function of the bin $i$ of a detector-level observable in the SR:
 
\begin{eqnarray}
\label{eqn:likelihood_diff}
N_{\text{signal,}i}^{\text{SR}} = \sum_j r_{ij} \cdot (1+f_i^{\text{nonfid}}) \cdot \sigma^{\mathrm{fid}}_{j} \cdot \mathcal{T}_i(D_{\text{VBF}}, D_{\text{top+VV}}) \cdot \mathcal{L}
\end{eqnarray}
 
where the index $j$ runs over all bins of the particle-level observable as defined in the fiducial phase space and $\sigma^{\mathrm{fid}}_{j}$ is the measured particle-level cross-section in bin $j$.
The $\mathcal{T}_i$ term is the signal probability density function of bin $i$ of the detector-level observable and is formulated in each bin of the two-dimensional $D_{\text{VBF}} - D_{\text{top+VV}}$ discriminant distribution.
The term $r_{ij}$ represents the detector response matrix that accounts for bin-to-bin migrations in the unfolding of the signal. It is calculated with \POWHEGva+\PYTHIAV{8} simulated signal samples, for events that are generated in the fiducial region and reconstructed in the SR\@. Its elements correspond to the probability that a signal event generated in the fiducial volume in bin $j$ of the observable is reconstructed in bin $i$ in the detector-level distribution. The correction $f_i^{\text{nonfid}}$ represents the fraction of events that are outside of the fiducial region but are reconstructed at detector-level in the SR\@. The $f_i^{\text{nonfid}}$ correction ranges from 7\% to 33\% depending on the observable and on the bin of the observable distribution.
 
To account for variations in the contribution of top-quark and $VV$ processes as a function of the value of the observable, multiple normalization parameters of this background ($\mu_{\text{top}+VV, \text{SR-}i}$) are determined from the fit to data in each measured kinematic distribution in a coarser binning than the one used in the differential cross-section measurements, i.e., it combines in a single bin the data of two adjacent bins.

The model dependence introduced via the unfolding method is reduced when using the response matrix, as opposed to the alternative bin-by-bin correction-factor method. Good closure is observed between unfolded pseudo-data samples and the corresponding particle-level distributions. Such tests show that, with a binning choice that reduces the bin-to-bin migrations while maintaining the signal sensitivity, the in-likelihood unfolding procedure without any additional regularization out-performs the in-likelihood unfolding with additional regularization. More specifically, while such regularization reduces the statistical uncertainty it also induces a systematic bias that overall increases the total uncertainty. Therefore, no regularization was applied in the unfolding.


\subsection{Systematic uncertainties}
\label{sec:syst}

\newcommand{\head}[2]{\multicolumn{1}{>{\centering\arraybackslash}p{#1}}{\textbf{#2}}}
Systematic uncertainties in the signal and control regions affect shape and normalization of the MC templates used in the fit for the signal and background processes, as well as the acceptance of the signal and bin-to-bin event migration. Each source of uncertainty is included as a Gaussian-constrained nuisance parameter in the likelihood.
The systematic uncertainties can be classified as originating from experimental or theoretical sources. Experimental uncertainties include those in lepton, jet, and missing transverse momentum reconstruction and calibration, while theoretical uncertainties are related to the modeling of the signal and background processes. The theoretical uncertainties and those with small impact on the cross-sections are symmetrized by taking the average of the upward and downward variations in each bin of a distribution.
 
Uncertainties in the leptons originate from the electron and muon reconstruction and identification efficiency, energy (for electrons) and momentum (for muons) scale and resolution, isolation efficiency~\cite{EGAM-2018-01,MUON-2018-03} as well as from trigger efficiency~\cite{Aad:2019wsl,Aad:2020uyd}, and are estimated by using tag-and-probe methods in $Z\rightarrow\ell\ell$ events.
 
The uncertainties in the jet energy resolution and scale are obtained from simulations and in situ measurements~\cite{PERF-2016-04}. The uncertainty in the suppression of jets originating from pile-up interactions~\cite{PERF-2014-03}, in the jet-vertex association as well as in the $b$-tagging efficiency and the mis-tag rate~\cite{FTAG-2018-01} are also considered.
A variation in the pile-up modeling in simulated events is included to account for the uncertainty in the ratio of the predicted and measured inelastic cross-sections.
The uncertainty in the $\met$ measurement is estimated by propagating the uncertainties in the transverse momenta of leptons and jets and by applying momentum scale and resolution uncertainties to the track-based soft term~\cite{PERF-2016-07}.
 
The uncertainty in the combined 2015–2018 integrated luminosity is 1.7\%~\cite{ATLAS-CONF-2019-021}, obtained using the LUCID-2 detector~\cite{Avoni_2018} for the primary luminosity measurements. The luminosity uncertainty is applied to background processes that are normalized to theoretical predictions and to the signal cross-section parameters in the fit.
 
Three sources of uncertainty related to the extrapolation factor used in the data-driven mis-Id background estimate are considered: the statistical uncertainty in the extrapolation factor itself, an uncertainty related to the subtraction of processes with two prompt leptons from the $Z/\gamma^*$+jets-enriched sample used to derive the extrapolation factor, and an uncertainty in the correction factor for the sample composition. Dedicated checks were carried out to assess the
dependence of the extrapolation factors on the number of jets reconstructed in the event, the distance between the lepton candidate and a nearby jet, and pile-up. No significant dependence was found. An MC-based test was also carried out, in which the same mis-Id background estimate technique as used in data for the measurements of the integrated and differential cross-sections is applied on MC $W$+jets samples using two different MC generators (see Section~\ref{sec:dataset_mc}). No bias was observed beyond statistical effects in the MC samples. Therefore, no systematic uncertainty was assigned to such an effect.

Finally, uncertainties may arise from the unfolding procedure used. Tests with MC simulations and ensemble tests with varying underlying cross-sections show that the method is successful in retrieving the particle-level distribution in the fiducial region from the reconstructed distribution in the signal region. This bias is studied using pseudo-experiments in which the underlying cross-section is varied and pseudo-data sets are produced and fitted using the nominal likelihood and matrix as used in the default fit to data. The underlying cross-sections are sampled from a multi-dimensional Gaussian distribution centered around the nominal expectation of each bin and uniformly uncorrelated variations of the cross-section in each bin. The difference of the fitted cross-section in each bin relative to the true cross-section is taken as a measure of the bias induced by the unfolding procedure and is determined as a function of the measured cross-section. This technique is illustrated in Ref.~\cite{HIGG-2018-29} and adapted for a multi-dimensional discriminant.
This bias was found to be negligible in the unregularized likelihood-based unfolding procedure compared to the leading systematic and modeling uncertainties.
Further uncertainties were considered in the unfolding method to include those arising from the theoretical assumptions used in deriving the detector response matrix. They are derived as described in the following paragraph. They are applied to all the elements of the detector response matrix and treated as fully correlated with the uncertainties of the same origin impacting the shape of the discriminants in the simulated MC templates.

Theoretical uncertainties for the signal and the major irreducible background processes include those originating from renormalization and factorization scale choices to account for missing higher-order QCD corrections, $\alpha_s$, and PDF~\footnote{An exception are the $t\bar{t}$ background samples, for which $\alpha_s$ variations are not included to avoid double-couting with other systematic variations.}, as detailed in Section~\ref{sec:dataset_mc}. The uncertainties associated with UE, PS, and hadronization modeling were estimated for signal and major background processes except the $Z/\gamma^*$+jets by comparing the MC templates and response matrices calculated using \PYTHIAV{8}\ and \HERWIG models. The uncertainties in the signal, top-quark, and $Z/\gamma^*$+jets production processes also account for differences in matrix element calculations and PS matching schemes. This is achieved by comparing the calculation by \POWHEGva+\HERWIGV{7} with the one by \MGFiveHw\, for the signal, \POWHEGva+\PYTHIAV{8}\ calculation with the one by \MADGRAPHFive\, interfaced to \PYTHIAV{8}\ for \ttbar, and \SHERPA\ calculation with the one by \MADGRAPHFive\ interfaced to \PYTHIAV{8} for $Z/\gamma^*$+jets.
 
For the predictions of the continuum $qq\rightarrow WW$, variations of the parameters that control the MC matching scale between matrix element and parton shower calculations, resummation scale as well as parton shower scheme and model (\PYTHIAV{8} and \HERWIG++) are considered, as described in Ref.~\cite{HIGG-2021-20}, whereas the matrix element uncertainty is assumed to be subdominant.

For backgrounds that are extracted from data,  the uncertainties are normalized to the integrated yield of events over the regions in which they are normalized.
 
Several tests were carried out on simulations to assess whether the analysis methodology introduces biases on the cross-section measurements. Pseudo-data sets were simulated using signal models with different predictions of yields and shapes for the observables that are used as inputs to the BDT trainings. For such tests, the SM \MGFiveHw\, prediction was used. In addition, models that implement a broad range of variations in anomalous interactions and model-independent parameterizations of new physics effects were included in such tests. No bias was observed in those tests as the analysis and fitting algorithms were able to reproduce the pseudo-data sets within the statistical uncertainty in the MC samples.
 
The impacts of the experimental and theoretical uncertainties in the measurements  after the maximum-likelihood fit are summarized in Table~\ref{tab:syst}. The dominant systematic uncertainties are theoretical. Among them, the largest uncertainties are those associated with the signal modeling, specifically the matrix element calculation, followed by uncertainties in the modeling of top quark, $WW$, ggF, and $Z/\gamma^*$+jets backgrounds. The largest source of experimental uncertainties affecting the integrated and differential cross-section measurements is the jet energy resolution.
 
\begin{table}[t!]
\centering
\small
\caption{Summary of the relative uncertainties in the measured integrated fiducial cross-section and ranges of relative uncertainties in groups of differential cross-sections. The different categories of systematic uncertainties include modeling uncertainties in signal and background processes, experimental uncertainties, luminosity as well as statistical uncertainties in MC samples. The total systematic uncertainties, obtained by combining the individual contributions and the statistical uncertainty in data, are also shown. }
\label{tab:syst}
\begin{tabular}{c|r|r|r|r|r|r}
\hline\hline
\noalign{\vspace{0.05cm}}
& Uncertainty [\%]                  & \multicolumn{5}{c}{Uncertainty range [\%]} \\
Source         &  $\sigma^{\mathrm{fid}}$  &     $\pT^{H}$      &   $\pT^{\ell\ell}$, $\pTlone$,       & $\mll$   &   $\pTjone$, $\pTjtwo$,   & $\mjj$  \\
&                                   &                   &   $\pTltwo$, |\dyll|,               &          &  |\dyjj|, \dphijj         &         \\
&                                   &                   &   |\dphill|, $\cos(\theta^*_{\eta})$  &          &                           &         \\
\noalign{\vspace{0.05cm}}
\hline\hline
\noalign{\vspace{0.05cm}}
Signal modeling         &  5    &  $<$ 1 -- 7     &   $<$ 1 -- 7     &  $<$ 1 -- 19     &  $<$ 1 -- 8    &  2 -- 7  \\
Signal parton shower     &  $<$ 1    &  $<$ 1 -- 2     &   $<$ 1 -- 1.8     &  $<$ 1 -- 10    &  $<$ 1 -- 1.8    &  $<$ 1 -- 7  \\
$\ttbar$ modeling       &  6    &  1.7 -- 30      &   3 -- 13      &  3 -- 80     &  3 -- 10    &  1.2 -- 70   \\
$WW$ modeling           &  4    &  $<$ 1 -- 12      &   3 -- 11      &  2 -- 90     &  3 -- 10     &  3 -- 40   \\
$Z/\gamma^*$+jets modeling       &  4    &  $<$ 1 -- 19      &   2 -- 18      &  4 -- 30     &  3 -- 13     &  2 -- 50   \\
ggF modeling            &  5    &  4.0 -- 28      &   3.4 -- 10      &  2.6 -- 12    &  2.3 -- 9.0    &  1.4 -- 86   \\
Mis-Id background       &  $<$ 1    &  $<$ 1 -- 12      &   1.1 -- 5     &  $<$ 1 -- 19     &  1 -- 3    &  $<$ 1 -- 40   \\
\hline
Jets \& Pile-up \& \met  &  5    &  8 -- 60      &   6 -- 30      &  6 -- 120    &  9 -- 30     &  9 -- 130  \\
$b$-tagging              &  $<$ 1    &  $<$ 1 -- 9     &   $<$ 1 -- 3     &  $<$ 1 -- 19     &  1.1 -- 3    &  $<$ 1 -- 40   \\
Leptons                  &  1.5    &  3 -- 17      &   2 -- 9     &  1.2 -- 13     &  1.7 -- 7    &  $<$ 1 -- 16   \\
Luminosity               &  1.5    &  1.7 -- 2     &   1.3 -- 1.9     &  $<$ 1 -- 4    &  1.5 -- 2    &  $<$ 1 -- 1.9  \\
MC statistics            &  5    &  10 -- 40      &   6 -- 30      &  6 -- 180    &  8 -- 30     &  7 -- 90   \\
\hline
Total systematics         &  13    &  19 -- 90      &   13 -- 60     &  12 -- 180     &  15 -- 50      &  15 -- 200   \\
Data statistics          &  20    &  50 -- 160     &   30 -- 110    &  30 -- 400     &  40 -- 100      &  50 -- 300   \\
Total uncertainty         &  23    &  50 -- 190     &   40 -- 120    &  30 -- 500     &  40 -- 100     &  50 -- 400   \\
\hline
\hline
\end{tabular}
\end{table}



\section{Results}
\label{sec:result}

\begin{figure}[htb!]
\centering
\includegraphics[width=0.70\linewidth]{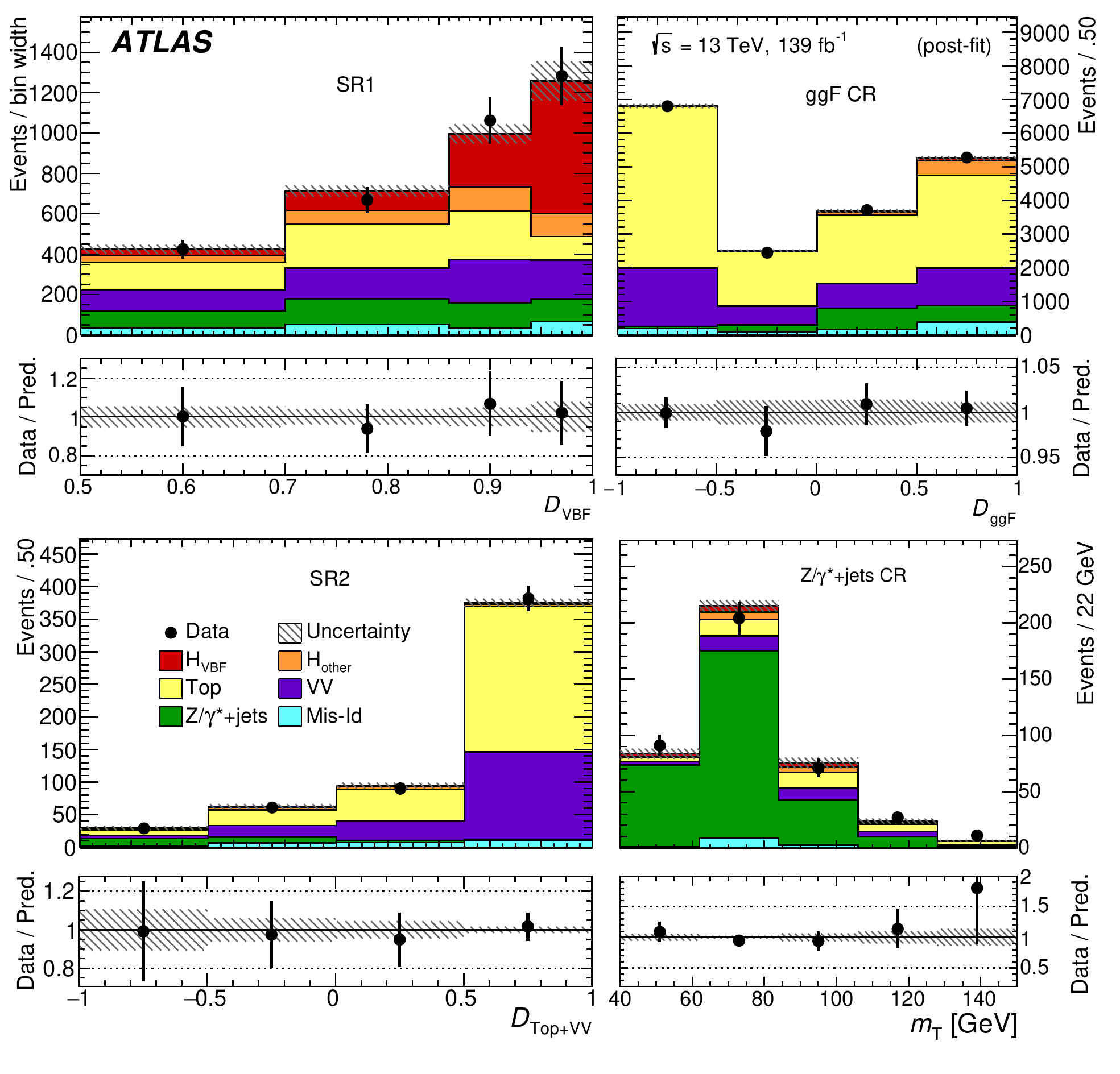}
\caption{The distributions of the discriminants in data and in the MC templates as results of the fit to data performed for the measurement of the integrated fiducial cross-section. The distributions include $D_{\text{VBF}}$ and $D_{\text{top+VV}}$ in SR1 (top left) and SR2 (bottom left), respectively, $\mT$ in the $Z/\gamma^*$+jets CR (bottom right), and $D_{\text{ggF-CR}}$ in the ggF CR (top right).}
\label{fig:discriminants_incl}
\end{figure}
\begin{figure}[htb!]
\centering
\includegraphics[width=0.50\linewidth]{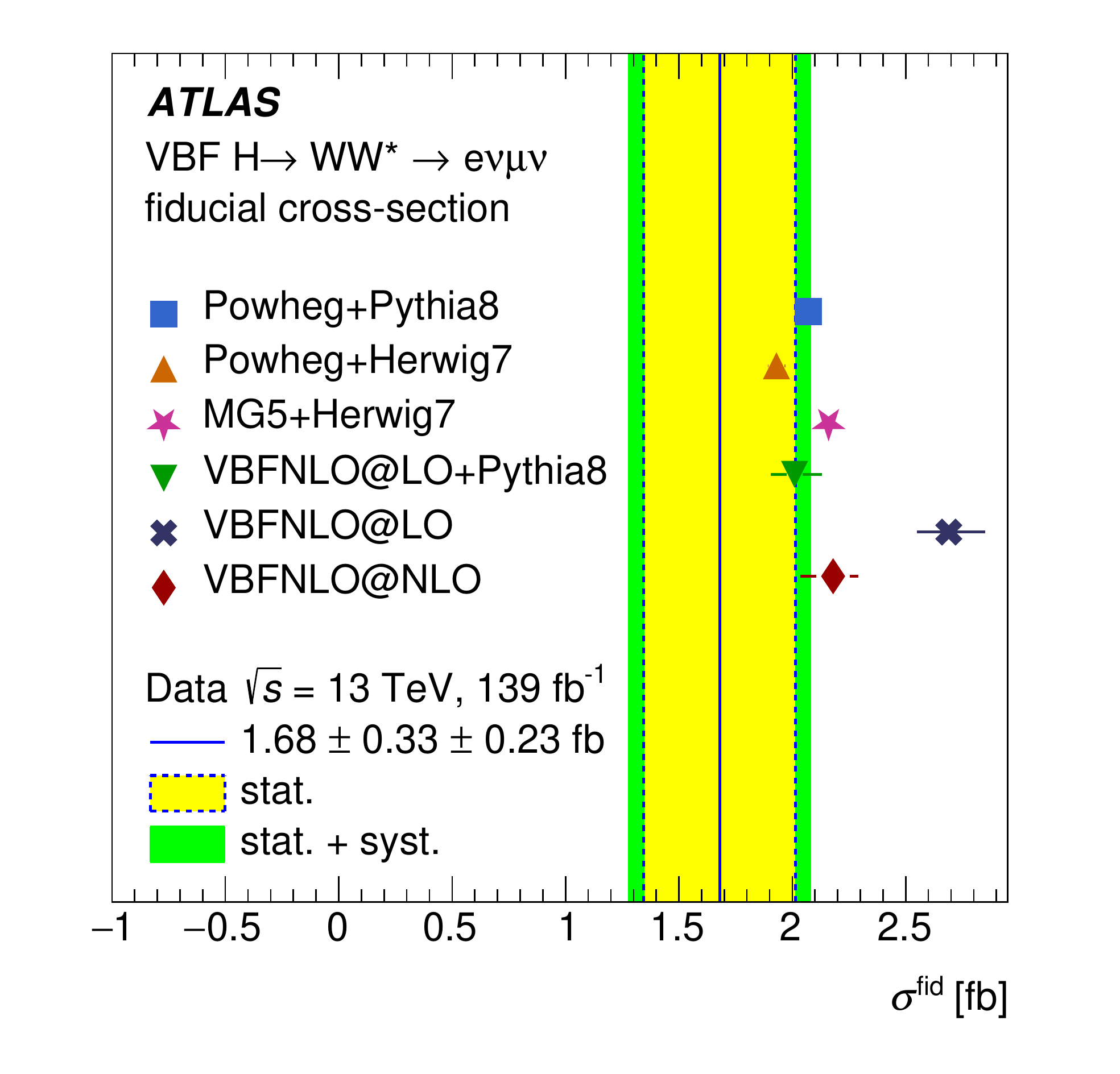}
\caption{The measured fiducial cross-section in comparison with the theoretical predictions in the fiducial phase space obtained from the parton shower generators \POWHEGva+\PYTHIAV{8}, \POWHEGva+\HERWIGV{7}, \MGFiveHw, and \vbfnlo$@$LO+\PYTHIAV{8} at particle level and from the parton-level generator \vbfnlo\, at LO and NLO\@. The vertical bands around the measurement indicate the statistical uncertainty (inner band) and the sum in quadrature of statistical, systematic, and luminosity uncertainties (outer band). Uncertainties are calculated for all predictions excluding \MGFiveHw, although for \POWHEGva+\PYTHIAV{8} they are not visible as they are compatible with the size of the marker.}
\label{fig:Result_InclXsec}
\end{figure}
The results include the measurements of the integrated fiducial cross-section  and the differential fiducial cross-section as a function of the variables presented in Section~\ref{sec:fid_phasespace_obs}. The experimental cross-sections are also compared with the SM predictions described in Section~\ref{sec:dataset_mc}.
 
\subsection{Integrated fiducial cross-section}
\label{sec:result_fiducial}

The distributions of the discriminants in data, compared with the templates as results of the fit to data, are shown in Figure~\ref{fig:discriminants_incl} for the case of the integrated fiducial cross-section measurement. The post-fit MC templates are in good agreement with the data in all kinematic ranges of the discriminant distributions in the SR1, SR2, and in the ggF CR and the $Z/\gamma^*$+jets CR\@.
 
The measured cross-section for VBF \hwwlnln production in the fiducial phase-space region, as defined in Table~\ref{tab:EvtSelection}, is
 
\begin{equation}
\sigma^{\mathrm{fid}} = 1.68 \pm 0.40 ~\mathrm{fb} = 1.68 \pm 0.33 ~\textrm{(stat)} \pm 0.23 ~\textrm{(syst)} ~\mathrm{fb}.
\label{eq:Result_xSectFid}
\end{equation}
 
The overall relative precision of the cross-section measurement is about 23\%, dominated by the statistical uncertainty in the data sample. The impact of individual sources of uncertainties is provided in Table~\ref{tab:syst}.
The normalization parameters of the backgrounds constrained by data in the control and signal regions are measured to be $0.91\pm 0.07$ for the top+$VV$ process in the SR, $0.85 \pm 0.07$ for the $Z/\gamma^*$+jets process in $Z/\gamma^*$+jets CR, $0.96 \pm 0.06$ and $0.83 \pm 0.15$ for the top+$VV$ and $Z/\gamma^*$+jets processes in ggF CR, respectively, and $1.13 \pm 0.41$ for the ggF process. They are summarized in Appendix~\ref{app:bkg_norm}.
The correlations between the measured cross-section and the background normalization factors are calculated including statistical and systematic uncertainties (see Appendix~\ref{app:corr_matrices}). The absolute values of the correlations between the parameters are smaller than $20\%$, with the exception of a positive correlation of $22\%$ between the signal cross-section and the top+$VV$ process in the SR, an anti-correlation of $33\%$  between the signal cross-section and the ggF Higgs boson normalization factor, and a correlation of $61\%$ between the normalization factors of the top+$VV$ processes in the SR and in the ggF CR\@.
 
The comparison between the measured cross-section and theoretical predictions is given in Figure~\ref{fig:Result_InclXsec}. The predictions are obtained from the parton shower generators \POWHEGva+\PYTHIAV{8}, \POWHEGva+\HERWIGV{7}, \MGFiveHw, and \vbfnlo$@$LO+\PYTHIAV{8} at particle level, as well as from the parton-level calculations \vbfnlo$@$LO and \vbfnlo$@$NLO, as described in Section~\ref{sec:dataset_mc} and summarized in Table~\ref{tab:MCSignalSamples}.
 
The theoretical uncertainties are estimated for the two predictions based on \POWHEGva\, generator and for the \vbfnlo$@$NLO and \vbfnlo$@$LO+\PYTHIAV{8} predictions, and are found to be smaller than the uncertainty in the experimental cross-section. The measured cross-section is overestimated by $15-28\%$ by the theoretical predictions calculated at NLO or at LO with the parton shower, but they are compatible at the level of approximately one standard deviation. The integrated fiducial cross-section predicted by \POWHEGva+\PYTHIAV{8} is $2.07\pm 0.04$ fb. The  uncertainty includes renormalization and factorization scale variations as well as PDF and $\alpha_s$ uncertainties. The \POWHEGva+\HERWIGV{7} prediction is $7\%$ lower than the one by \POWHEGva+\PYTHIAV{8}, while the predictions by \vbfnlo$@$NLO and  \MGFiveHw\, are similar and $4\%-5\%$ greater than the one by \POWHEGva+\PYTHIAV{8}. The uncertainty in the \vbfnlo$@$NLO is estimated to be $^{+5\%}_{-7\%}$, which makes this prediction compatible with the one by \POWHEGva+\PYTHIAV{8}.
The prediction by \vbfnlo$@$LO+\PYTHIAV{8} is also in agreement with the one by \POWHEGva+\PYTHIAV{8} and is only $3\%$ lower. This result is to be compared with the fixed-order calculation by \vbfnlo$@$LO, which overestimates \vbfnlo$@$LO+\PYTHIAV{8} and \POWHEGva+\PYTHIAV{8} predictions by approximately $24\%$.
The predicted integrated fiducial cross-section has a small dependence on the PDF set, i.e., less than $3\%$, when comparing the \vbfnlo$@$NLO predictions using the NLO \texttt{CT14}, \texttt{MMHT14}, and \nnpdfnlo PDF sets.
 
\begin{figure}[tb!]
\centering
\includegraphics[width=0.80\linewidth]{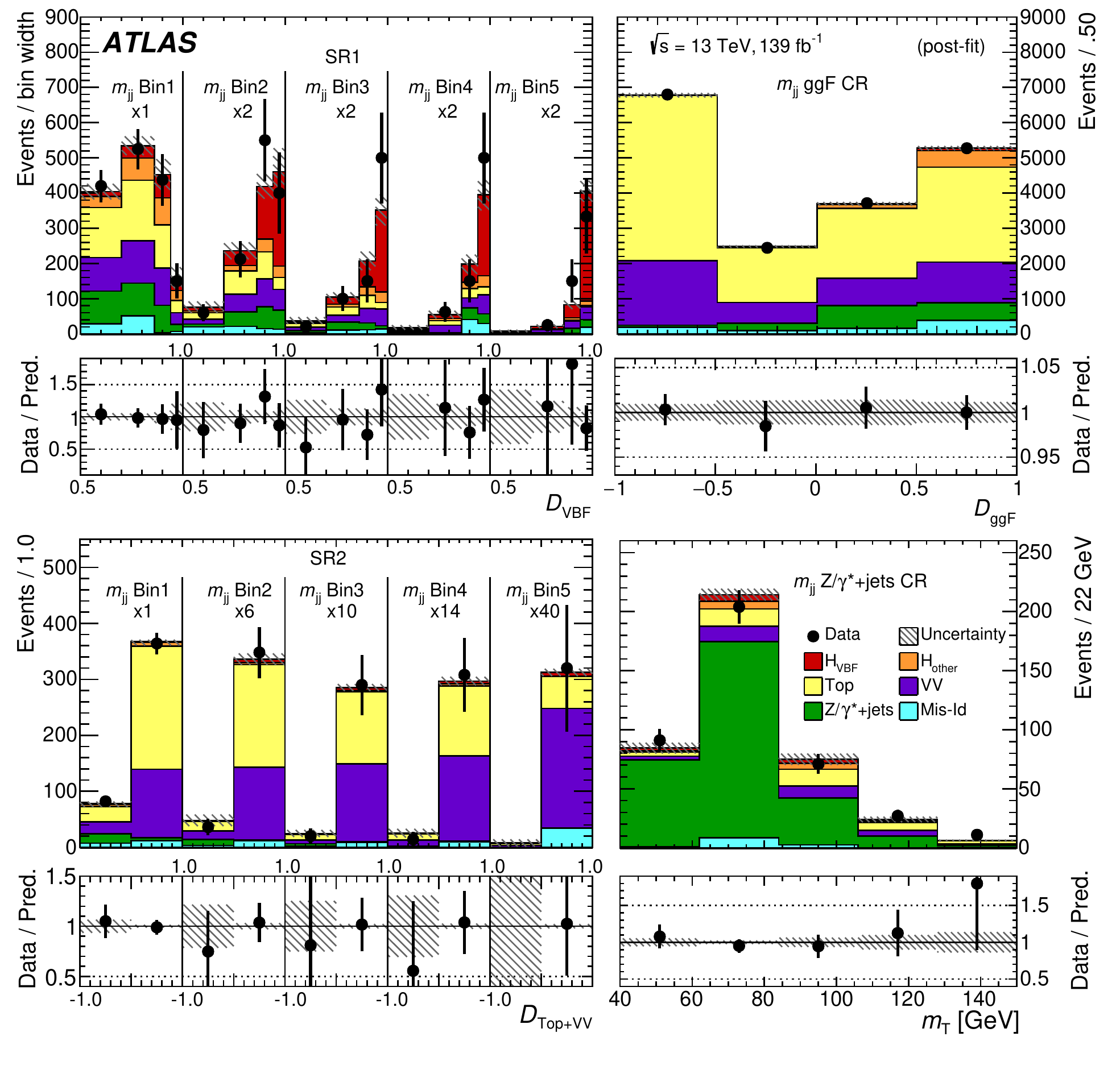}
\caption{The distributions of the discriminants in data and in the MC templates as results of the fit to data performed for the measurement of the \mjj differential cross-section. The distributions of $D_{\text{VBF}}$  and $D_{\text{top+VV}}$ discriminants in SR1 (top left) and SR2 (bottom left), respectively, are shown for each \mjj bin in the ranges of [0.5, 1.0] for $D_{\text{VBF}}$ and [-1.0, 1.0] for $D_{\text{top+VV}}$, while \mT and $D_{\text{ggF-CR}}$ are shown in the $Z/\gamma^*$+jets CR (bottom right) and in the ggF CR (top right), respectively. For visibility, the event yield in each bin of a kinematic distribution is multiplied by the factor specified in the legend.}
\label{fig:discriminants_mjj}
\end{figure}
\begin{figure}[tb!]
\centering
\subfloat[]{\includegraphics[width=0.4\linewidth]{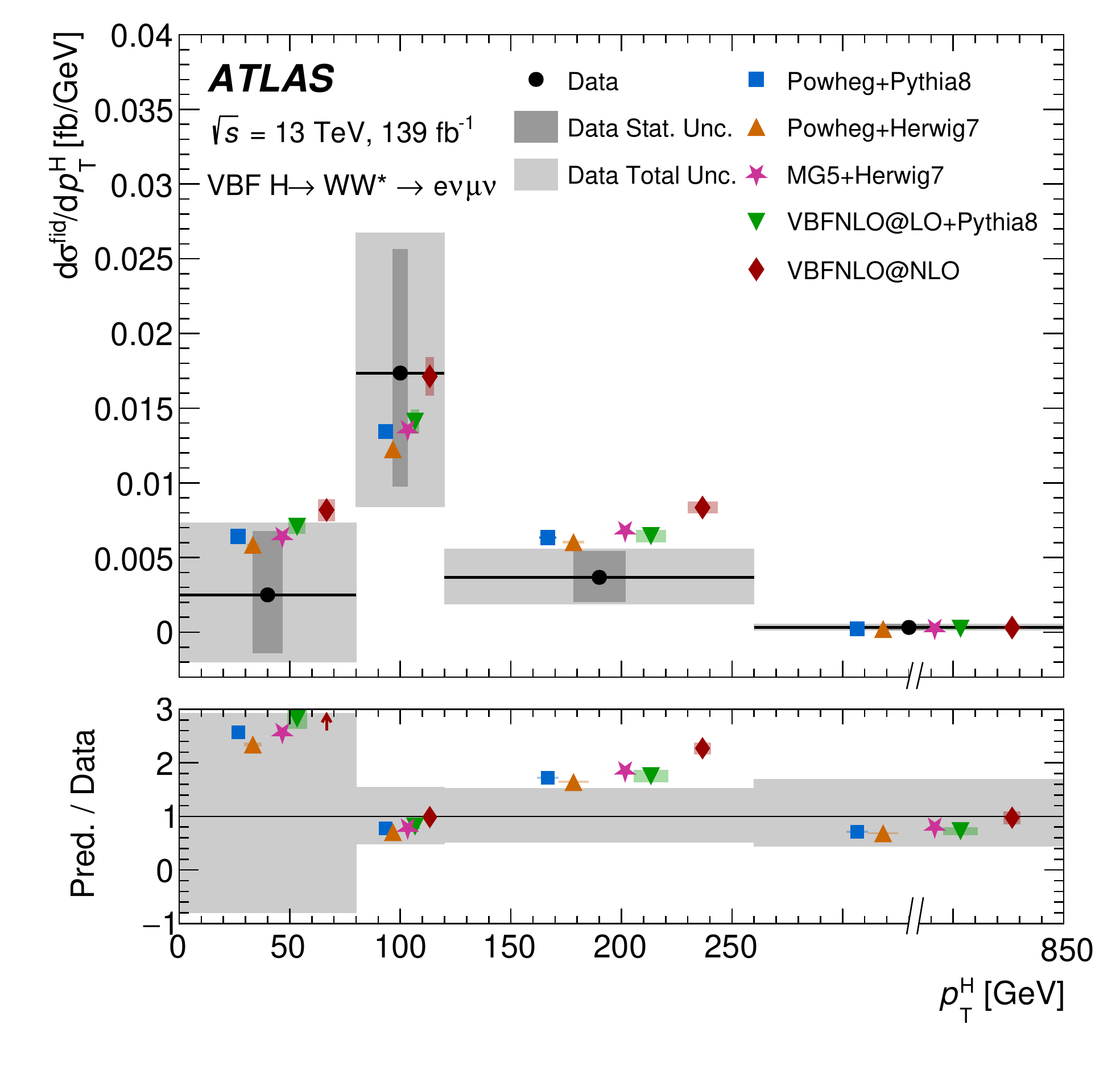}\label{fig:diff_pt_H}}
\subfloat[]{\includegraphics[width=0.4\linewidth]{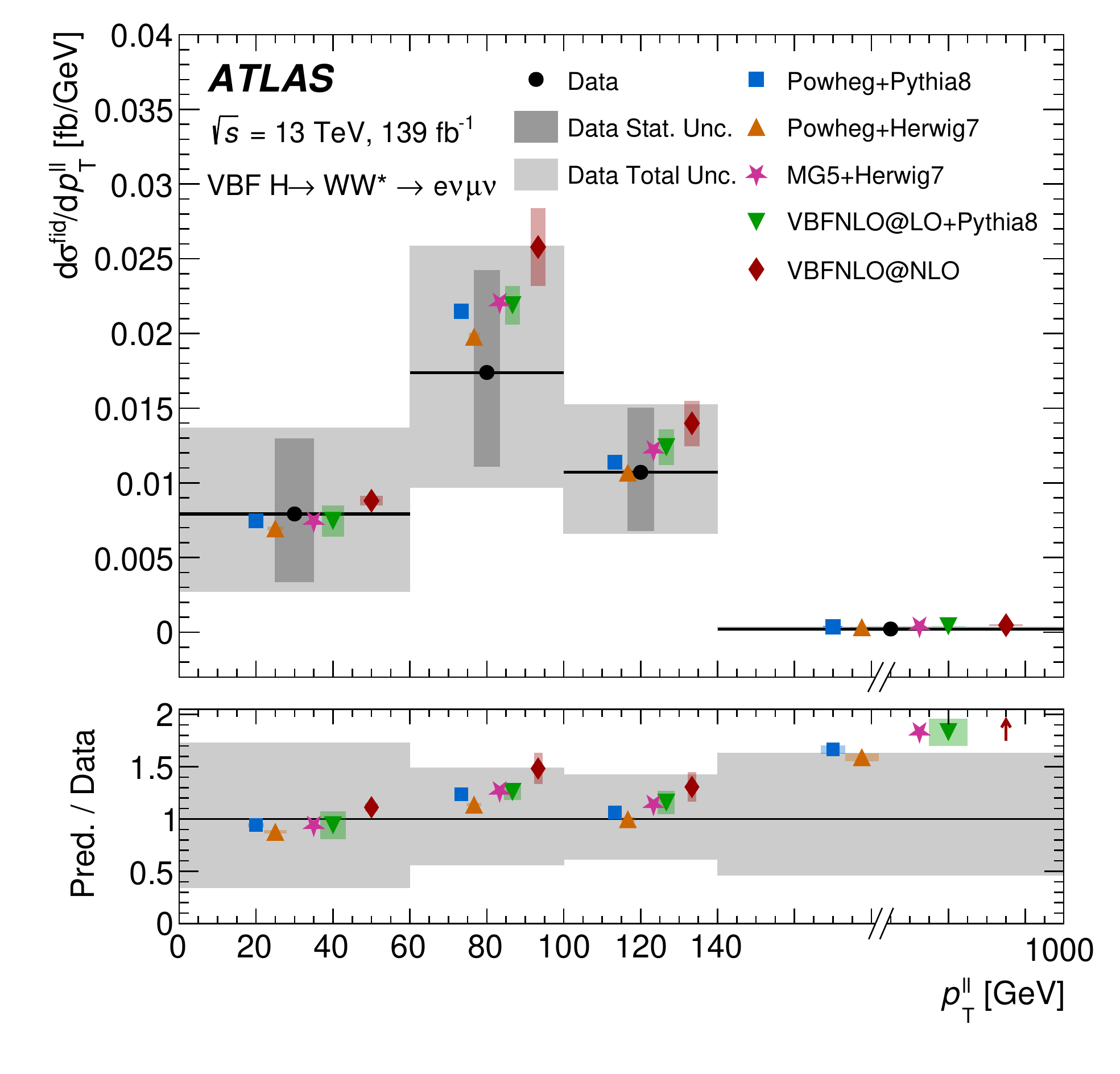}\label{fig:diff_Ptll}}
\caption{The measured differential fiducial cross-sections for \protect\subref{fig:diff_pt_H} $\pTH$ and \protect\subref{fig:diff_Ptll} $\pTll$, in comparison with the theoretical predictions from the parton shower generators \POWHEGva+\PYTHIAV{8}, \POWHEGva+\HERWIGV{7}, \MGFiveHw, and \vbfnlo$@$LO+\PYTHIAV{8} at particle level and from the parton-level generator \vbfnlo$@$NLO\@. The inner boxes on the data points show the statistical uncertainties, while the total uncertainties are indicated by the outer boxes. The error bars on the expected cross-sections indicate the PDF and scale systematic uncertainties, calculated as described in Section~\ref{sec:dataset_mc}. The bottom panels show the ratios of different predictions to the data and the bands represent the total uncertainties of the measurement.}
\label{fig:Result_diffXsect_pTH-pTll}
\end{figure}

\subsection{Differential fiducial cross-sections}
\label{sec:result_diff}
 
The distributions of the discriminants in data, compared with the MC templates as results of the fits to data, are shown in Figure~\ref{fig:discriminants_mjj} for the case of the \mjj differential cross-section measurement.
The post-fit MC templates are in good agreement with the data in all kinematic ranges of the discriminant distributions in the signal and control regions for all differential cross-section measurements.
 
The bin boundaries for the measured distributions in the signal region were determined, before the unblinding of the data, based on the \POWHEGva+\PYTHIAV{8} prediction scaled to the expected data yield. In some distributions, bins with low statistics in data are merged with a neighbor if the relative uncertainty per unit of bin width is three times larger than the average uncertainty per units of bin width as calculated from all the bins of that distribution. The definitions of the measured kinematic distributions are given in Section~\ref{sec:fid_phasespace_obs}. The differential fiducial cross-sections as a function of the transverse momentum of the Higgs boson and the \pT of the dilepton system are shown in Figure~\ref{fig:Result_diffXsect_pTH-pTll}, while those as a function of variables that probe the kinematics of the Higgs boson decay products, such as \pTlone, \pTltwo, \mll, |\dyll|, |\dphill|, and $\cos(\theta^*_{\eta})$, are shown in Figure~\ref{fig:Result_diffXsect_leptons}. Differential cross-sections as a function of variables that probe the jet activity in VBF events follow in Figure~\ref{fig:Result_diffXsect_jets}, which includes the \pTjone, \pTjtwo, \mjj, |\dyjj|, and \dphijj.
\begin{figure}[ph!]
\centering
\subfloat[]{\includegraphics[width=0.35\linewidth]{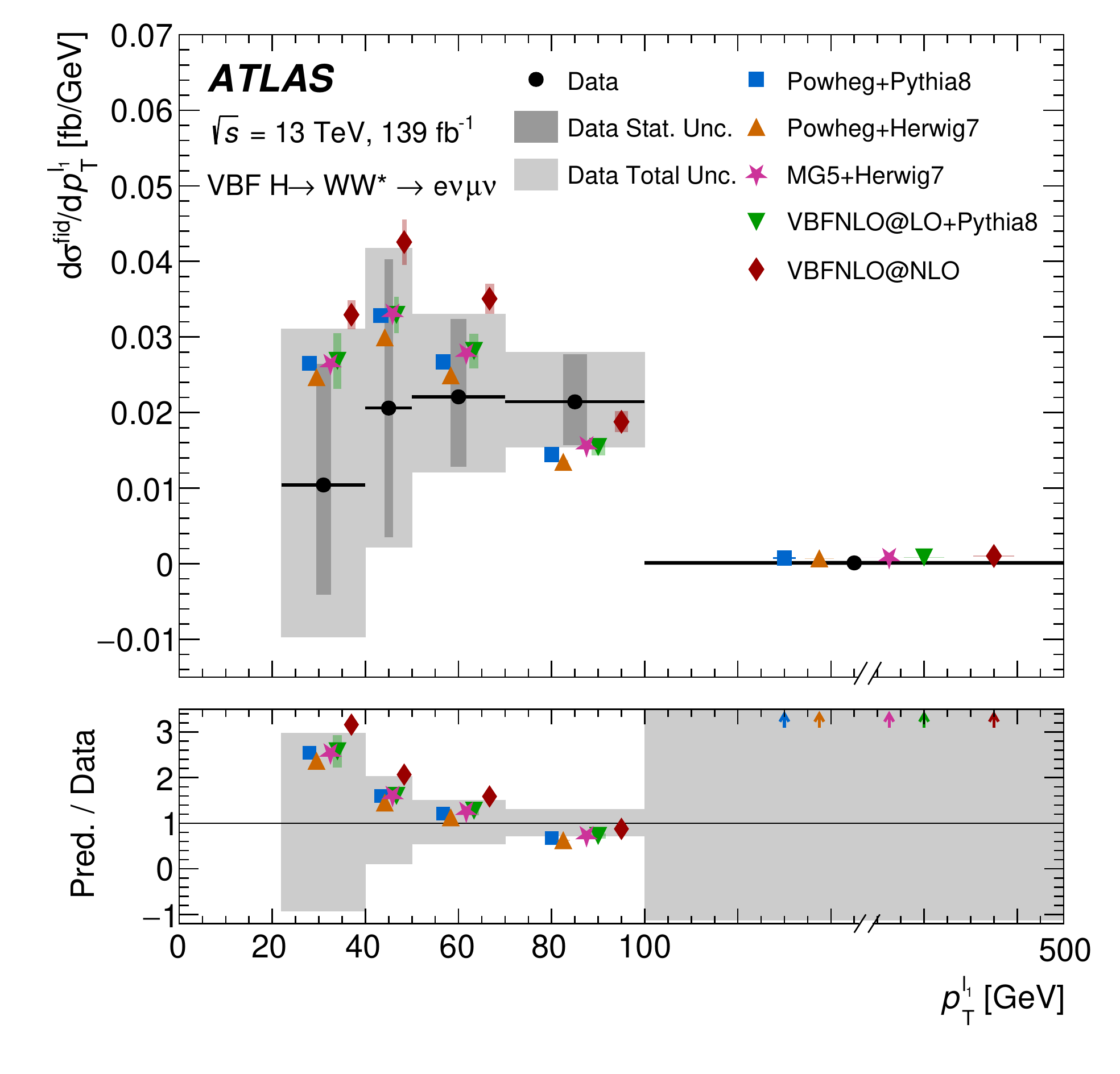}\label{fig:diff_lep0_pt}}
\subfloat[]{\includegraphics[width=0.35\linewidth]{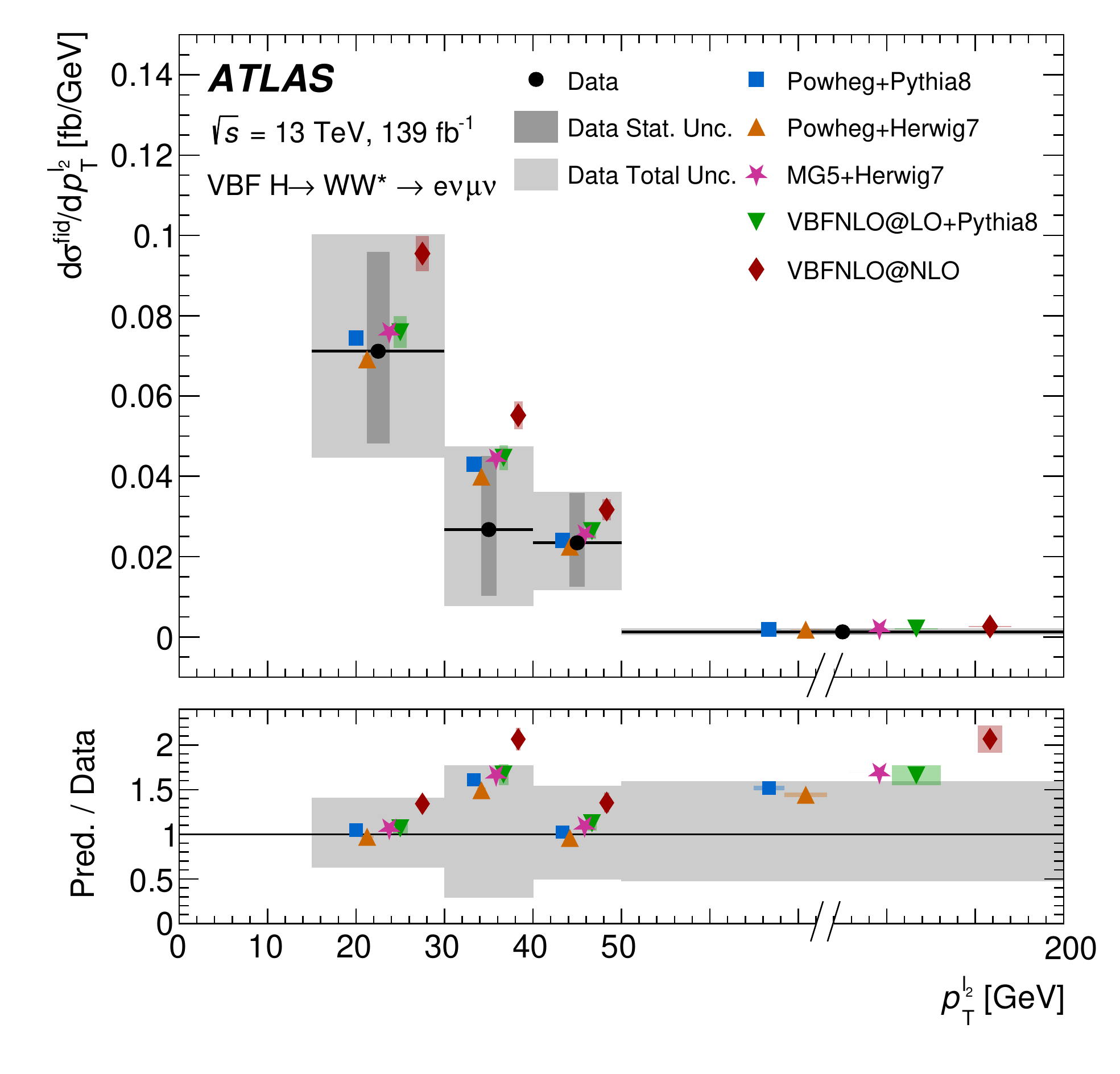}\label{fig:diff_lep1_pt}} \\
\subfloat[]{\includegraphics[width=0.35\linewidth]{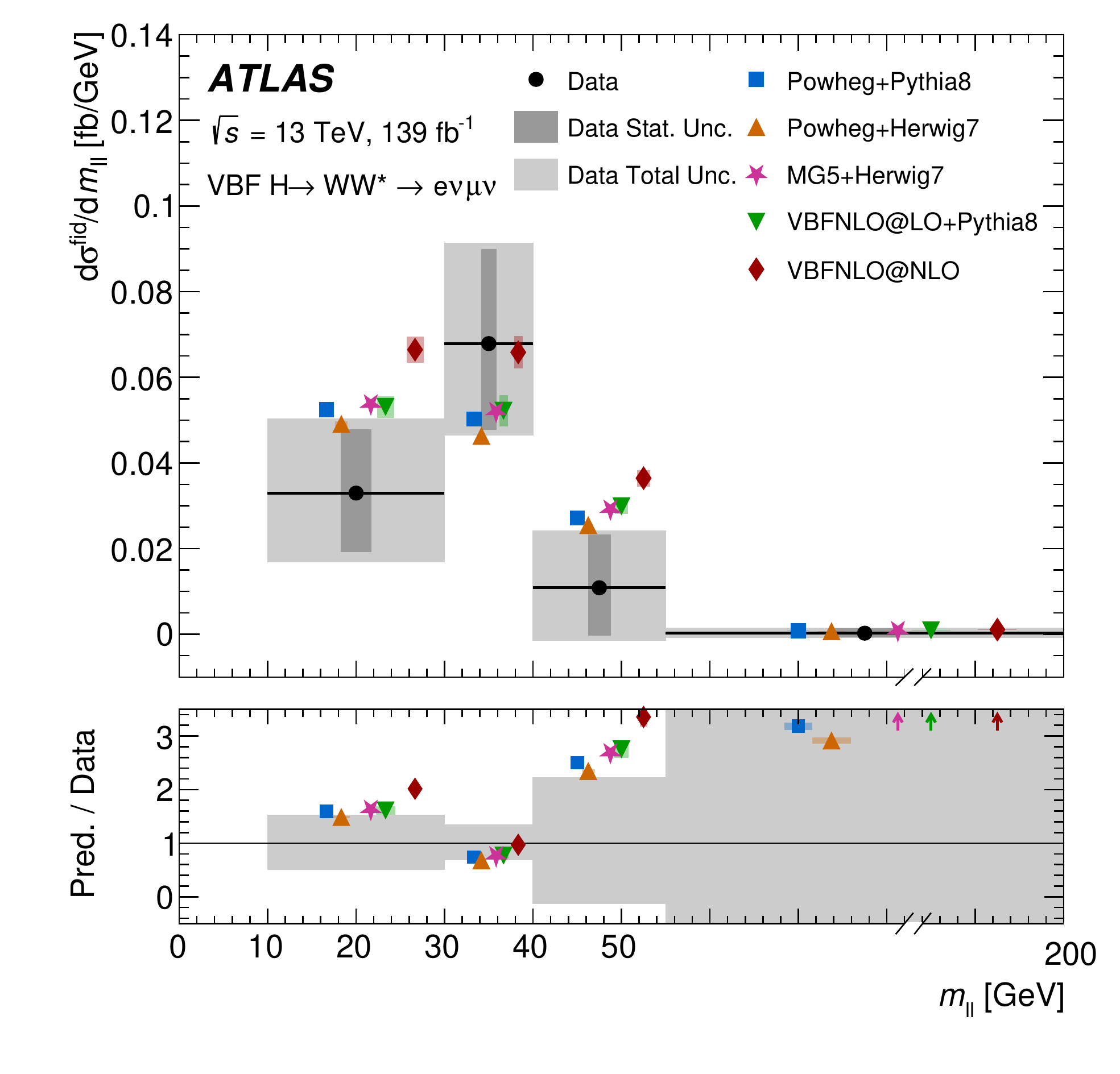}\label{fig:diff_mll}}
\subfloat[]{\includegraphics[width=0.35\linewidth]{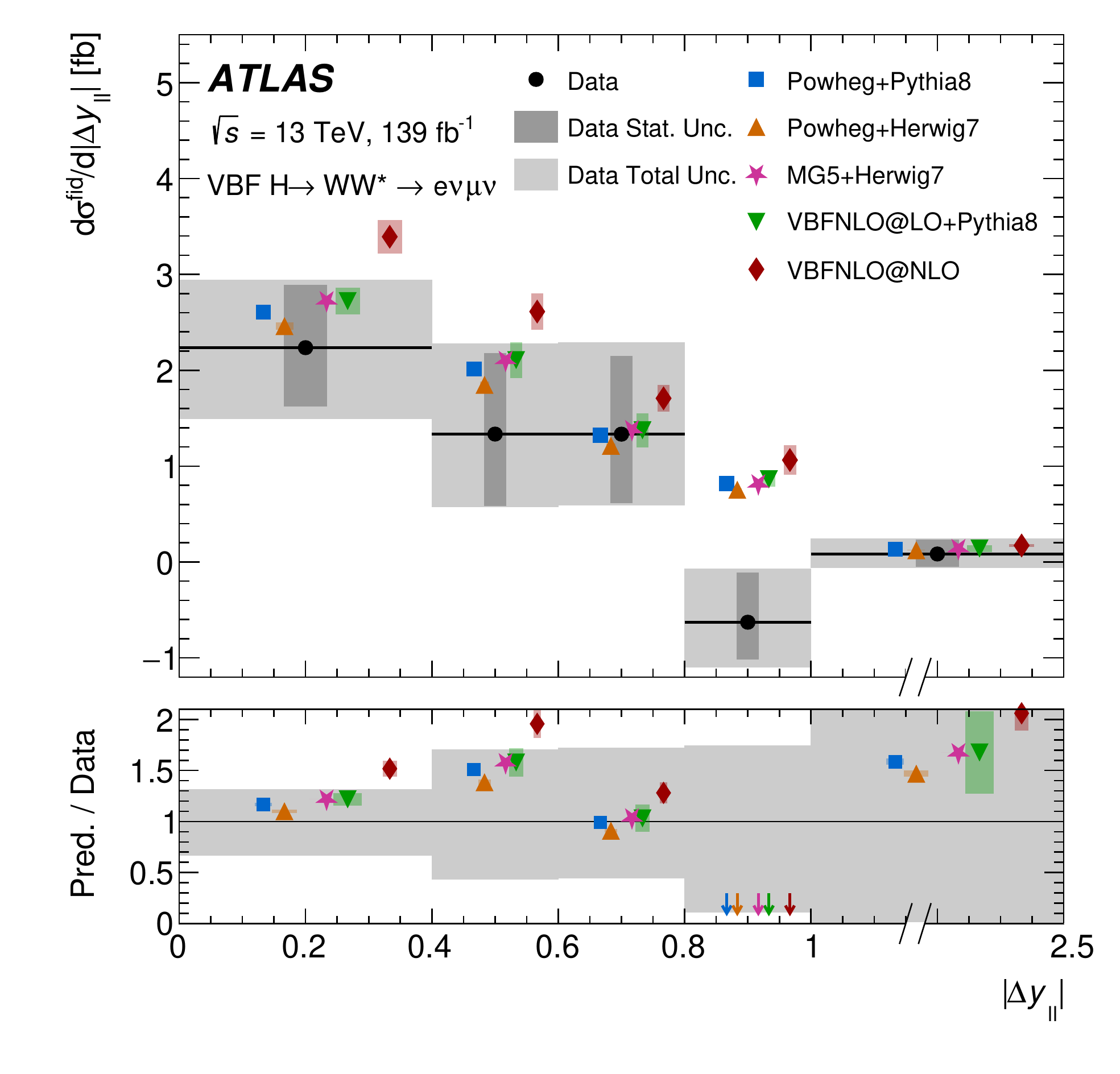}\label{fig:diff_dyll}} \\
\subfloat[]{\includegraphics[width=0.35\linewidth]{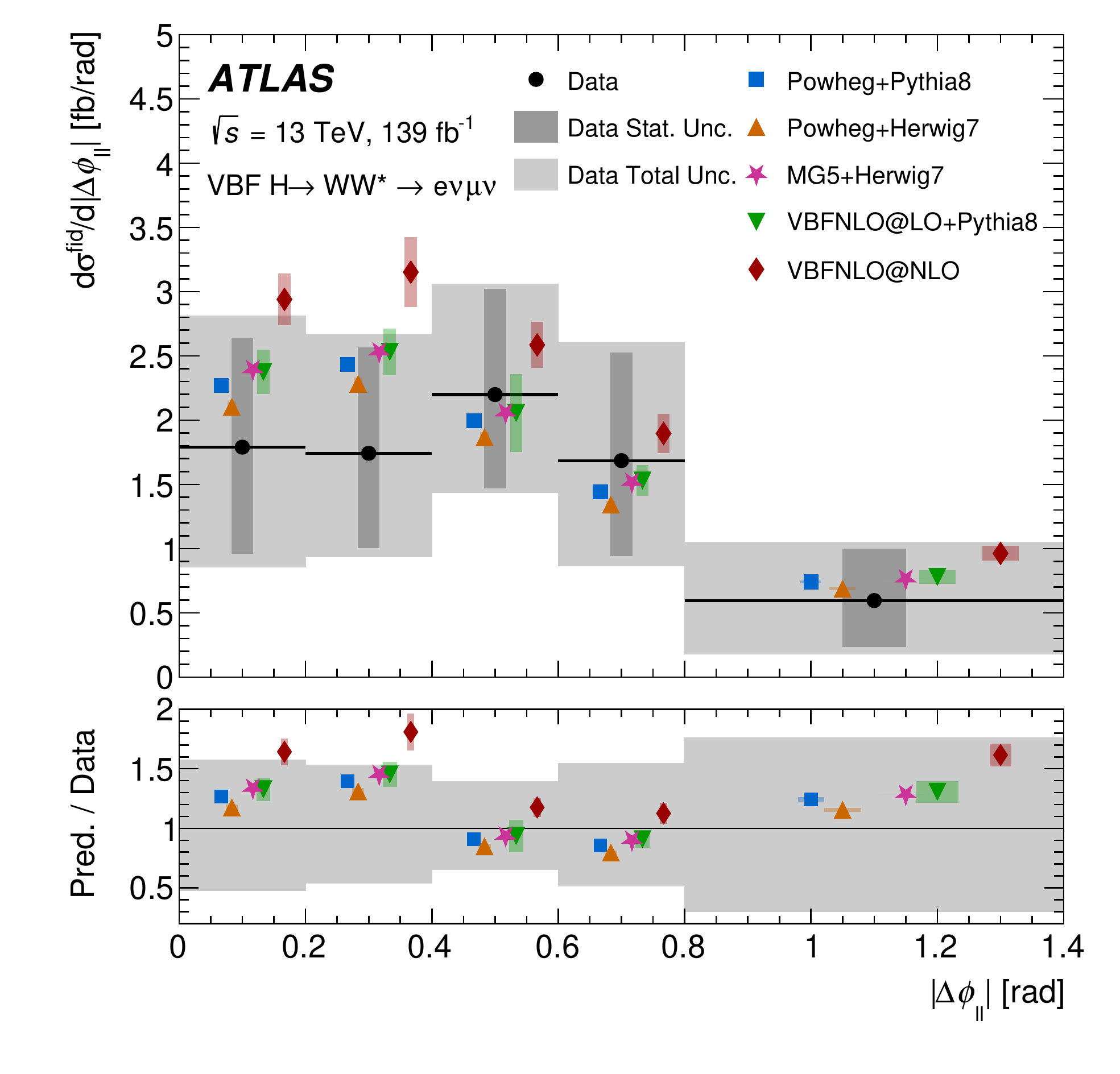}\label{fig:diff_dphill}}
\subfloat[]{\includegraphics[width=0.35\linewidth]{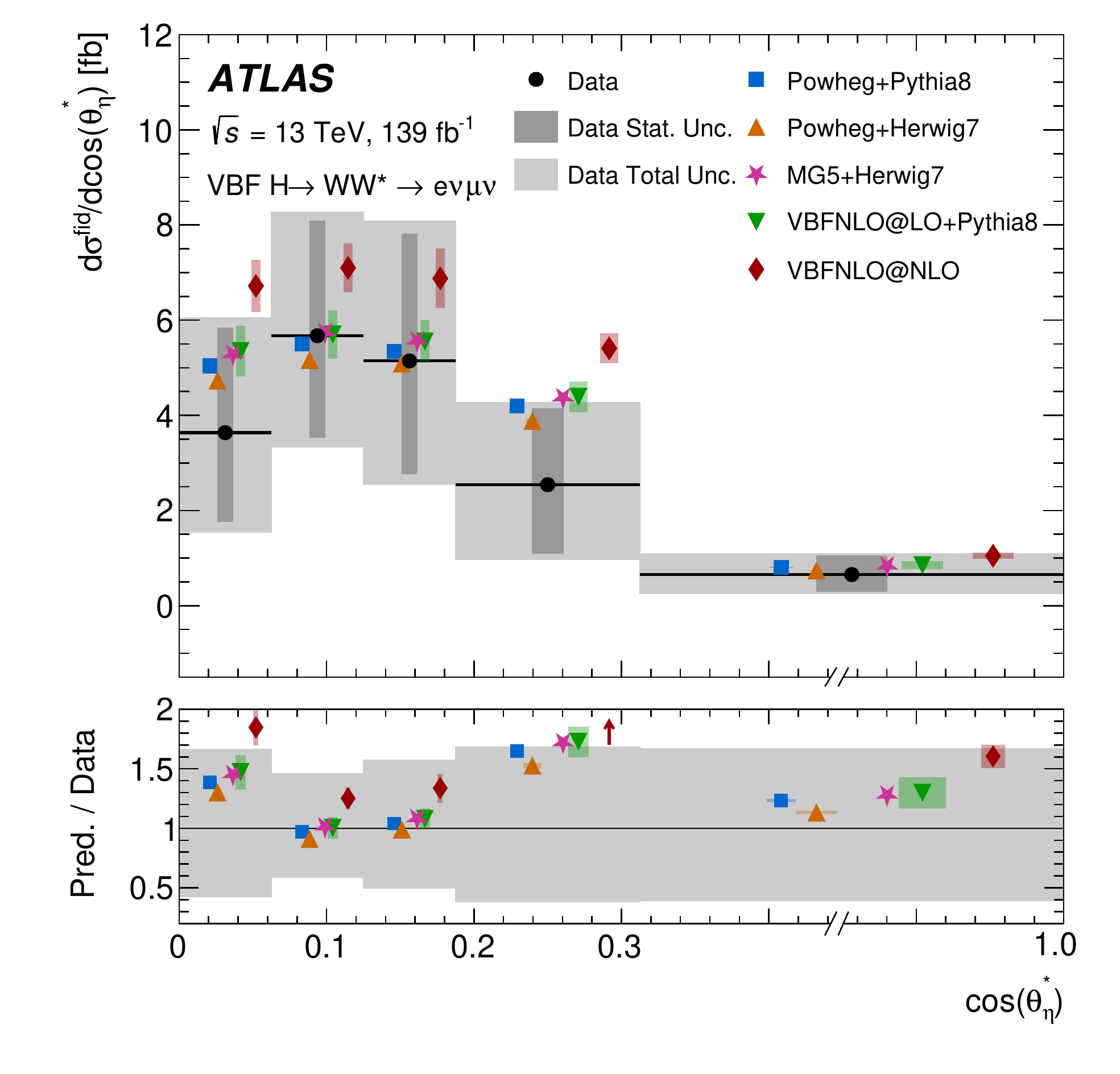}\label{fig:diff_costhetastar}}
\caption{The measured differential fiducial cross-sections for \protect\subref{fig:diff_lep0_pt} $\pTlone$, \protect\subref{fig:diff_lep1_pt} $\pTltwo$, \protect\subref{fig:diff_mll} $\mll$, \protect\subref{fig:diff_dyll} $|\dyll|$, \protect\subref{fig:diff_dphill} $|\dphill|$, and \protect\subref{fig:diff_costhetastar} $\cos(\theta^*_{\eta})$. The experimental results are compared with theoretical predictions from the parton shower generators \POWHEGva+\PYTHIAV{8}, \POWHEGva+\HERWIGV{7}, \MGFiveHw, and \vbfnlo$@$LO+\PYTHIAV{8} at particle level and from the parton-level generator \vbfnlo$@$NLO\@. The inner boxes on the data points show the statistical uncertainties, while the total uncertainties are indicated by the outer boxes. The error bars on the expected cross-sections indicate the PDF and scale systematic uncertainties, calculated as described in Section~\ref{sec:dataset_mc}. The bottom panels show the ratios of different predictions to the data and the bands represent the total uncertainties of the measurement.}
\label{fig:Result_diffXsect_leptons}
\end{figure}
\begin{figure}[ph!]
\centering
\subfloat[]{\includegraphics[width=0.35\linewidth]{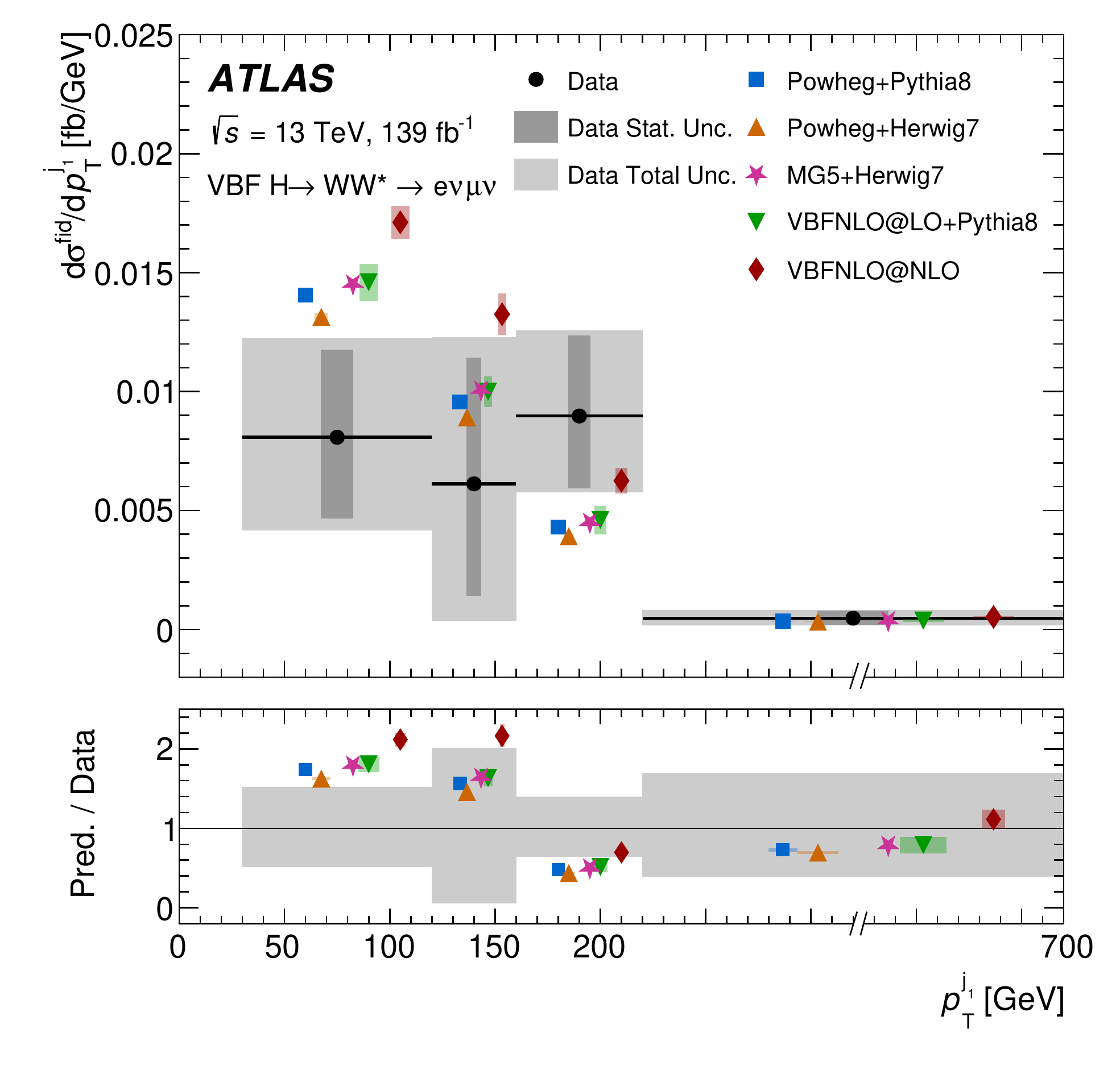}\label{fig:diff_jet0_pt}}
\subfloat[]{\includegraphics[width=0.35\linewidth]{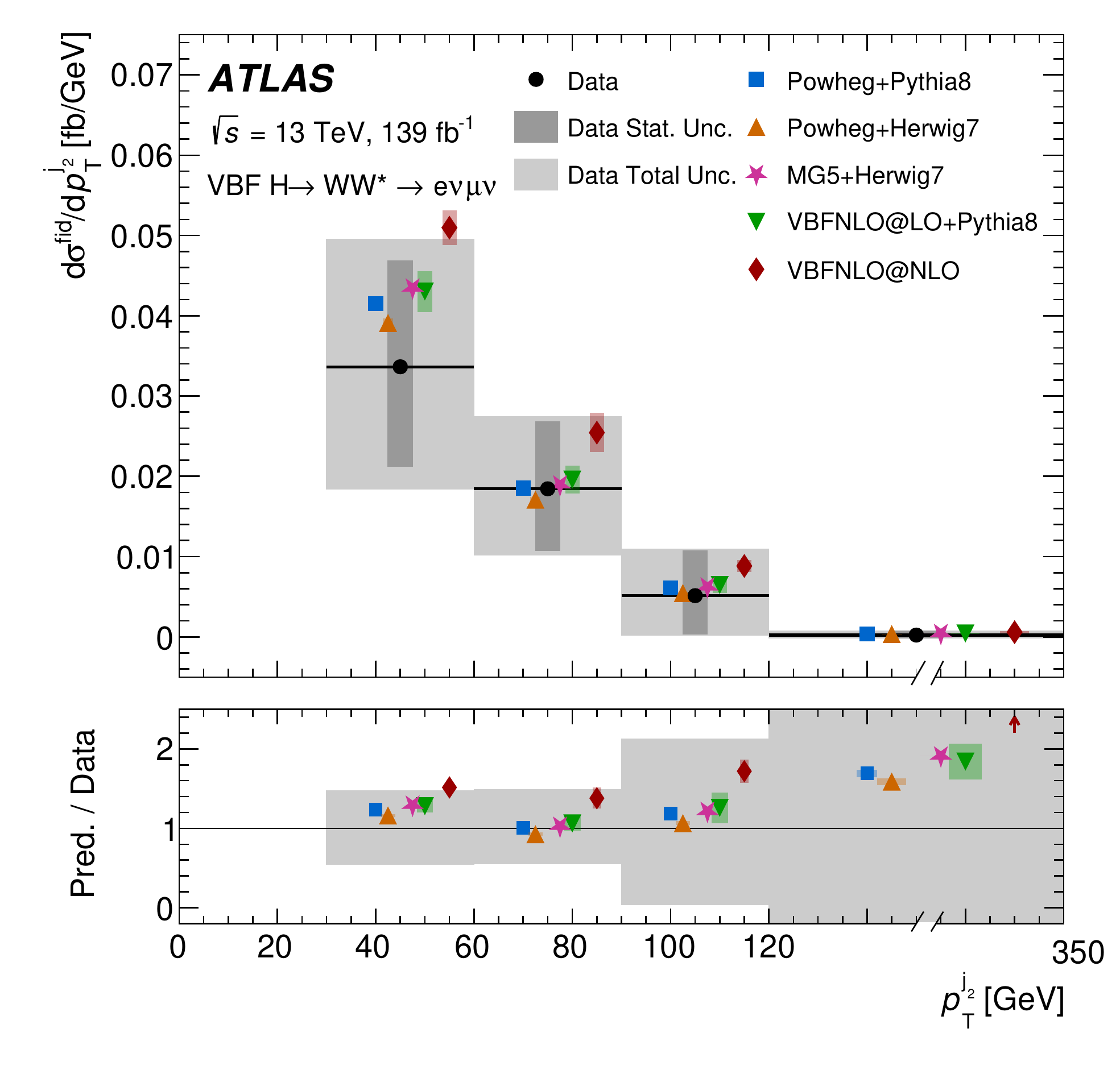}\label{fig:diff_jet1_pt}} \\
\subfloat[]{\includegraphics[width=0.35\linewidth]{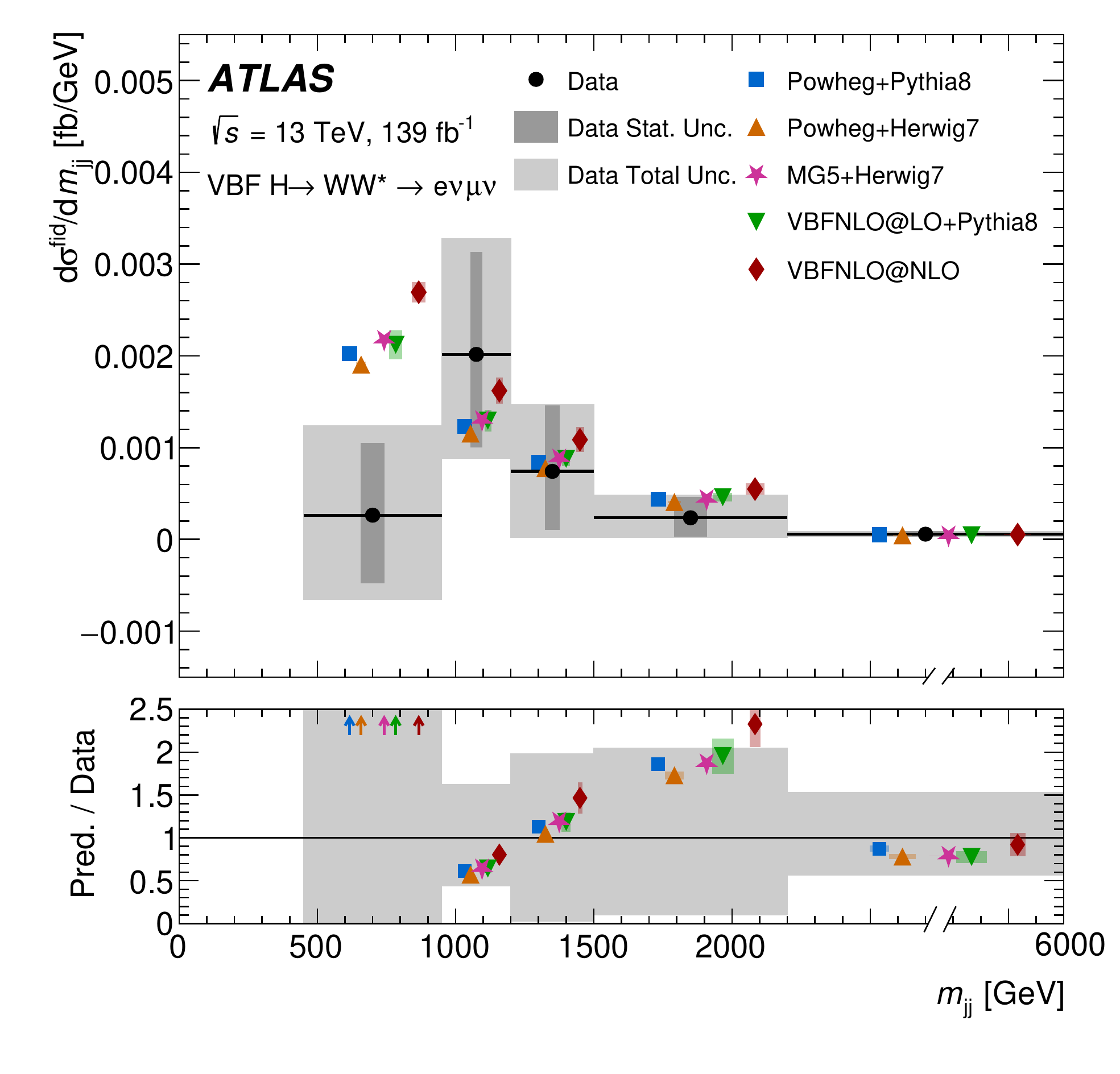}\label{fig:diff_mjj}}
\subfloat[]{\includegraphics[width=0.35\linewidth]{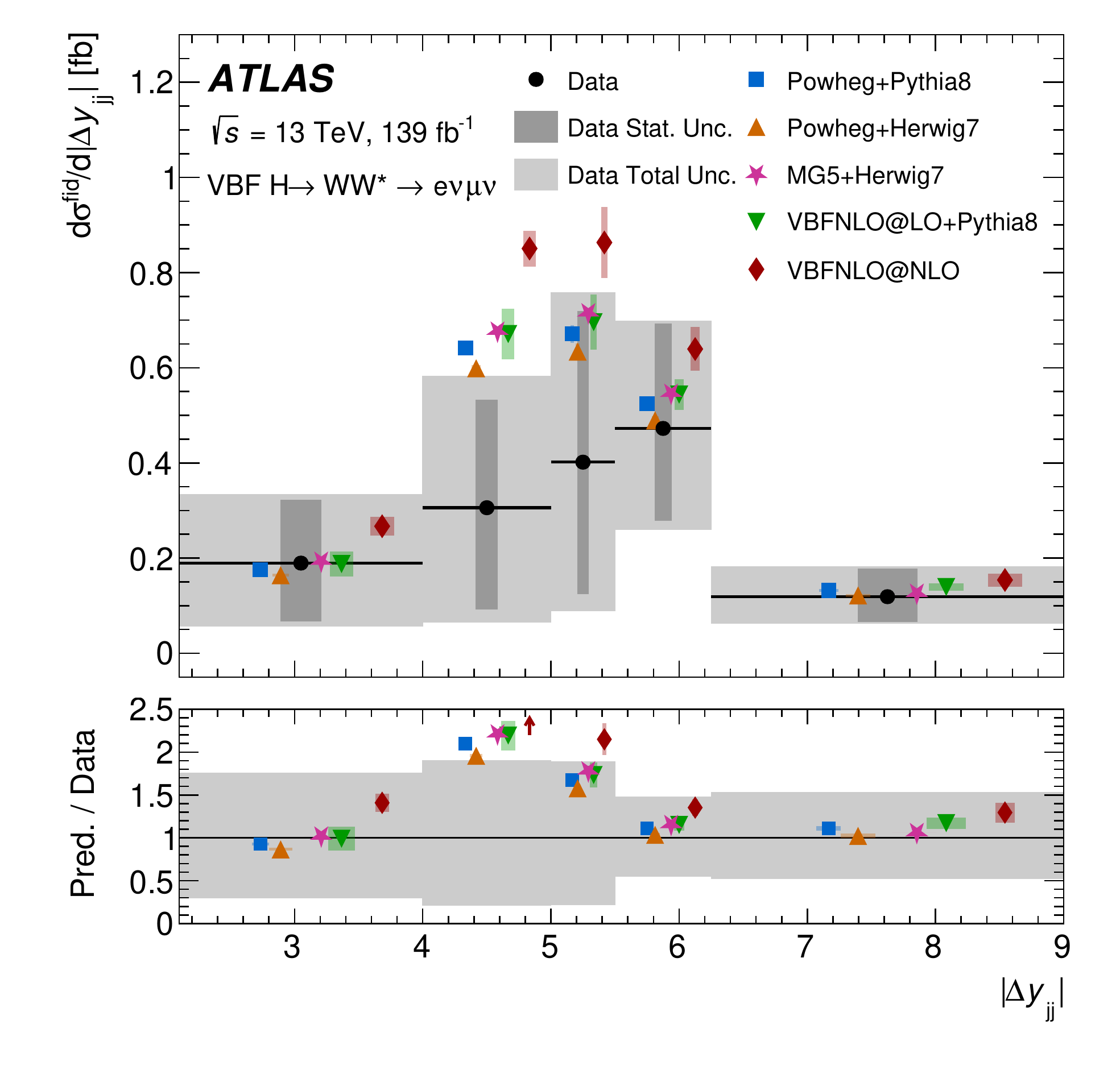}\label{fig:diff_dyjj}} \\
\subfloat[]{\includegraphics[width=0.35\linewidth]{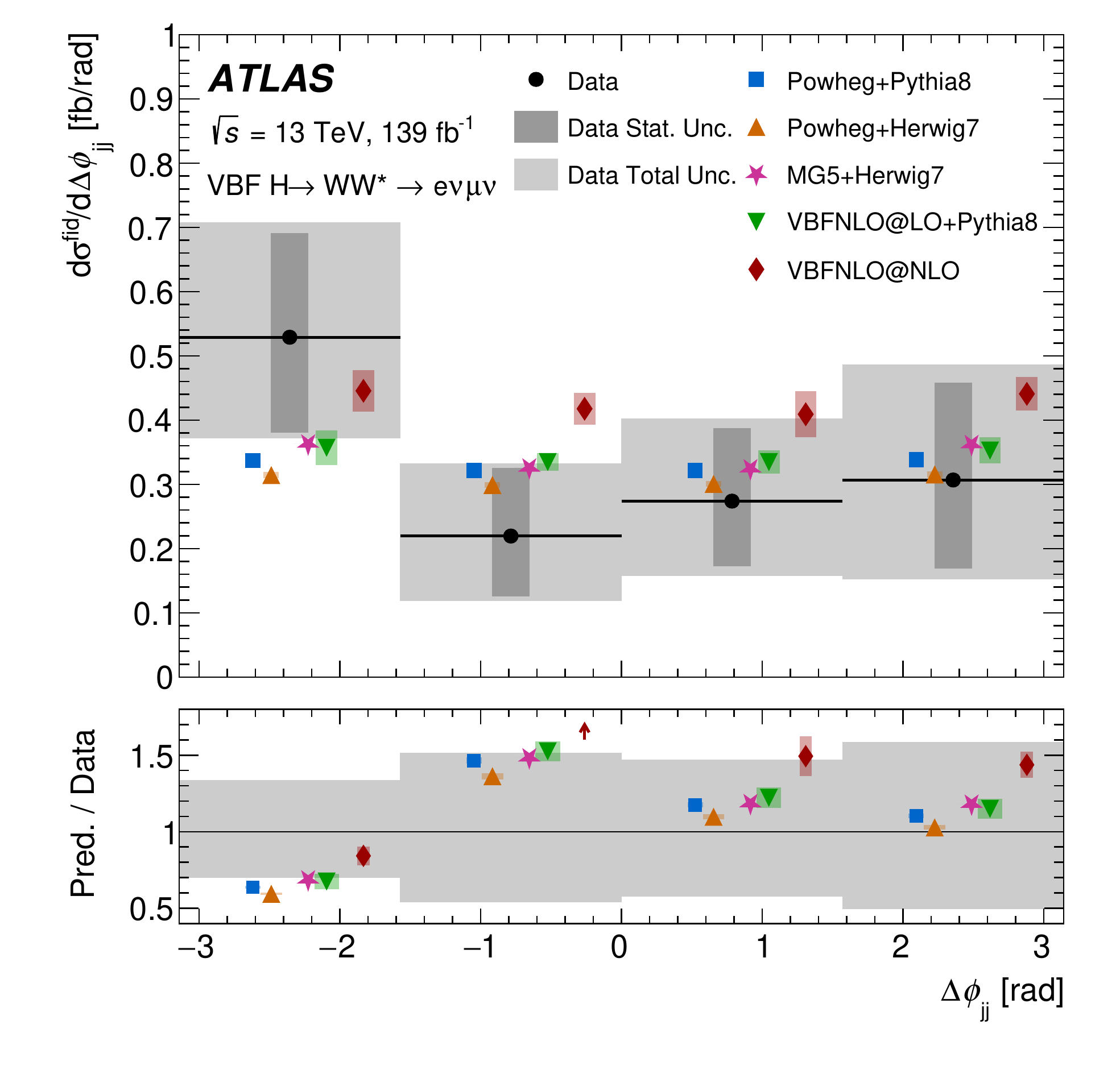}\label{fig:diff_dphijj}}
\caption{The measured differential fiducial cross-sections for \protect\subref{fig:diff_jet0_pt} $\pTjone$, \protect\subref{fig:diff_jet1_pt} $\pTjtwo$, \protect\subref{fig:diff_mjj} $\mjj$, \protect\subref{fig:diff_dyjj} $|\dyjj|$, and \protect\subref{fig:diff_dphijj} $|\dphijj|$. The experimental results are compared with theoretical predictions from the parton shower generators \POWHEGva+\PYTHIAV{8}, \POWHEGva+\HERWIGV{7}, \MGFiveHw, and \vbfnlo$@$LO+\PYTHIAV{8} at particle level and from the parton-level generator \vbfnlo$@$NLO\@. The inner boxes on the data points show the statistical uncertainties, while the total uncertainties are indicated by the outer boxes. The error bars on the expected cross-sections indicate the PDF and scale systematic uncertainties, calculated as described in Section~\ref{sec:dataset_mc}. The bottom panels show the ratios of different predictions to the data and the bands represent the total uncertainties of the measurement.}
\label{fig:Result_diffXsect_jets}
\end{figure}

The per-bin relative precision of the differential cross-section measurements is typically in the range of $35-80\%$, but in a few bins with poor statistics it can reach or exceed $200-300\%$. Statistical uncertainties dominate the experimental precision in all bins of the distributions and the impact of individual sources of uncertainties is shown in Table~\ref{tab:syst}.
The normalization parameters of the backgrounds constrained by data in the signal and control regions are measured independently for each differential cross-section and are in the range of $0.70-1.09$ for the top+$VV$ normalization factors in the SR, $0.83-0.87$ for the $Z/\gamma^*$+jets normalization factor in the $Z/\gamma^*$+jets CR, $0.93-1.20$ for the ggF process normalization, and $0.79-0.86$ and $0.95-0.98$ for the $Z/\gamma^*$+jets and top+$VV$ normalization parameters, respectively, in
the ggF CR. These normalization factors are compatible across the differential cross-section measurements and with those measured for the integrated cross-section, as summarized in Appendix~\ref{app:bkg_norm}.

The correlations between the measured cross-section values in each bin of the distributions and the background normalization factors are calculated in the likelihood fits performed independently for each differential cross-section (see Appendix~\ref{app:corr_matrices}). The (anti-)correlations among parameters are generally less than $20\%$, with a few exceptions: anti-correlations ($10\%-50\%$) are observed between the yields of adjacent bins of the cross-sections, between the top+$VV$
normalization factors and the cross-section bins in which the top+$VV$ background is fit (up to about $35\%$), between the ggF Higgs boson and the cross-section bins with the highest ggF contamination (up to about $35\%$); correlations (up to about $60\%$) are observed between the normalization factor of top+$VV$ in the ggF CR and those in SR2 bins with high top+$VV$ contamination, as well as between the top+$VV$ normalizatiton factors in different SR2 bins (up to about $40\%$).

The data are compared with SM expectations calculated by the parton-shower generators \POWHEGva+\PYTHIAV{8}, \POWHEGva+\HERWIGV{7}, \MGFiveHw, and \vbfnlo$@$LO+\PYTHIAV{8} at particle level and the parton-level generator \vbfnlo$@$NLO\@. The comparison with the \vbfnlo$@$LO prediction is not presented because of the significant discrepancy observed in the integrated cross-section measurement, which illustrates the inadequacy of fixed-order LO calculations with no parton shower to describe the data with this kinematic selection. Similarly to the integrated cross-section, the uncertainties on the theoretical predictions are smaller than those on the experimental distributions. The presented calculations predict differential cross-sections that are numerically close and with the available statistics in data it is difficult to discriminate between them in a conclusive way. Overall,  there is good agreement between measured cross-sections and \POWHEGva+\PYTHIAV{8} prediction. Some discrepancies are visible in the jet \pT distribution and in the first bin of the \mjj distribution. The differential cross-sections predicted by \POWHEGva+\HERWIGV{7} are $5\%-10\%$ lower than those predicted by \POWHEGva+\PYTHIAV{8} in all regions of the phase space, showing a small impact of the parton shower model. The cross-sections predicted by \MGFiveHw\ are close to those predicted by \POWHEGva+\PYTHIAV{8} and typically about by $15\%-20\%$ larger than those by \POWHEGva+\HERWIGV{7}. The cross-sections predicted by \vbfnlo$@$NLO  are $5\%-10\%$ greater than those by \POWHEGva+\PYTHIAV{8} in all regions of phase space and beyond their estimated uncertainties, which range from $5\%$ to $20\%$.
In several kinematic regions, \vbfnlo$@$NLO overestimates the measured cross-sections beyond the experimental uncertainties, with discrepancies that range between one and two standard deviations. The prediction by \vbfnlo$@$LO+\PYTHIAV{8} is close to those by the other MC generators that implement parton showering, despite the LO approximation in the matrix elements, showing the importance of the parton shower for a good description of the data.

Finally, correlations between cross-sections and background normalization parameters in pairs of differential cross-section measurements are calculated using a bootstrap procedure~\cite{ATLAS:2021kho}. Such correlations will allow pairs of differential cross-section measurements to be used simultaneously in future re-interpretations of the experimental results, for example in combined fits to place limits on new-physics models. The bootstrap method uses a set of pseudo-experiments (also known as replicas) of the nominal data sample, derived by introducing statistical fluctuations by Poisson perturbations and a randomization of the nuisance parameters associated with the systematic uncertainties. The random numbers for the pseudo-experiments are simulated using unique seeds and each replica is then analysed in the same way as the nominal data sample to arrive at a set of replica measurements.  The correlations are then extracted from these replica measurements after a full fit is performed on each replica. Figure~\ref{fig:XS_correlation} summarizes the correlations between cross-section bins in pairs of measured distributions. The correlation values together with the results of the fits used to extract the integrated and differential cross-sections are made available in the HepData repository~\cite{HepData}.
Such a technique is also used to assess if the fit results are compatible across the measured differential cross-sections. Pulls between pairs of measured quantities, performed by using the ensembles from the bootstrapped tests, showed a good pairwise agreement for all differential measurements with a mean global $\chi^2/$ndf of about 1.3.
\begin{figure}[ht]
\centering
\includegraphics[width=0.65\linewidth]{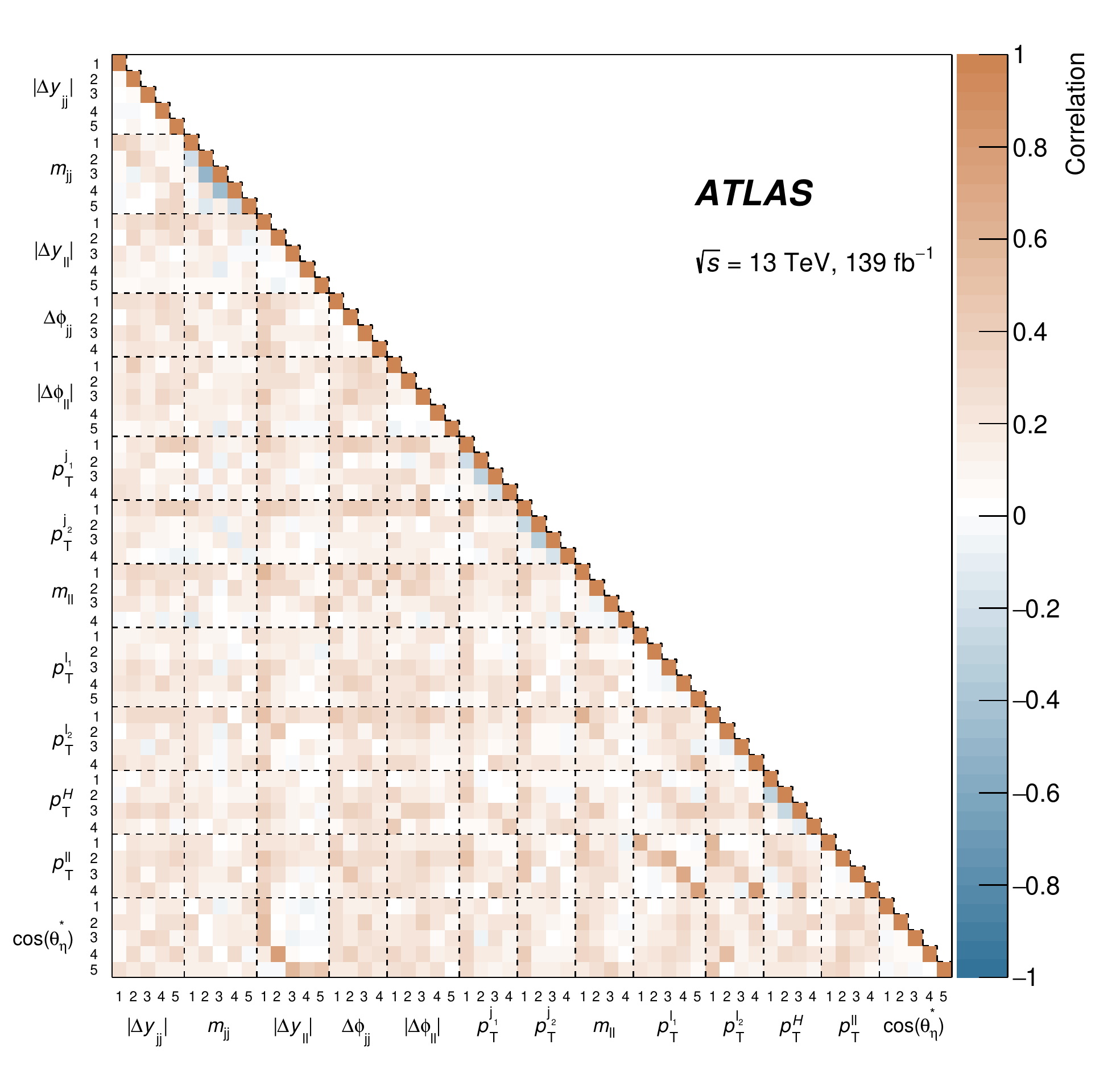}
\caption{Summary of correlations between the measured differential cross-sections using a bootstrap technique. The binning of the distributions corresponds to the binning used in the differential cross-section measurements presented in Figures~\ref{fig:Result_diffXsect_pTH-pTll}-\ref{fig:Result_diffXsect_jets}.}
\label{fig:XS_correlation}
\end{figure}


\section{Constraints on anomalous interactions}
\label{sec:eft}


The measurements of the differential fiducial cross-sections are used to constrain extensions to the SM that introduce anomalous interactions of the Higgs boson to SM particles using an EFT approach.
In this approach, which exploits the exclusive kinematical region of the VBF Higgs boson production and the \hwwlnln decay phase space, the anomalous interactions are introduced in an EFT Lagrangian, $\mathcal{L}_{\textrm{EFT}}$, via additional higher-dimensional operators $\mathcal{O}_i^{(d)}$ of mass dimension $d$ that supplement the SM Lagrangian $\mathcal{L}_{\textrm{SM}}$ as follows:
 
\begin{equation}
\mathcal{L}_{\textrm{EFT}} = \mathcal{L}_{\textrm{SM}} + \displaystyle\sum_d \displaystyle\sum_i \frac{c_i^{(d)}}{\Lambda^{(d-4)}} \mathcal{O}_i^{(d)}, ~\textrm{for } d > 4.
\end{equation}
 
The parameters $c_i^{(d)}$ specify the strength of the anomalous interactions induced by the corresponding operators and are known as the Wilson coefficients, while $\Lambda$ is the scale of new physics. Only dimension-six operators are considered, since the dimension-five and dimension-seven operators violate lepton and baryon number conservation and the impact of higher-dimensional operators is expected to be suppressed by higher powers of the new-physics scale $\Lambda$~\cite{Hays:2018zze}. For energies less than the scale of new physics, only the ratio $c_i^{(d=6)} / \Lambda^{2}$ can be constrained by the data.
 
Constraints are set on the Wilson coefficients defined in the Standard Model Effective Field Theory (SMEFT) formalism~\cite{Brivio:2017btx} in the Warsaw basis ~\cite{Grzadkowski:2010es}. The measurements in the VBF \hwwlnln channel do not provide sensitivity for simultaneous constraints on the full set of these coefficients. To reduce the number of relevant parameters, a minimal flavor-violating scenario is assumed and only operators affecting the tree-level Higgs boson couplings are considered.
The sensitivity of the differential cross-sections to several operators were studied. The Wilson coefficients were then constrained one at a time using the differential distribution that is most sensitive to the corresponding operator. The constraints were obtained for different charge-parity states, i.e., seven CP-even and three CP-odd operators. 
The CP-even operators describing interactions between the Higgs boson and vector bosons are associated with the Wilson coefficients $\cHW$, $\cHB$, and $\cHWB$, while the corresponding CP-odd operators are associated with the $\cHWtil$, $\cHBtil$, and $\cHWBtil$ Wilson coefficients. The CP-even Higgs boson interactions with quarks that are associated with the $\cHq1$, $\cHq3$, $\cHu$,  and $\cHd$ Wilson coefficients are also considered.
 
The constraints on the Wilson coefficients can be derived by comparing the measured differential fiducial cross-section with the one predicted by SMEFT\@. For that purpose, the corresponding expected signal production cross-sections, the branching ratio and the signal acceptances are parameterized in terms of the Wilson coefficients. The amplitude for the signal process is split into a SM part, $\mathcal{M}_{\textrm{SM}}$, and a dimension-six term that contains the anomalous interactions in the EFT amplitudes $\mathcal{M}_i$. Therefore, the dependence of differential cross-sections on the EFT parameters can be written as follows:
 
\begin{equation}
\sigma\propto|\mathcal{M}_{\textrm{EFT}}|^2 = |\mathcal{M}_{\textrm{SM}} + \displaystyle\sum_i \frac{c_i}{\Lambda^2} \mathcal{M}_i|^2  = |\mathcal{M}_{\textrm{SM}}|^2 + 2 \displaystyle\sum_i \frac{c_i}{\Lambda^2} Re(\mathcal{M}_{\textrm{SM}}^* \mathcal{M}_i) + \displaystyle\sum_{i,j} \frac{c_i c_j}{\Lambda^4} Re(\mathcal{M}_i^* \mathcal{M}_j).
\label{eq:EFTamplitute_full}
\end{equation}
 
The second term on the right-hand side represents the interference between the SM and dimension-six EFT amplitudes, also known as the \textit{linear term}, while the third term, referred to as the \textit{quadratic term}, comprises the contribution from pure anomalous interactions from dimension-six EFT operators alone.
 
The constraints from the VBF \hwwlnln channel on the relevant Wilson coefficients allow a rather large range of parameter values in which the quadratic term cannot be neglected even though its contribution is suppressed by $\Lambda^4$. Such dimension-six quadratic terms are therefore included in the EFT parameterization and the constraints on the dimension-six operators presented in this section are derived with and without the pure dimension-six terms
included in the theoretical prediction. Since the linear terms from dimension-eight operators are suppressed by the same factor, they could also give non-negligible contributions. The comparison of the constraints obtained from linear-only and linear-plus-quadratic EFT prediction provides a qualitative measure of the uncertainties associated to missing terms in the EFT operator expansion.

\begin{figure}[htb!]
\centering
\subfloat[]{\includegraphics[width=0.45\linewidth]{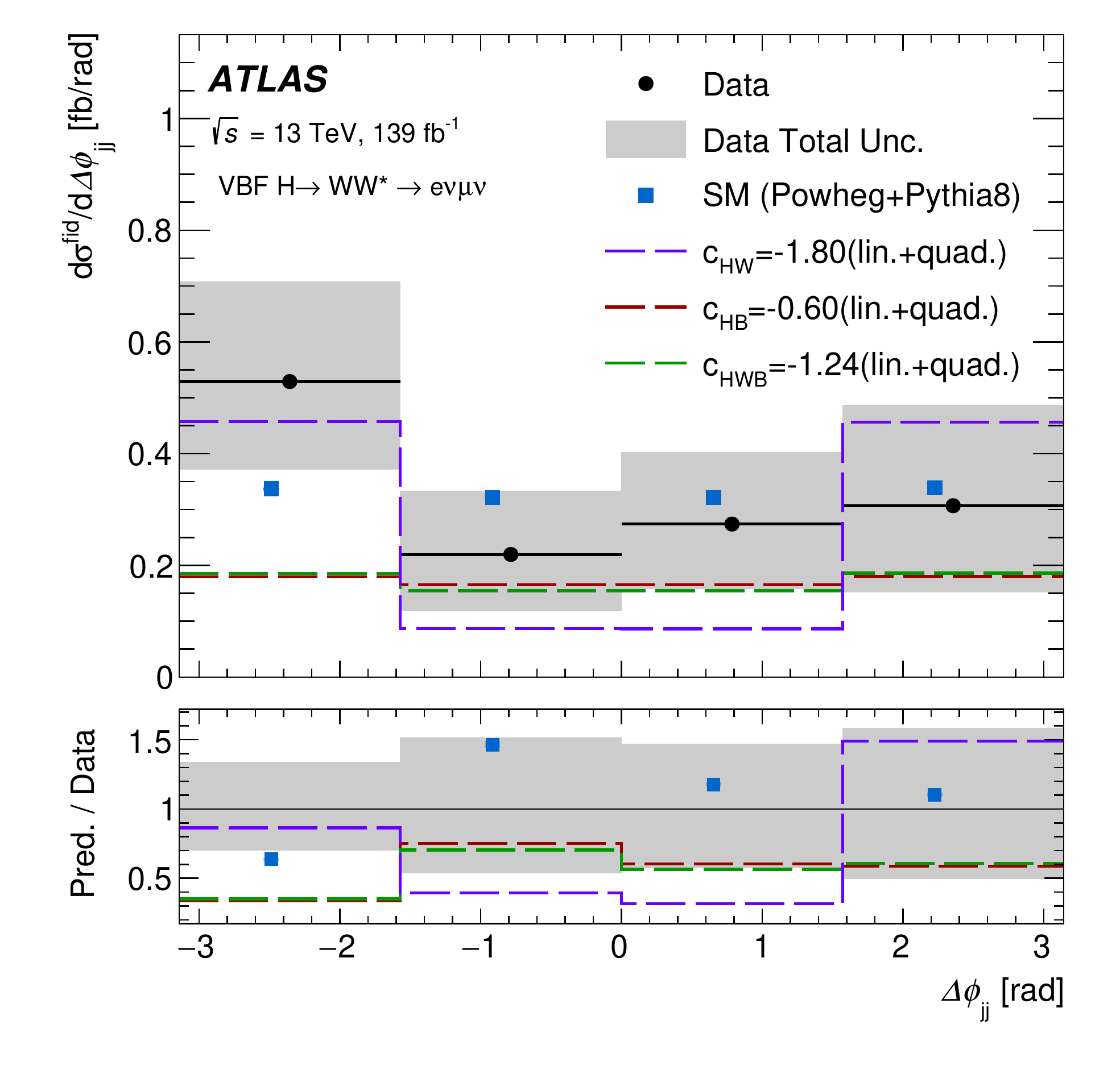}\label{fig:eft_dphijj_chw-linquad}}
\subfloat[]{\includegraphics[width=0.45\linewidth]{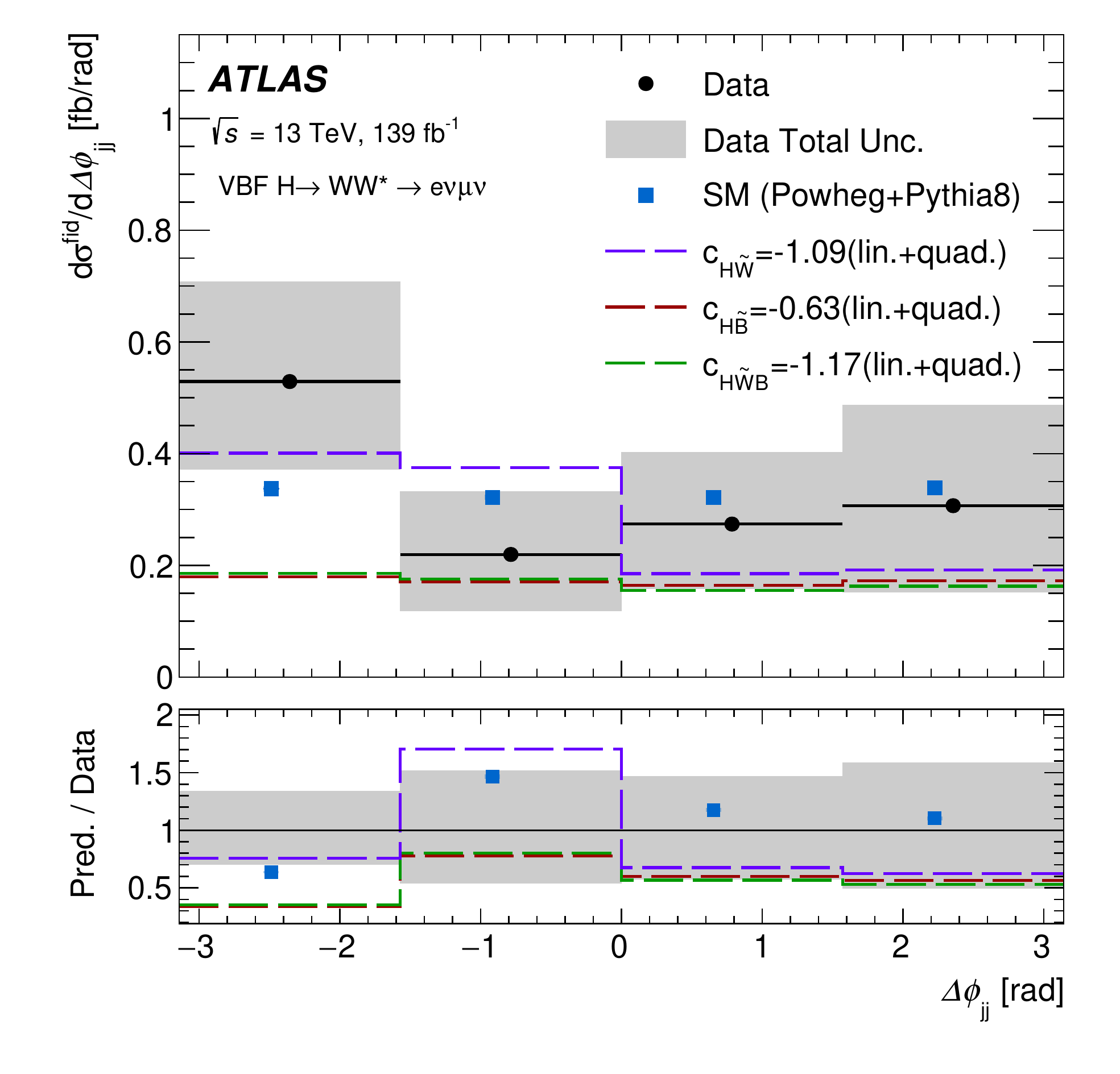}\label{fig:eft_dphijj_chwtil_linquad}} \\
\subfloat[]{\includegraphics[width=0.45\linewidth]{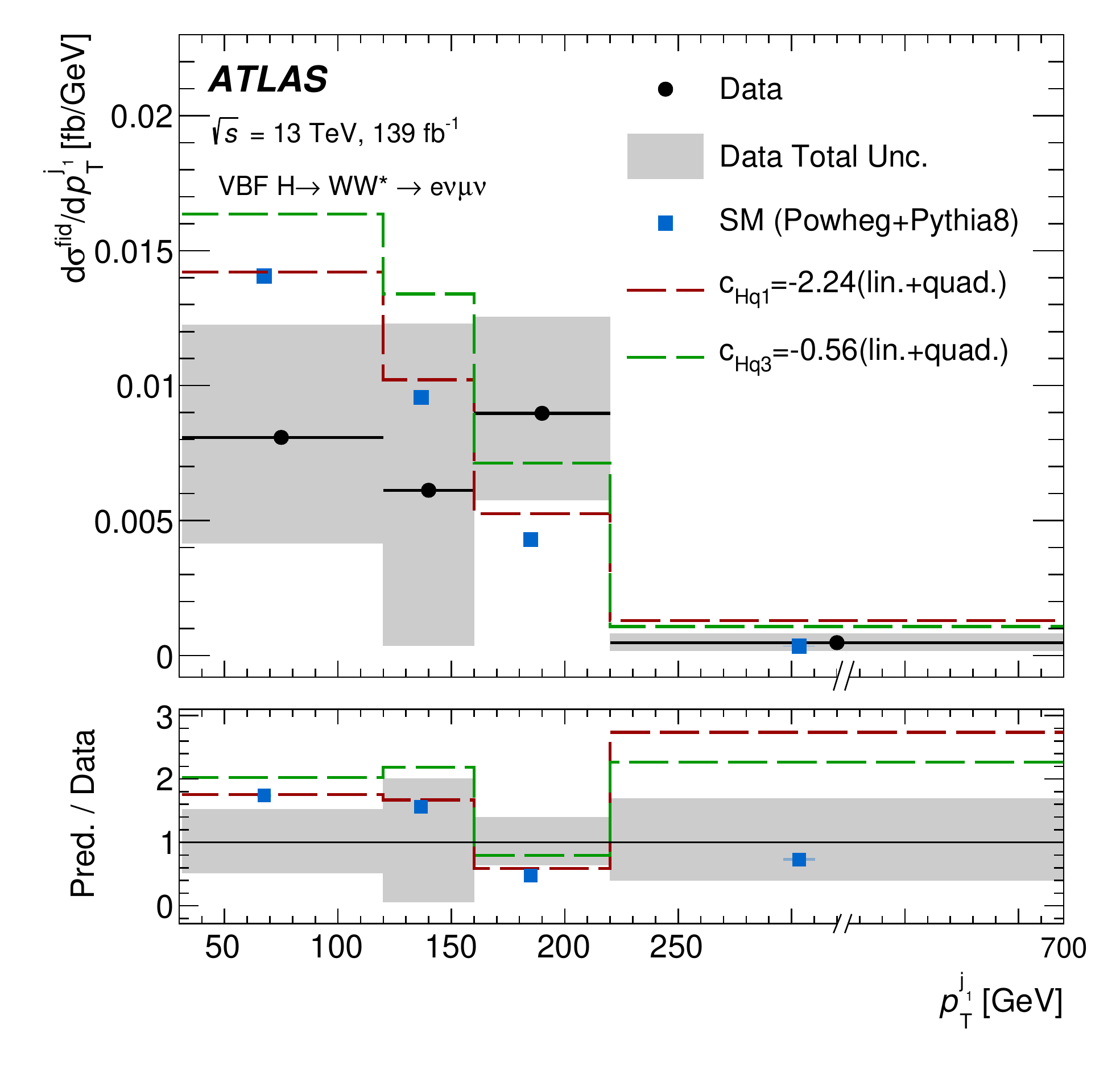}\label{fig:eft_jet0_pt_chq3q1_linquad}}
\subfloat[]{\includegraphics[width=0.45\linewidth]{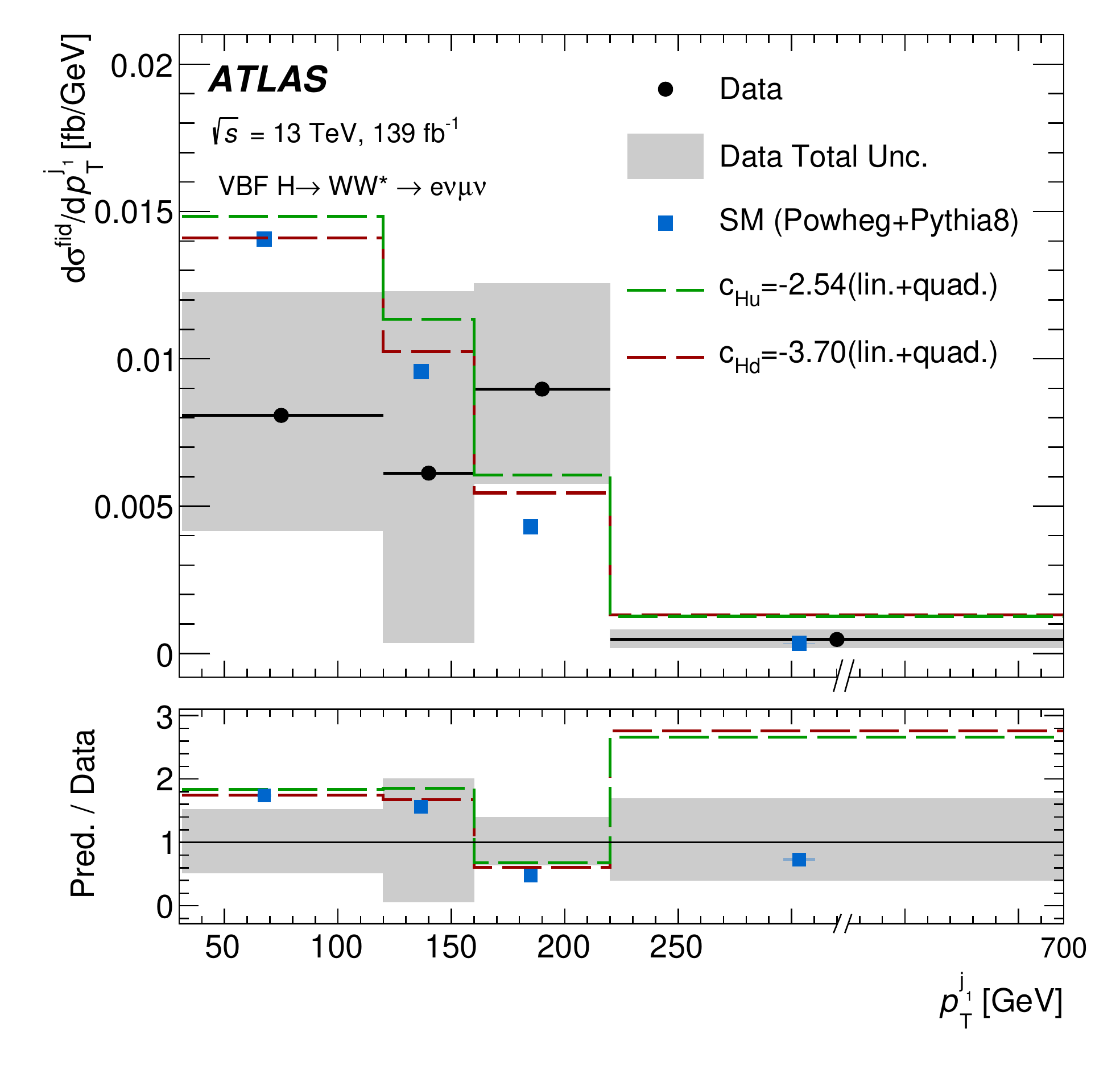}\label{fig:eft_jet0_pt_chuhd_linquad}}
\caption{The experimental VBF \hwwlnln  differential cross-section as a function of \protect\subref{fig:eft_dphijj_chw-linquad}-\protect\subref{fig:eft_dphijj_chwtil_linquad} $\dphijj$ and \protect\subref{fig:eft_jet0_pt_chq3q1_linquad}-\protect\subref{fig:eft_jet0_pt_chuhd_linquad} $\pTjone$ is shown together with the expected SM cross-section (\POWHEGva+\PYTHIAV{8}) and linear plus quadratic (lin.+quad.) EFT model. The parameter values correspond to the negative limits set at 95\% confidence level obtained from the statistical interpretation of data with the lin.+quad. parameterizations. The contributions from the sum of the interference and the pure dimension-six terms are represented as dashed lines.}
\label{fig:EFT_impact}
\end{figure}
 
The differential cross-sections can be described by combining the parameterization of the VBF Higgs boson production cross-section, the branching ratio of the Higgs boson to two \W bosons and the signal acceptance. The procedure for deriving the parameterization of the differential fiducial cross-sections follows Ref.~\cite{HIGG-2019-13}.
The parameterizations for \cHWBtil and \cHu are found to be poorly described by a linear and a linear plus quadratic function of the Wilson coefficients for values beyond the sensitivity of the measurement, i.e., outside the limit ranges. This effect is due to a dependence of the fiducial selection efficiency on the EFT parameters for extreme values of these couplings and not to a data unfolding bias. The associated bias was studied and its effect was assigned as an uncertainty to the EFT parameterization for those couplings. It was found to have negligible impact on the limits, i.e., about an order of magnitude smaller than the impact of systematic uncertainties assigned to the signal modeling.

The contributions of the pure-SM, the interference and the pure dimension-six terms to the differential cross-sections in Eq. (\ref{eq:EFTamplitute_full}) are generated at leading-order in perturbative QCD and EW processes using the \MADGRAPHFive\,2.7.2 generator. The anomalous interactions from the dimension-six operators are provided by the SMEFTsim package~\cite{Brivio:2017btx,Brivio:2020onw} in the $U(3)^5$ flavor symmetric limit with the cut-off scale set to $\Lambda$ = 1 \TeV. The events are simulated using the \texttt{NNPDF30NLO} PDF set~\cite{Ball:2014uwa} and the $G_F$ EW scheme. The \MADGRAPHFive\, samples are interfaced to the \PYTHIAV{8.230} program with the dipole recoil approximation and the A14 set of tuned parameters for parton showering, hadronization, and multiple parton scattering. The pure-SM prediction was validated against \vbfnlo$@$LO+\PYTHIAV{8}.
An assumption is made that higher-order QCD corrections, applied in a multiplicative way, are similar for the SM prediction and the one for anomalous interactions. Therefore, the parameterizations of the cross-sections as a function of each Wilson coefficient are expected to have small effects associated to missing higher-order corrections~\cite{Degrande:2016dqg}. The higher-order calculation is taken to be the prediction from \POWHEGva+\PYTHIAV{8}, see Section~\ref{sec:dataset_mc_signal}.
 
The impact of the sum of the pure SM, the interference, and the pure dimension-six contributions to the differential fiducial cross-section is shown in Figure~\ref{fig:EFT_impact} in comparison to the pure SM contribution and the measured distributions. The parameter values correspond to the negative limits set at 95\% confidence level obtained from the statistical interpretation of data with the linear-plus-quadratic parameterizations.
The \dphijj observable is found to be sensitive to anomalous couplings between the Higgs boson and vector bosons, described by the parameters $\cHW$, $\cHB$, and $\cHWB$ and their CP-odd counterparts. For the CP-even operators that account for anomalous interactions between the Higgs boson and quarks, the leading jet \pT observable shows the largest kinematic dependencies, most noticeably at high energy scales.

\begin{table}[!tb]
\centering
\caption{ Summary of EFT operators in the SMEFT formalism that are probed with differential cross-section measurements in the VBF \hwwlnln channel. The corresponding structure in terms of the SM fields in the Warsaw basis (second column) is shown together with the associated Wilson coefficient (first column). The Higgs boson doublet field $H$ and its complex conjugate are denoted by $H$ and $H^{\dag}$, respectively. The left-handed quark doublets (the right-handed up-type or down-type quarks) are denoted by $q$ ($u$ or $d$), while $V_{\mu\nu}$ ($\tilde{V}_{\mu\nu}$ =$\epsilon^{\mu\nu\rho\sigma} V_{\rho\sigma}$) is the (dual) field strength tensor for a given gauge field of the electroweak interactions with $V$ = $B$, $\W^n$ ($n=\{1,2,3 \}$), and $\tau^n$ are the Pauli matrices. The bosonic operators with (without) a dual field strength tensor are CP-odd (CP-even). For details of the formalism used see Ref.~\cite{Brivio:2020onw}.
The expected and observed 95\% confidence interval for the Wilson coefficients, using fits to the differential cross-section measured as a function of the observable that is indicated in the $3^{rd}$ column (labeled as ``Fit distr.''), are shown in the $5^{th}$ and $6^{th}$ columns, respectively. Results are presented when excluding (lin.) or including (lin. + quad.) the pure dimension-six contributions to the EFT prediction, as indicated in the $4^{th}$ column.
}
\label{tab:EFT_results}
\begin{tabular}{cccrrr}
\hline
\hline
\noalign{\vspace{0.05cm}}
Wilson & Operator  & Fit    & Param. & \multicolumn{2}{c}{95\% confidence interval [\TeV$^{-2}$]} \\
Coeff. & Structure & distr. & Order    &   Expected     &  Observed                             \\
\hline
\noalign{\vspace{0.1cm}}
$\cHW$ &  $H^{\dag} H\, W^n_{\mu\nu}W^{n\mu\nu}$   & $\dphijj$ & lin.  & [-1.7, 1.6] &  [-2.6, 0.60] \\
&                                            &           & lin. + quad.  & [-1.4, 1.4] &  [-1.8, 0.61] \\
$\cHB$ &  $H^{\dag} H\, B_{\mu\nu}B^{\mu\nu}$       & $\dphijj$ & lin.   & [-5.9, 6.4] &  [-6.7, 4.6] \\
&                                             &           & lin. + quad.  & [-0.59, 0.66] &  [-0.60, 0.66] \\
$\cHWB$ &  $H^{\dag} \tau^n H\, W^n_{\mu\nu} B^{\mu\nu}$   & $\dphijj$ & lin.   & [-10, 9] &  [-14, 5.9] \\
&                                            &           & lin. + quad.  & [-1.2, 1.1] &  [-1.2, 1.1] \\
$\cHq1$ & $(H^{\dag}i\overleftrightarrow{D}_{\mu}H )(\bar{q}\gamma^{\mu}q)$ &  $p_{\textrm{T}}^{j1}$ & lin.  & [-12, 15] &  [-6.9, 22] \\
&                                                                        &                & lin. + quad.  & [-1.9, 1.7] &  [-2.2, 2.0] \\
$\cHq3$ & $(H^{\dag}i\overleftrightarrow{D}^n_{\mu}H )(\bar{q}\tau^n\gamma^{\mu}q)$ & $p_{\textrm{T}}^{j1}$ & lin. & [-0.56, 0.47] &  [-0.74, 0.30] \\
&                                                                                 &                    & lin. + quad.  & [-0.43, 1.2] &  [-0.56, 0.43] \\
$\cHu$ & $(H^{\dag}i\overleftrightarrow{D}_{\mu}H )(\bar{u}\gamma^{\mu}u)$ &  $p_{\textrm{T}}^{j1}$ & lin.  & [-8.3, 6.9] &  [-11, 4.2] \\
&                                                                         &                    & lin. + quad.  & [-2.0, 2.6] &  [-2.5, 3.1] \\
$\cHd$ & $(H^{\dag}i\overleftrightarrow{D}_{\mu}H )(\bar{d}\gamma^{\mu}d)$ &  $p_{\textrm{T}}^{j1}$ & lin. & [-21, 25] &  [-13, 33] \\
&                                                                          &                    & lin. + quad.  & [-3.0, 2.7] &  [-3.7, 3.4] \\
\noalign{\vspace{0.1cm}}
\hline
\noalign{\vskip 1.5mm}
$\cHWtil$ &   $H^{\dag}H\, \tilde{W}^n_{\mu\nu}W^{n\mu\nu}$   & $\dphijj$ & lin.   & [-1.7, 1.7] &  [-1.8, 1.3] \\
&                                            &           & lin. + quad.  & [-1.4, 1.4] &  [-1.1, 1.4] \\
$\cHBtil$ &  $H^{\dag}H\, \tilde{B}_{\mu\nu}B^{\mu\nu}$       & $\dphijj$ & lin.   & [-28, 28] &  [-32, 22] \\
&                                             &           & lin. + quad.  & [-0.62, 0.62] &  [-0.63, 0.63] \\
$\cHWBtil$ &  $H^{\dag} \tau^n H\, \tilde{W}^n_{\mu\nu} B^{\mu\nu}$   & $\dphijj$ & lin.   & [-15, 15] &  [-17, 12] \\
&                                            &           & lin. + quad.  & [-1.2, 1.1] &  [-1.2, 1.1] \\
\noalign{\vspace{0.1cm}}
\hline
\hline
\end{tabular}
\end{table}

The measured differential cross-section at particle-level as a function of \dphijj and \pTjone, as well as the corresponding EFT-dependent theoretical predictions are used to define a likelihood function, following the same inference approach as described in Section~\ref{sec:method_diff}. Statistical correlations amongst the bins of a measured distribution, the experimental and the theoretical systematic uncertainties are included in the likelihood function using the same formalism as presented in Section~\ref{sec:method}.
The constraints are set on one of the Wilson coefficients at a time, while the values of the remaining coefficients are assumed to be equal to zero.
 
The confidence level at each value of a Wilson coefficient is calculated using the same profile-likelihood test statistic~\cite{Feldman:1997qc} as used in Section~\ref{sec:method}, which is assumed to be distributed according to a $\chi^2$ distribution with one degree of freedom following from Wilks’ theorem~\cite{10.1214/aoms/1177732360}. This technique allows the 95\% confidence intervals to be constructed for each Wilson coefficient.
The expected and observed 95\% confidence intervals on the dimension-six operators are shown in Table~\ref{tab:EFT_results} and in Figure~\ref{fig:EFT_results}. The measurements are dominated by the statistical uncertainty and the parameterization uncertainty is included as a systematic contribution. For each Wilson coefficient, confidence intervals are shown when including or not-including the pure dimension-six contribution in the theoretical prediction.
For several Wilson coefficients, the 95\% confidence intervals are affected by the addition of the pure dimension-six contribution to the linear term. Among them, the $\cHB$, $\cHWB$, $\cHBtil$, $\cHWBtil$, $\cHq1$, $\cHu$,  and $\cHd$ parameters are the most impacted by the quadratic term, which considerably improves their constraints.
 
\begin{figure}[htb!]
\centering
\includegraphics[width=0.5\linewidth]{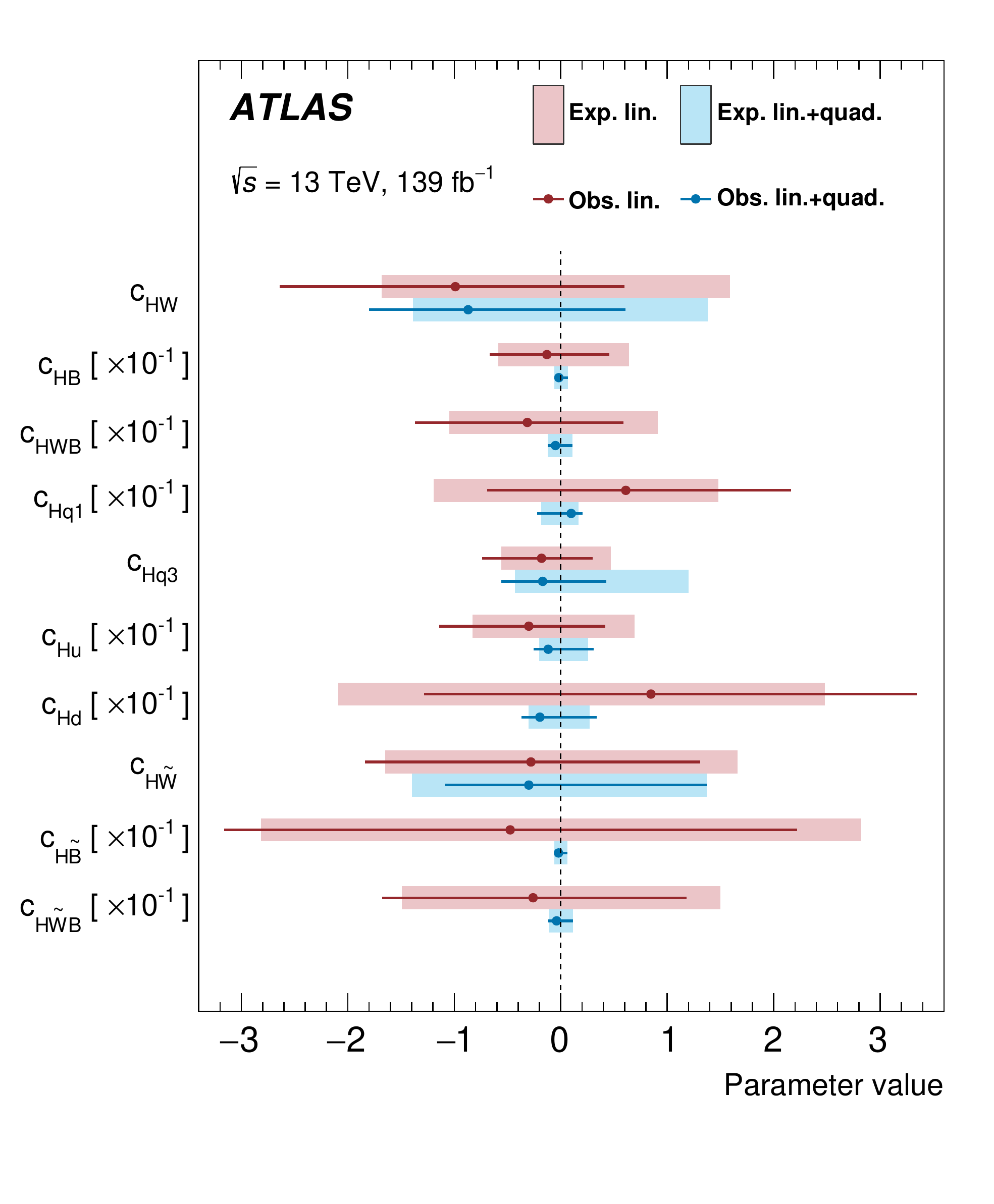}
\caption{The observed and expected values of SMEFT Wilson coefficients from CP-even and CP-odd operators obtained with only one Wilson coefficient left floating at a time in the fit to data  while all others are set to zero.
The horizontal bands (bars) represent the expected (observed) error with a confidence level of 95\% using the parameterizations excluding ('lin.') and including ('lin. + quad.') the pure dimension-six contributions, respectively. For the observed parameters, the data points represent the best-fit values. The values of \cHB, \cHWB, \cHq1, \cHu, \cHd,  \cHBtil, and \cHWBtil are scaled by a factor of 0.1.
}
\label{fig:EFT_results}
\end{figure}
 
The results are in good agreement with the SM predictions. Using the approximation that includes only the linear term, the  most stringent limits are derived for the $\cHq3$, $\cHW$, and $\cHWtil$. However, more stringent constraints are provided for all Wilson coefficients when the quadratic term is added to the parameterization. In this case, the most stringent constraints, with absolute values of the upper and lower limits smaller than 1.0, are obtained for $\cHq3$, $\cHWB$, and $\cHBtil$, but good constraints are also provided for $\cHW$, $\cHWB$, $\cHWtil$, and $\cHWBtil$. Such results shows the significant impact of the quadratic term and in turn the sensitivity to missing higher dimensional operators in the EFT expansion.
 
The sensitivities to the $\cHW$ and $\cHWtil$ parameters is driven by the anomalous couplings at the VBF production and the \hww decay vertices, whereas the CP even $\cHWB$ and $\cHB$ parameters affect other decay modes more significantly, e.g., $H\rightarrow \gamma\gamma$~\cite{HIGG-2019-13}, which reflects in the EFT effect on the branching ratio of the Higgs boson. For the CP-odd operators, since the interference contribution is zero in parity-even observables, the constraints set on these parameters are induced by the asymmetric shape effects in the \dphijj observable, which has CP-odd discrimination power. In the quadratic order, the branching ratio is affected by CP-odd operators, resulting in significant quadratic effect for the constraints of the $\cHWBtil$ and $\cHBtil$ parameters.
\begin{figure}[tb!]
\centering
\includegraphics[width=0.5\linewidth]{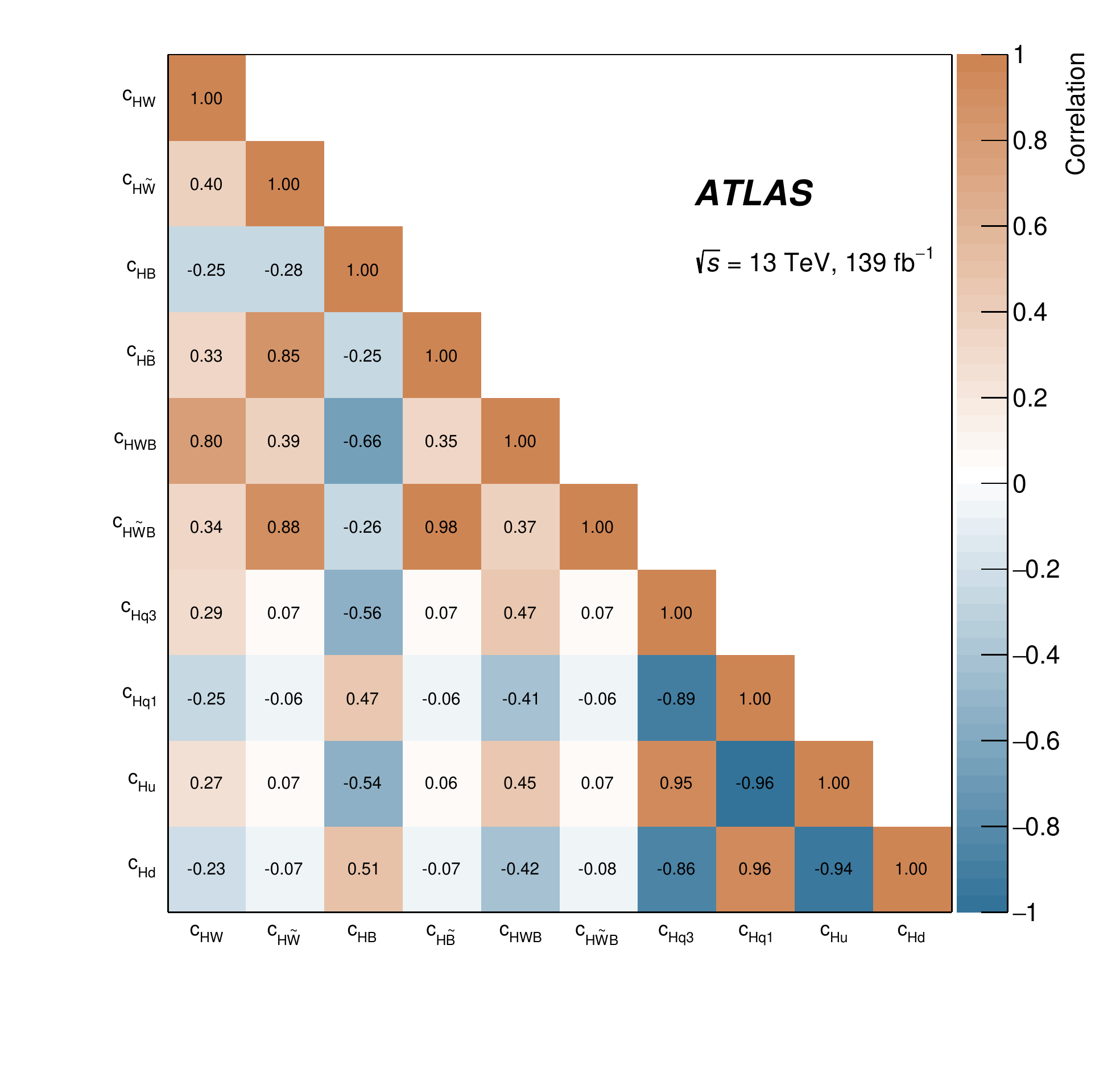}
\caption{Correlations between the measured SMEFT Wilson coefficients.}
\label{fig:EFT_correlation}
\end{figure}
 
Correlations between pairs of SMEFT Wilson coefficients are measured using the bootstrapping technique presented in Section~\ref{sec:result_diff}, and are shown in Figure~\ref{fig:EFT_correlation}. A large positive correlation of $80\%$ is observed between $\cHW$ and $\cHWB$.  Large positive correlations ($85\%-98\%$) are also seen among CP-odd coefficients, which in turn have moderate correlations (approx. $\pm 30\%$ to $\pm 40\%$) with the CP-even coefficients associated with the Higgs boson interacting with vector bosons ($\cHW$, $\cHB$, and $\cHWB$) and very small correlations with coefficients associated with the Higgs boson interacting with quarks ($\cHq1$, $\cHq3$, $\cHu$,  and $\cHd$). The latter coefficients show large (positive or negative) correlations among each other and moderate (positive or negative) correlations (approx. $\pm 30\%$ to $\pm 60\%$) with the CP-even $\cHW$, $\cHB$, and $\cHWB$ coefficients. Moderate correlations are observed between $\cHB$ and $\cHWB$, and between $\cHB$ and $\cHW$.
 
Since correlations between pairs of differential cross-sections are provided, as detailed in Section~\ref{sec:result_diff}, pairs of measurements can be used to set constraints on new and different models of physics beyond the SM, for example in future global fits.


\FloatBarrier

\section{Conclusion}
\label{sec:conclusion}

Integrated and differential fiducial cross-section measurements for the vector-boson fusion production of the Higgs boson in the $\hwwlnln$ final state are presented using proton–proton collision data collected by the ATLAS experiment at a center-of-mass energy of $\sqrt{s}= 13$ \TeV\ and with an integrated luminosity of 139 \ifb. The data are selected by requiring one electron and one muon from the decay of \ensuremath{W}\xspace bosons along with two energetic jets in the final state and are corrected to particle level.
The measured integrated fiducial cross-section is $\sigma^{\textrm{fid}} =1.68 \pm 0.33 ~\textrm{(stat)} \pm 0.23 ~\textrm{(sys)}$ fb. It is found to be consistent with SM predictions that include NLO QCD and NLO electroweak corrections and with LO QCD calculations that implement parton showering.
The fiducial cross-section is also measured differentially as a function of kinematic and angular variables of the final-state charged leptons and jets that are sensitive to the Higgs boson production and decay. The uncertainties in the differential cross-section measurements are driven by the data statistical uncertainty. Correlations between the differential fiducial cross-section measurements are provided to allow their simultaneous use in future studies of new physics signatures and tuning of SM predictions. Simulations at LO and NLO QCD interfaced to parton shower routines provide a good description of the data in most regions of phase space, while a fixed-order NLO calculation overestimates the experimental results in several kinematic regions at the level of one to two standard deviations.
An EFT framework that includes the interference between the SM and dimension-six amplitudes as well as pure dimension-six terms is used to set limits on CP-even and CP-odd dimension-six operators.
Stringent constraints are set on several CP-even and CP-odd EFT parameters that are sensitive to the interactions between the Higgs boson and vector bosons as well as CP-even operators that are sensitive to the interactions between the Higgs bosons and quarks. The most stringent constraints are obtained when the quadratic term is added to the EFT parameterization, indicating that those limits are sensitive to neglected contributions of higher-dimensional operators in the EFT expansion.


\section*{Acknowledgements}


We thank CERN for the very successful operation of the LHC, as well as the
support staff from our institutions without whom ATLAS could not be
operated efficiently.
 
We acknowledge the support of
ANPCyT, Argentina;
YerPhI, Armenia;
ARC, Australia;
BMWFW and FWF, Austria;
ANAS, Azerbaijan;
CNPq and FAPESP, Brazil;
NSERC, NRC and CFI, Canada;
CERN;
ANID, Chile;
CAS, MOST and NSFC, China;
Minciencias, Colombia;
MEYS CR, Czech Republic;
DNRF and DNSRC, Denmark;
IN2P3-CNRS and CEA-DRF/IRFU, France;
SRNSFG, Georgia;
BMBF, HGF and MPG, Germany;
GSRI, Greece;
RGC and Hong Kong SAR, China;
ISF and Benoziyo Center, Israel;
INFN, Italy;
MEXT and JSPS, Japan;
CNRST, Morocco;
NWO, Netherlands;
RCN, Norway;
MEiN, Poland;
FCT, Portugal;
MNE/IFA, Romania;
MESTD, Serbia;
MSSR, Slovakia;
ARRS and MIZ\v{S}, Slovenia;
DSI/NRF, South Africa;
MICINN, Spain;
SRC and Wallenberg Foundation, Sweden;
SERI, SNSF and Cantons of Bern and Geneva, Switzerland;
MOST, Taiwan;
TENMAK, T\"urkiye;
STFC, United Kingdom;
DOE and NSF, United States of America.
In addition, individual groups and members have received support from
BCKDF, CANARIE, Compute Canada and CRC, Canada;
PRIMUS 21/SCI/017 and UNCE SCI/013, Czech Republic;
COST, ERC, ERDF, Horizon 2020, ICSC-NextGenerationEU and Marie Sk{\l}odowska-Curie Actions, European Union;
Investissements d'Avenir Labex, Investissements d'Avenir Idex and ANR, France;
DFG and AvH Foundation, Germany;
Herakleitos, Thales and Aristeia programmes co-financed by EU-ESF and the Greek NSRF, Greece;
BSF-NSF and MINERVA, Israel;
Norwegian Financial Mechanism 2014-2021, Norway;
NCN and NAWA, Poland;
La Caixa Banking Foundation, CERCA Programme Generalitat de Catalunya and PROMETEO and GenT Programmes Generalitat Valenciana, Spain;
G\"{o}ran Gustafssons Stiftelse, Sweden;
The Royal Society and Leverhulme Trust, United Kingdom.
 
The crucial computing support from all WLCG partners is acknowledged gratefully, in particular from CERN, the ATLAS Tier-1 facilities at TRIUMF (Canada), NDGF (Denmark, Norway, Sweden), CC-IN2P3 (France), KIT/GridKA (Germany), INFN-CNAF (Italy), NL-T1 (Netherlands), PIC (Spain), ASGC (Taiwan), RAL (UK) and BNL (USA), the Tier-2 facilities worldwide and large non-WLCG resource providers. Major contributors of computing resources are listed in Ref.~\cite{ATL-SOFT-PUB-2023-001}.


\clearpage
\appendix
\part*{Appendix}
\addcontentsline{toc}{part}{Appendix}
\section{Background normalization factors}
\label{app:bkg_norm}

The values of the background normalization parameters for all measured differential cross-sections, as results of the fit to data, are shown in Figure~\ref{fig:norm_param} for the integrated and differential fiducial cross-section measurements.
\begin{figure}[ht]
\centering
\subfloat[]{\includegraphics[width=0.40\linewidth]{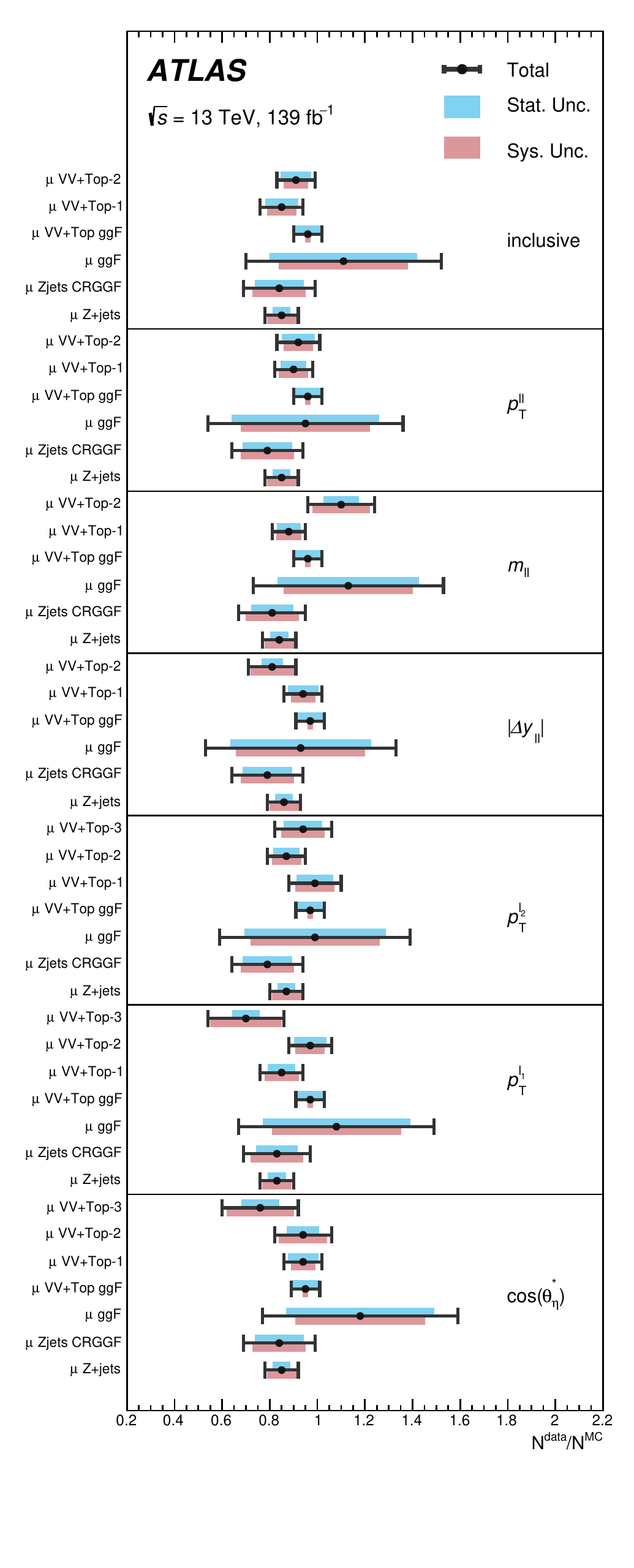}}\hfill
\subfloat[]{\includegraphics[width=0.40\linewidth]{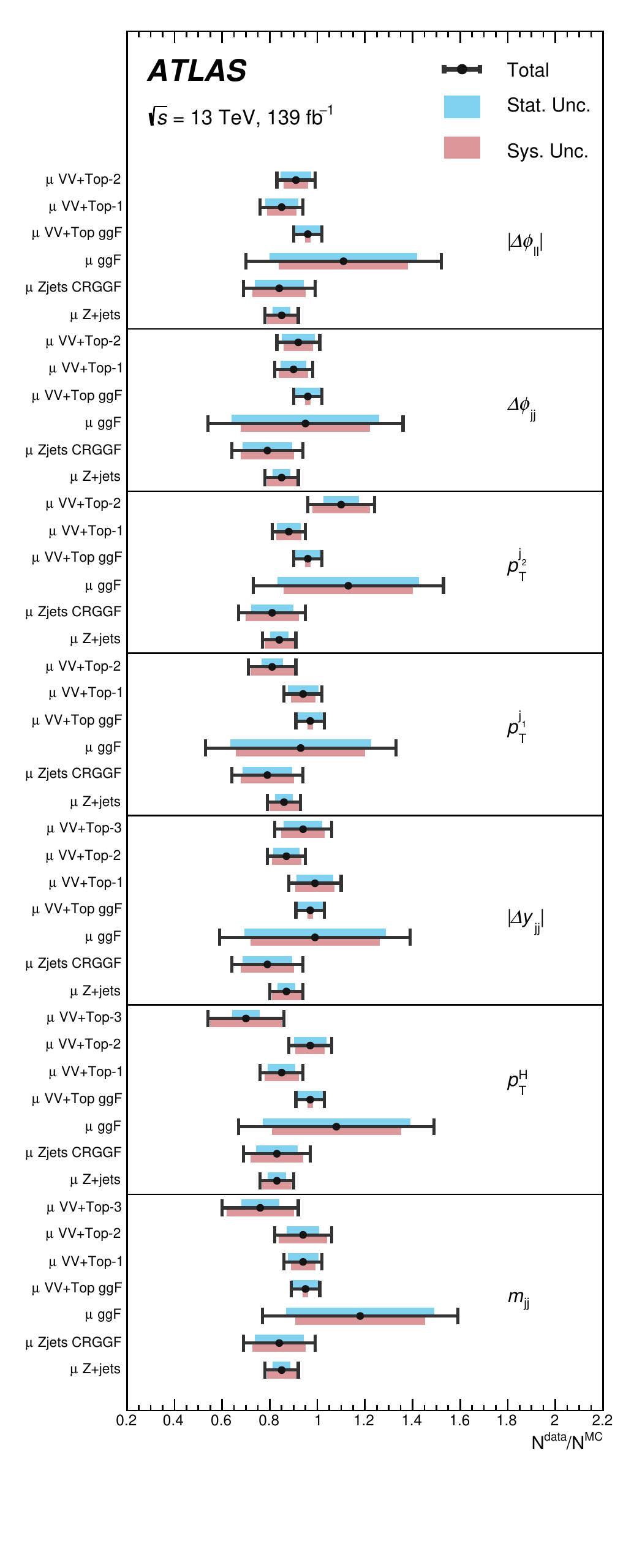}}
\caption{Background normalization factors for the measurements of integrated and differential fiducial cross-sections as results of their respective fits to data. The normalization factors for the integrated and differential fiducial cross-sections as a function of $\pT^{\ell\ell}$,  $\mll$, $|\dyll|$, $\pTlone$, $\pTltwo$, and $\cos(\theta^*_{\eta})$ are shown in (a), while those for the differential fiducial cross-sections as a function of $|\dphill|$, $\dphijj$, $\pTjone$, $\pTjtwo$, $|\dyjj|$, $p_T^H$, and $\mjj$ are shown in (b).}
\label{fig:norm_param}
\end{figure}


\clearpage
\section{Correlation Matrices}
\label{app:corr_matrices}

The correlation matrix between the measured integrated cross-section and the background normalization factors is shown in Figure~\ref{fig:Result_CorrMatrixt_Incl} while the correlation matrices for the differential cross-sections are shown in Figures~\ref{fig:Result_CorrMatrixt_pTH-pTll},~\ref{fig:Result_CorrMatrixt_leptons} and ~\ref{fig:Result_CorrMatrixt_jets}.
\begin{figure}[ht]
\centering
\includegraphics[width=0.40\linewidth]{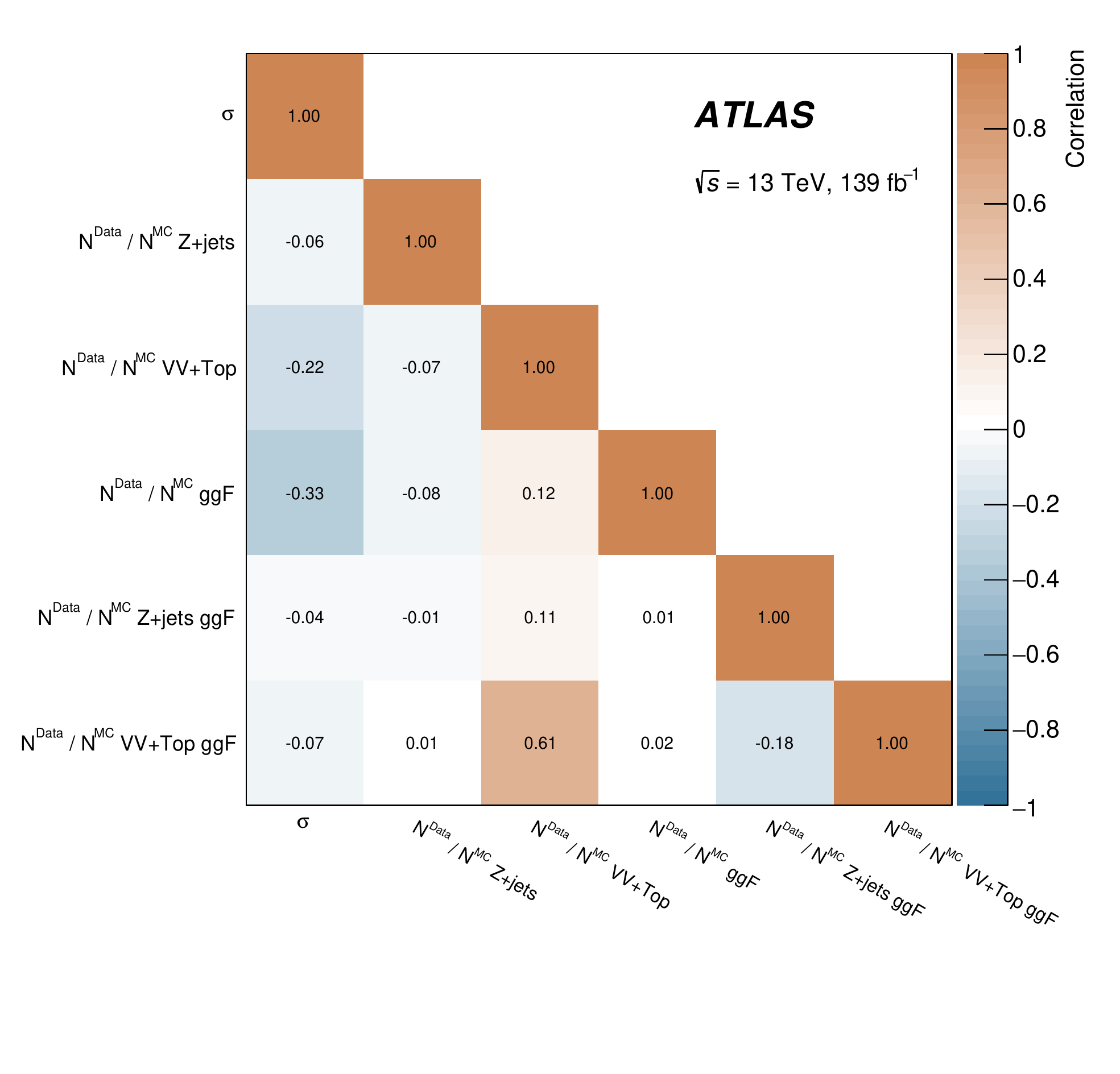}
\caption{The correlation matrix between the measured integrated cross-section and the background normalization factors.}
\label{fig:Result_CorrMatrixt_Incl}
\end{figure}
\begin{figure}[ht]
\centering
\subfloat[]{\includegraphics[width=0.40\linewidth]{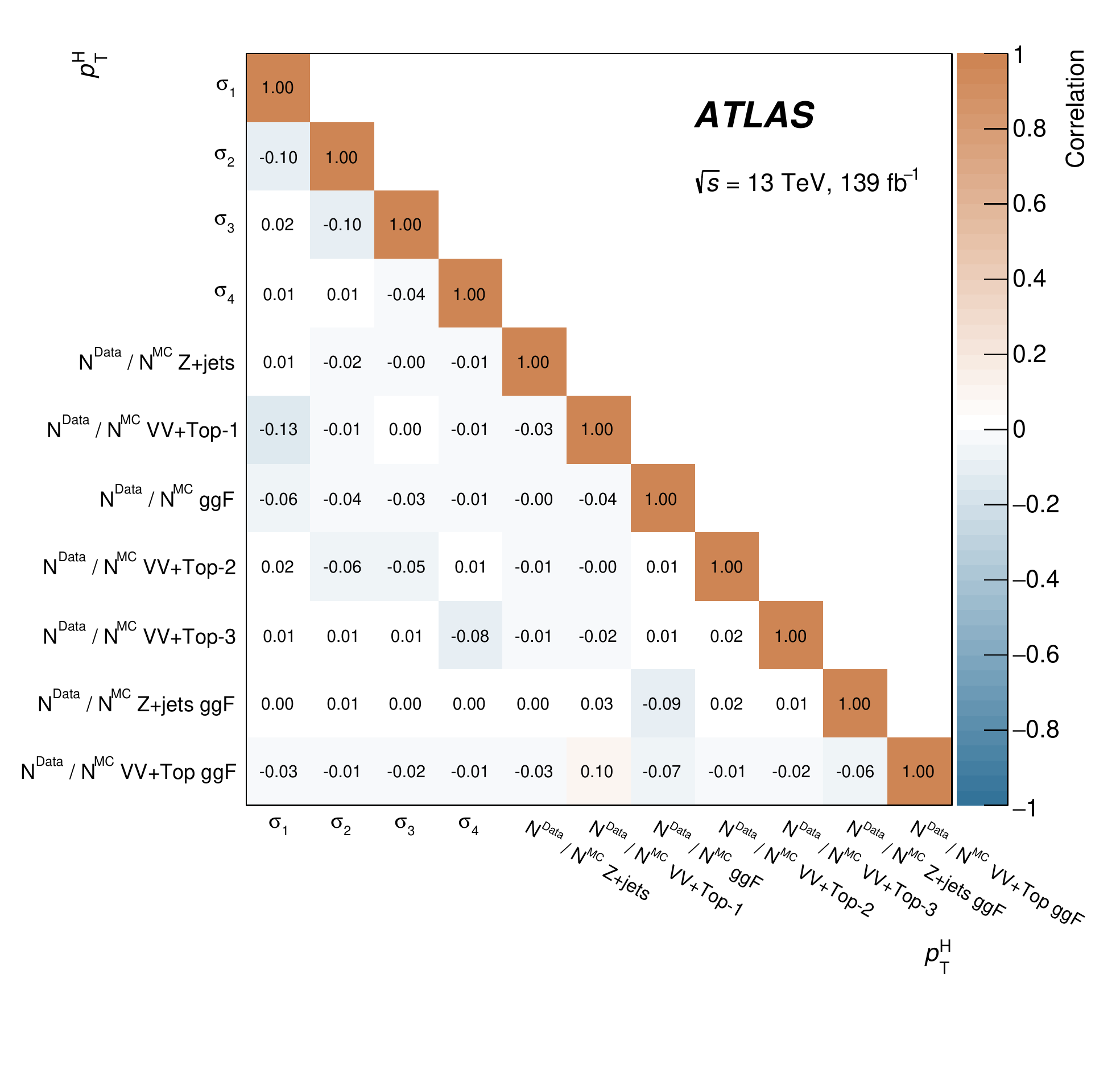}\label{fig:corMatrix_pt_H}}
\subfloat[]{\includegraphics[width=0.40\linewidth]{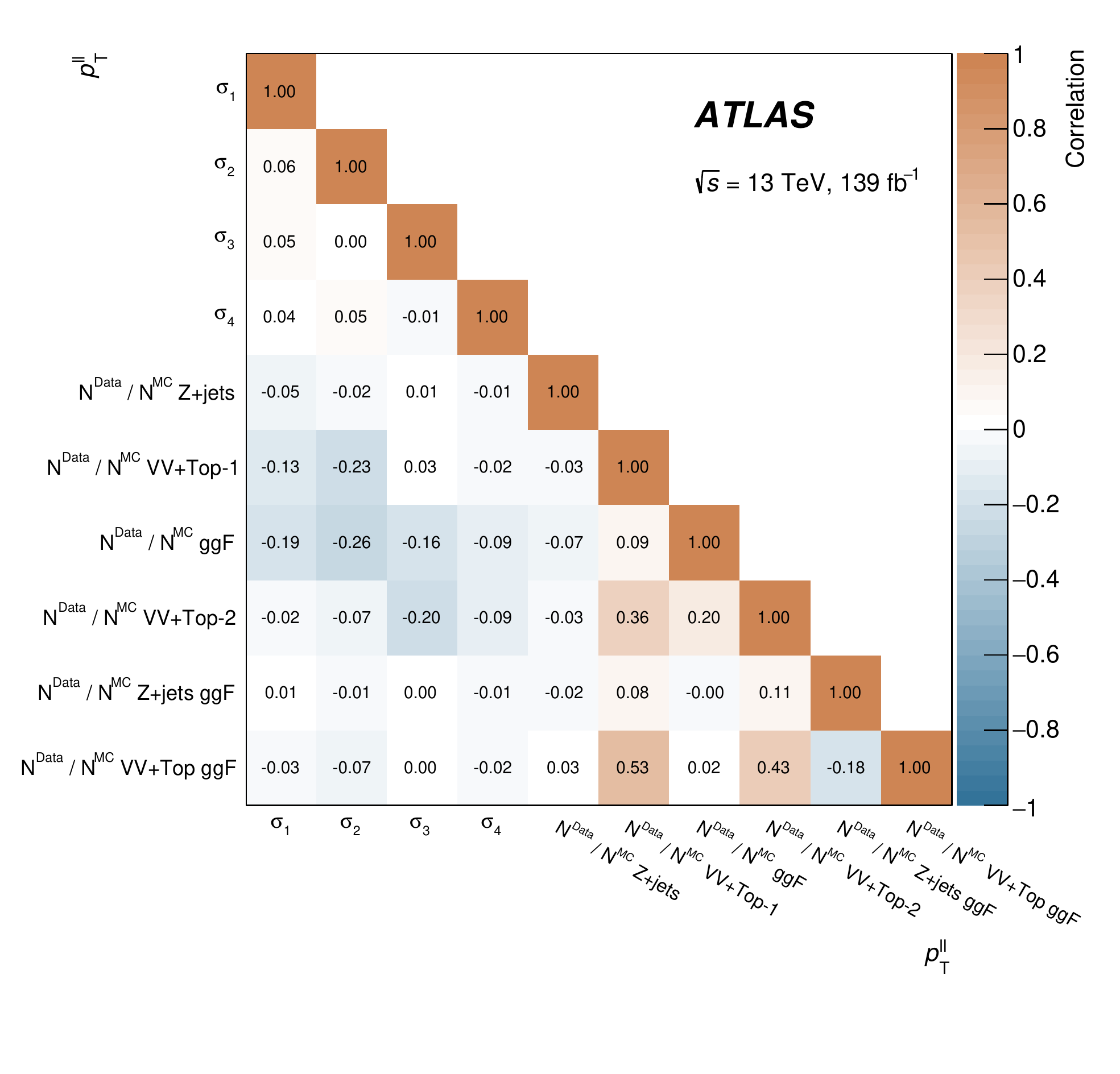}\label{fig:corMatrix_Ptll}}
\caption{The correlation matrices between the background normalization factors and the measured differential cross-sections in bins of the \protect\subref{fig:corMatrix_pt_H} $\pT^{H}$ and \protect\subref{fig:corMatrix_Ptll} $\pT^{\ell\ell}$ variables. The symbol $\sigma_{i}$ is the cross-section in bin-$i$. }
\label{fig:Result_CorrMatrixt_pTH-pTll}
\end{figure}
\begin{figure}[ht]
\centering
\subfloat[]{\includegraphics[width=0.40\linewidth]{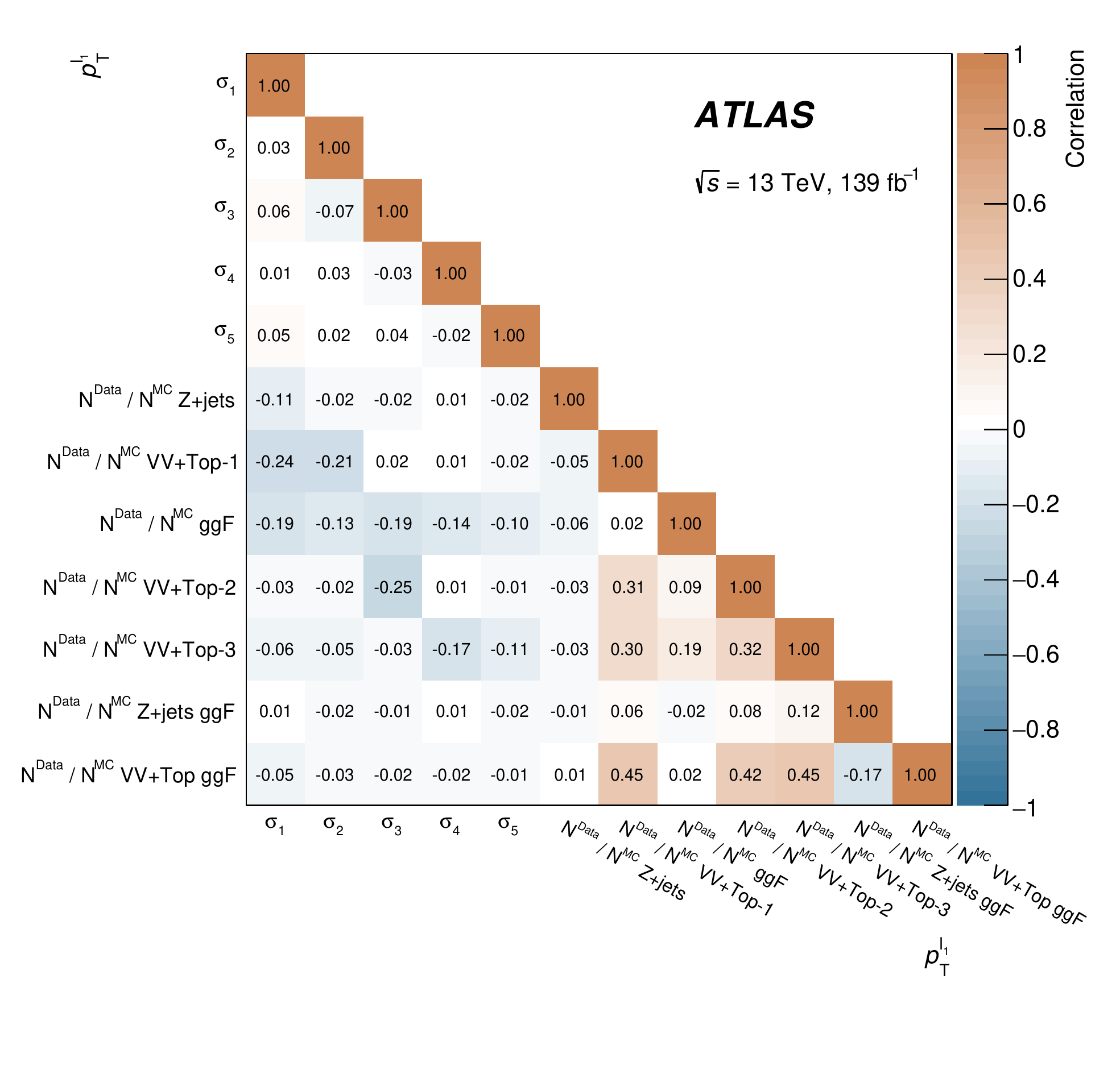}\label{fig:corMatri_lep0_pt}}
\subfloat[]{\includegraphics[width=0.40\linewidth]{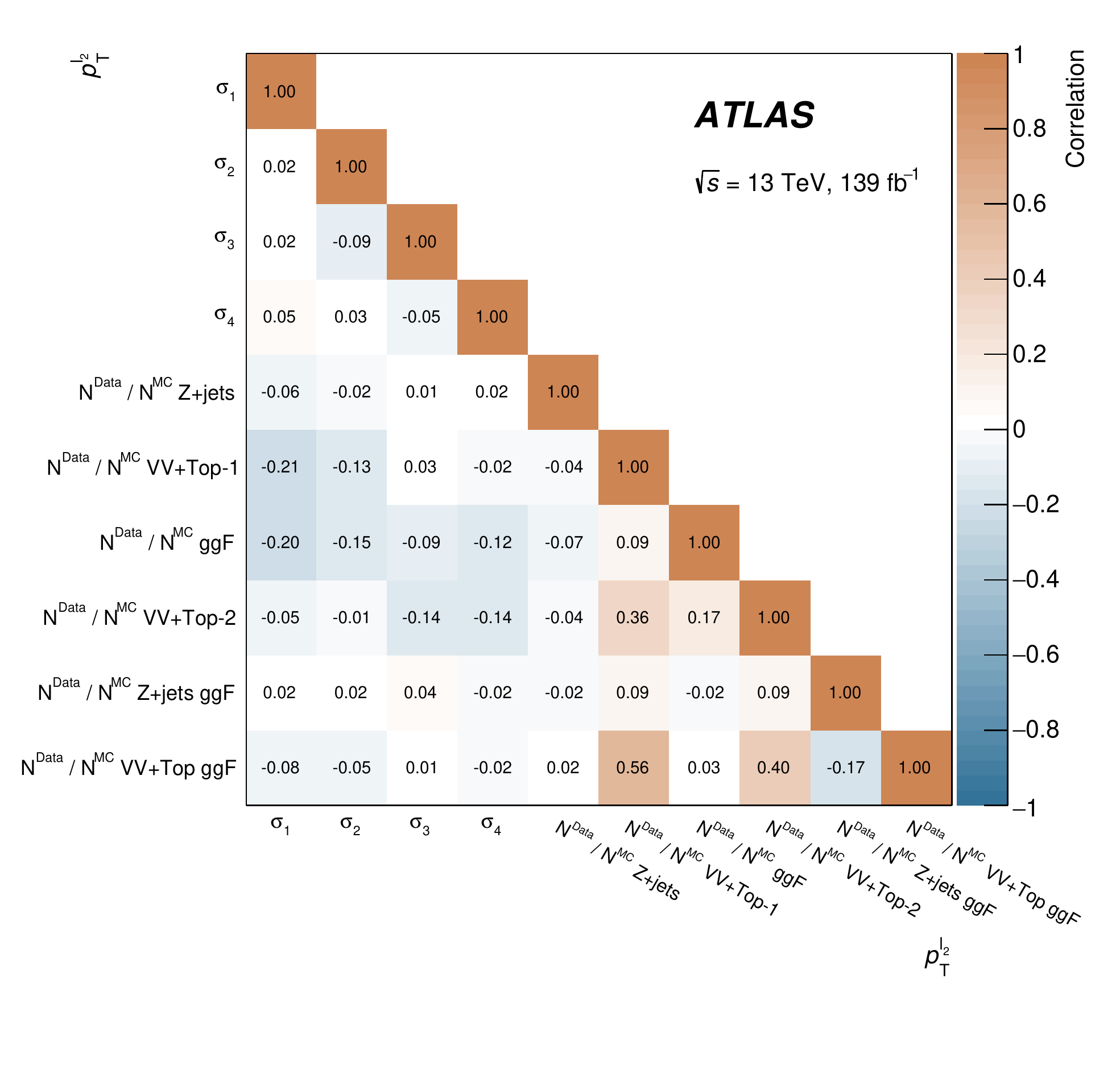}\label{fig:corMatrix_lep1_pt}} \\
\subfloat[]{\includegraphics[width=0.40\linewidth]{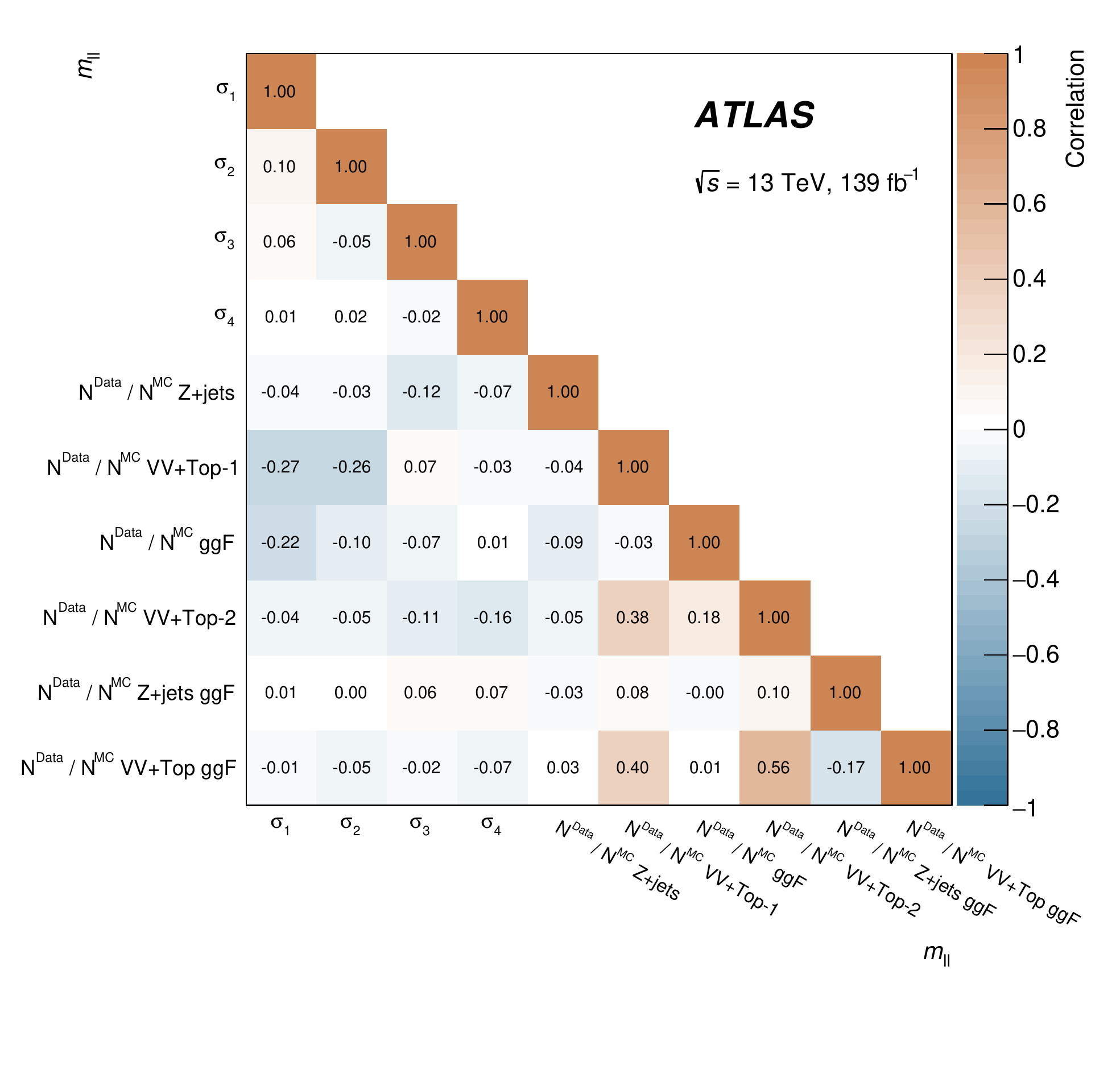}\label{fig:corMatrix_mll}}
\subfloat[]{\includegraphics[width=0.40\linewidth]{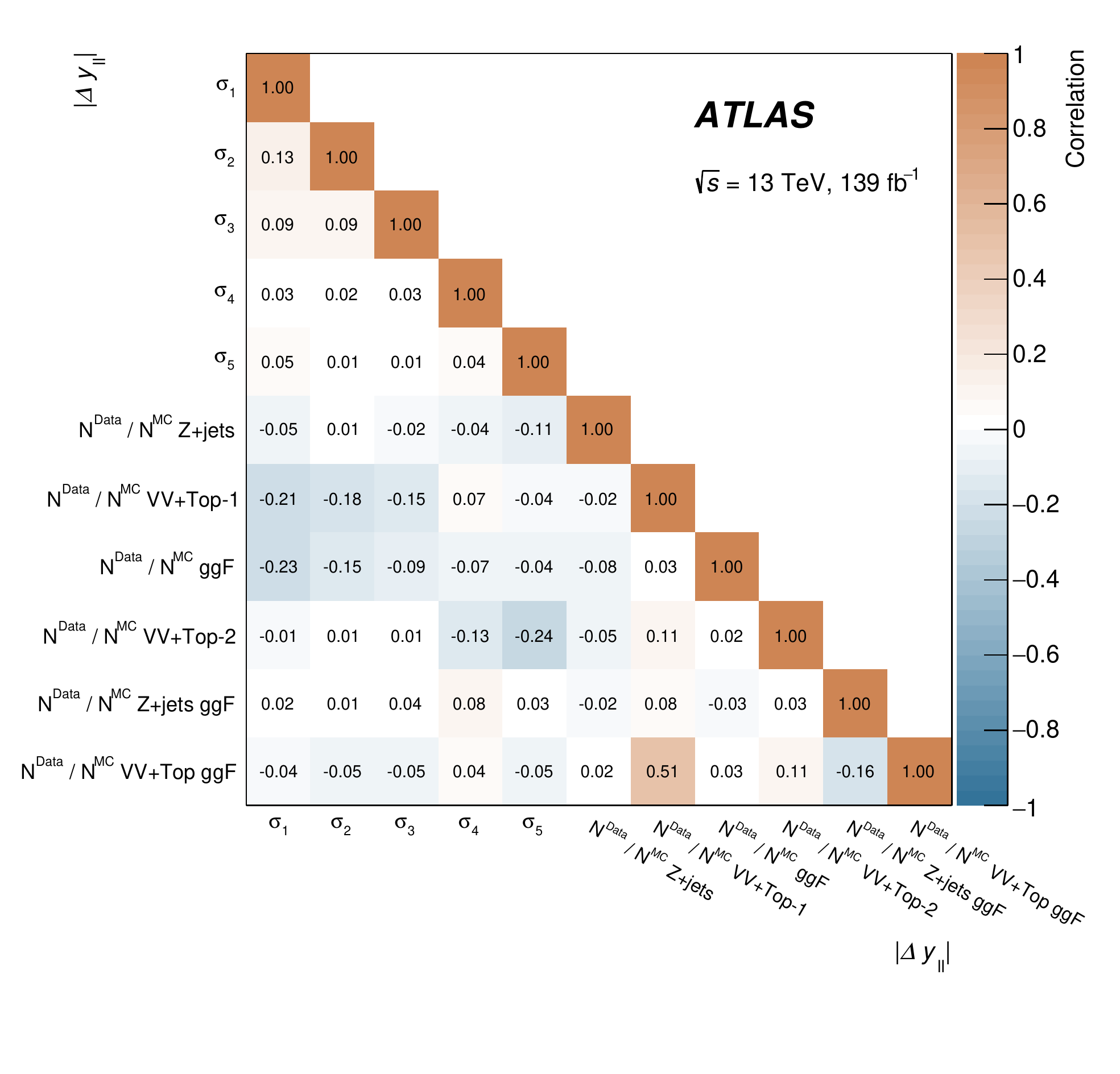}\label{fig:corMatrix_dyll}} \\
\subfloat[]{\includegraphics[width=0.40\linewidth]{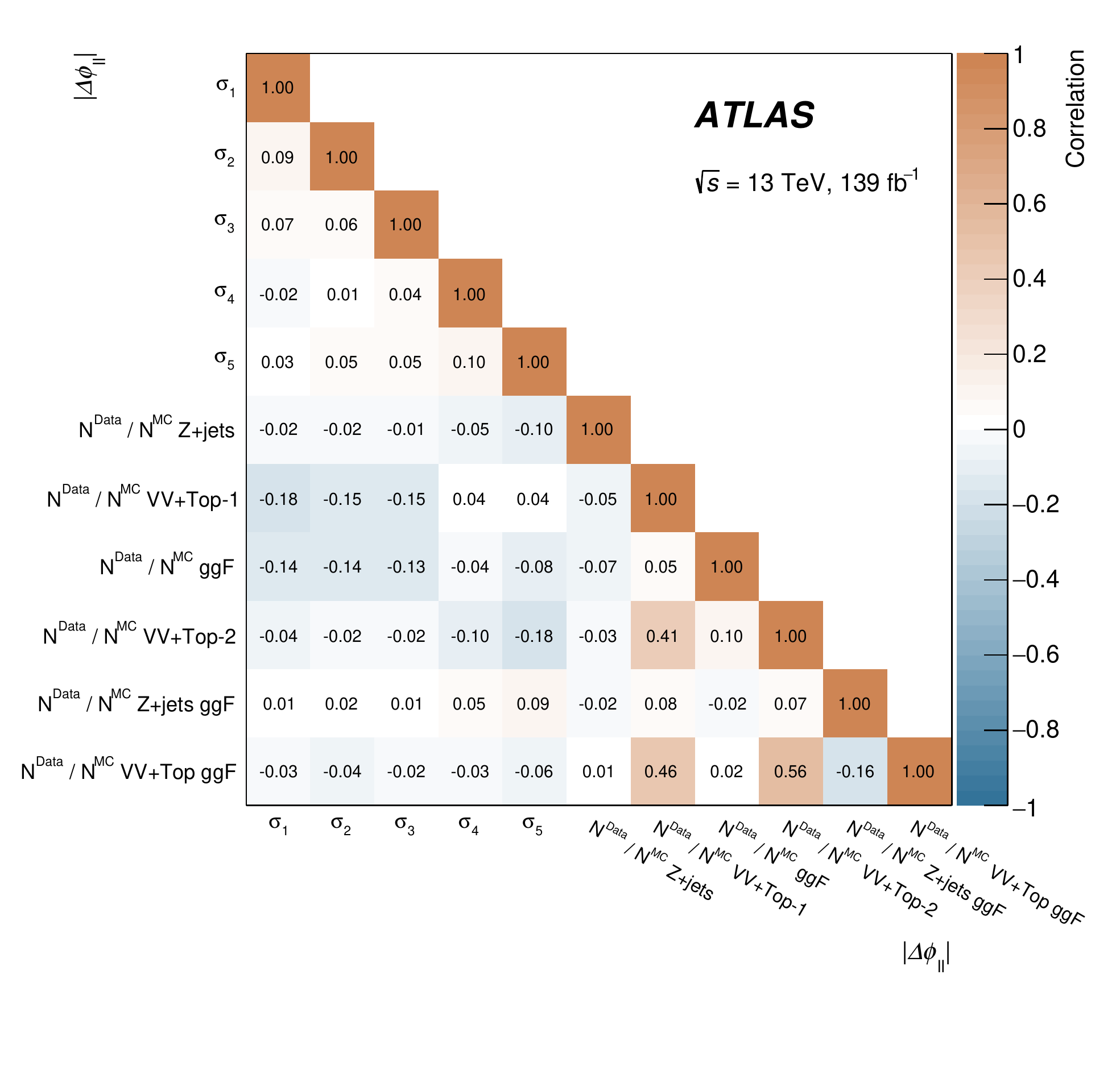}\label{fig:corMatrix_dphill}}
\subfloat[]{\includegraphics[width=0.40\linewidth]{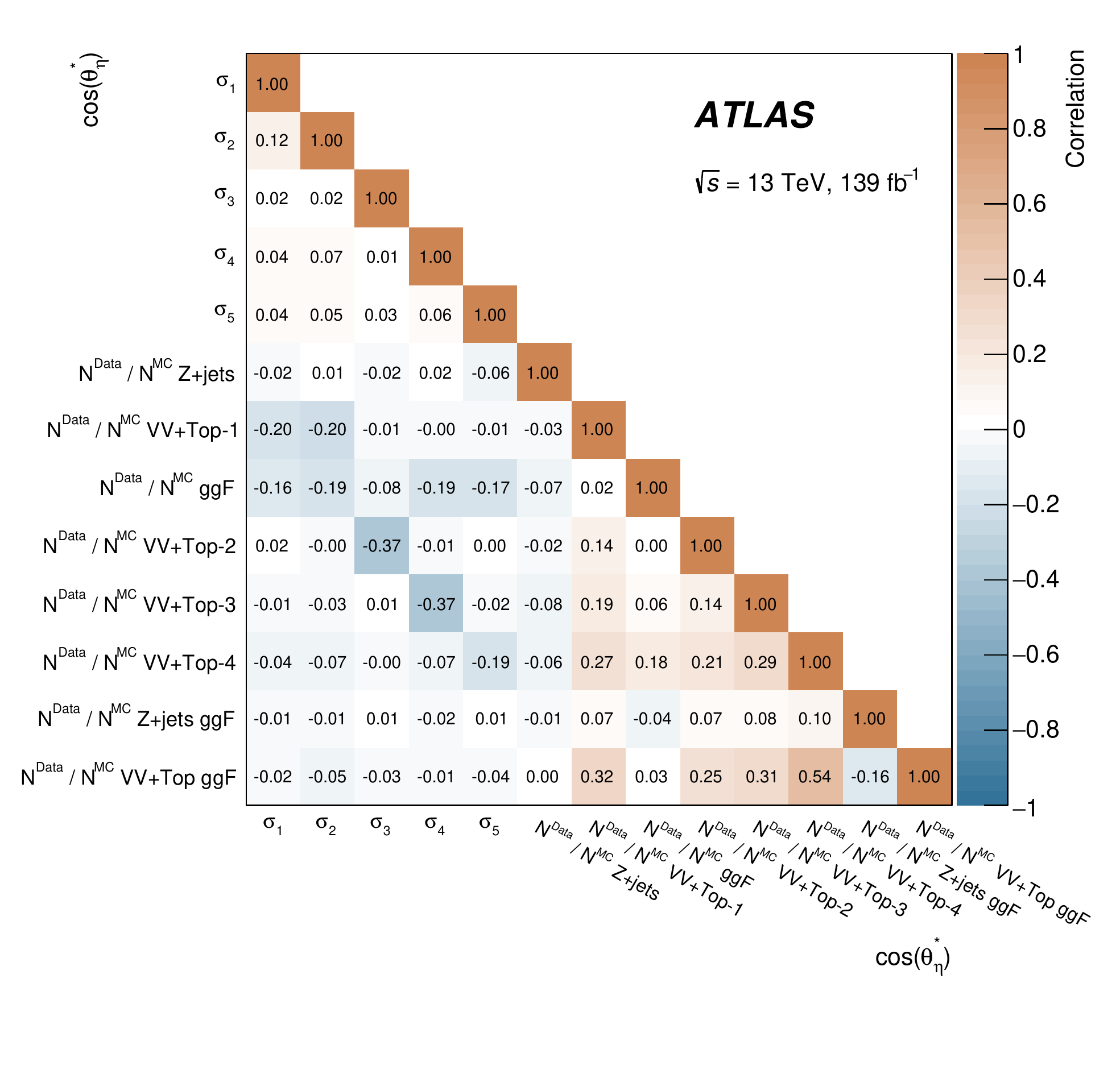}\label{fig:corMatrix_costhetastar}}
\caption{The correlation matrices between the background normalization factors and the measured differential cross-sections in bins of lepton kinematic variables such as \protect\subref{fig:corMatri_lep0_pt} $\pTlone$, \protect\subref{fig:corMatrix_lep1_pt} $\pTltwo$, \protect\subref{fig:corMatrix_mll} $\mll$, \protect\subref{fig:corMatrix_dyll} $|\dyll|$ , \protect\subref{fig:corMatrix_dphill} $|\dphill|$, and \protect\subref{fig:corMatrix_costhetastar} $\cos(\theta^*_{\eta})$. The symbol $\sigma_{i}$ is the cross-section in bin-$i$. }
\label{fig:Result_CorrMatrixt_leptons}
\end{figure}
\begin{figure}[ht]
\centering
\subfloat[]{\includegraphics[width=0.40\linewidth]{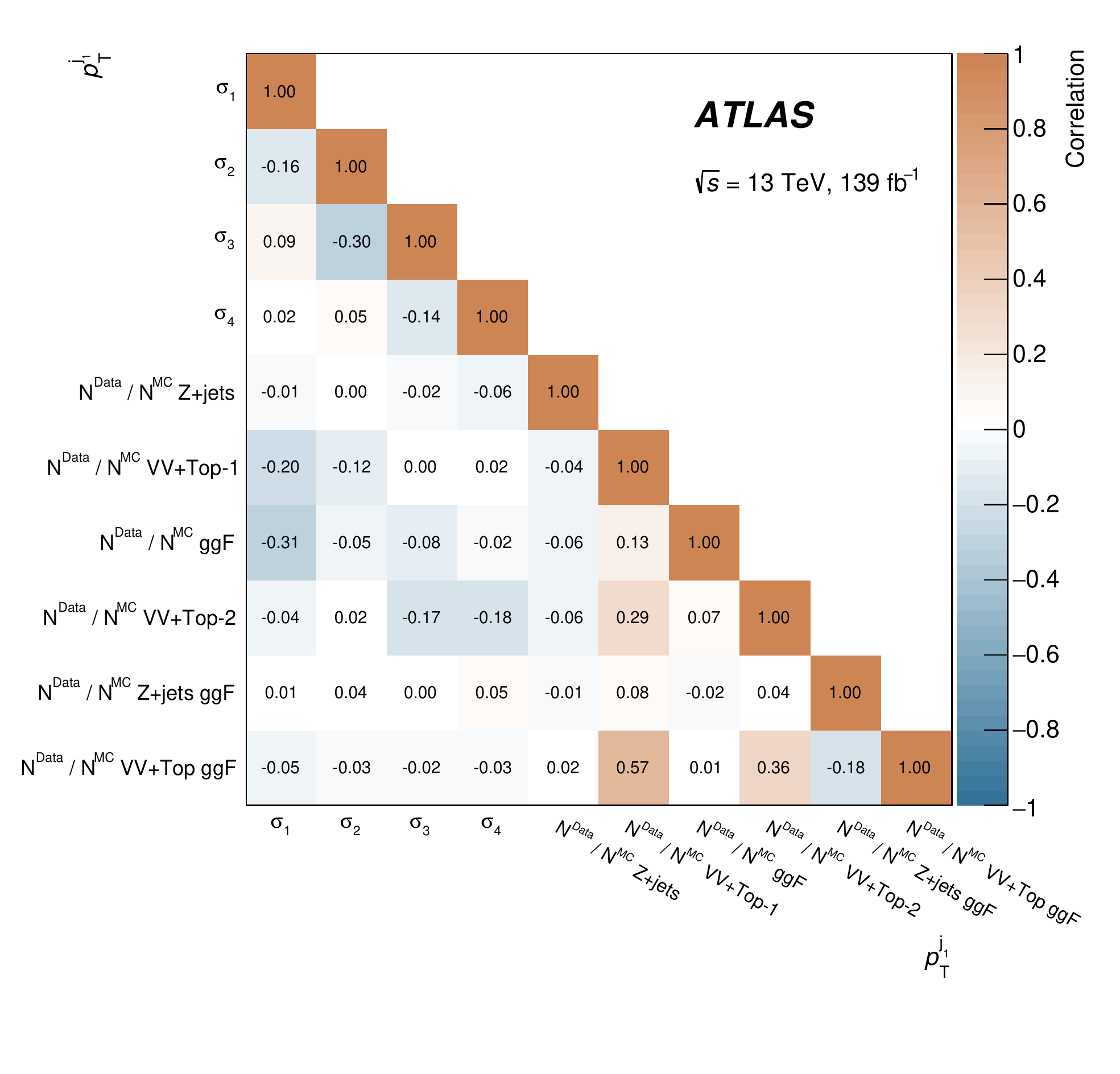}\label{fig:corMatri_jet0_pt}}
\subfloat[]{\includegraphics[width=0.40\linewidth]{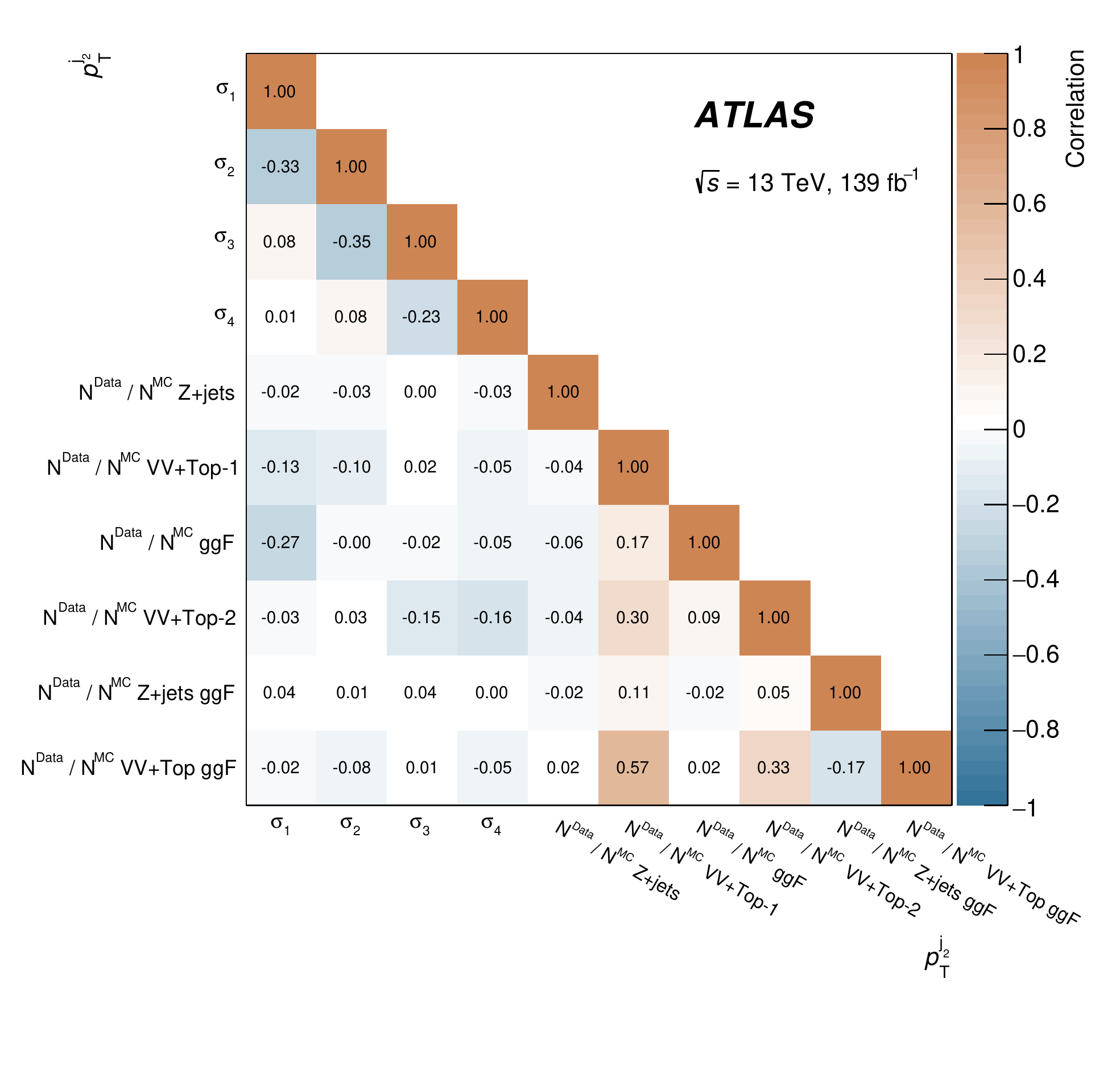}\label{fig:corMatri_jet1_pt}} \\
\subfloat[]{\includegraphics[width=0.40\linewidth]{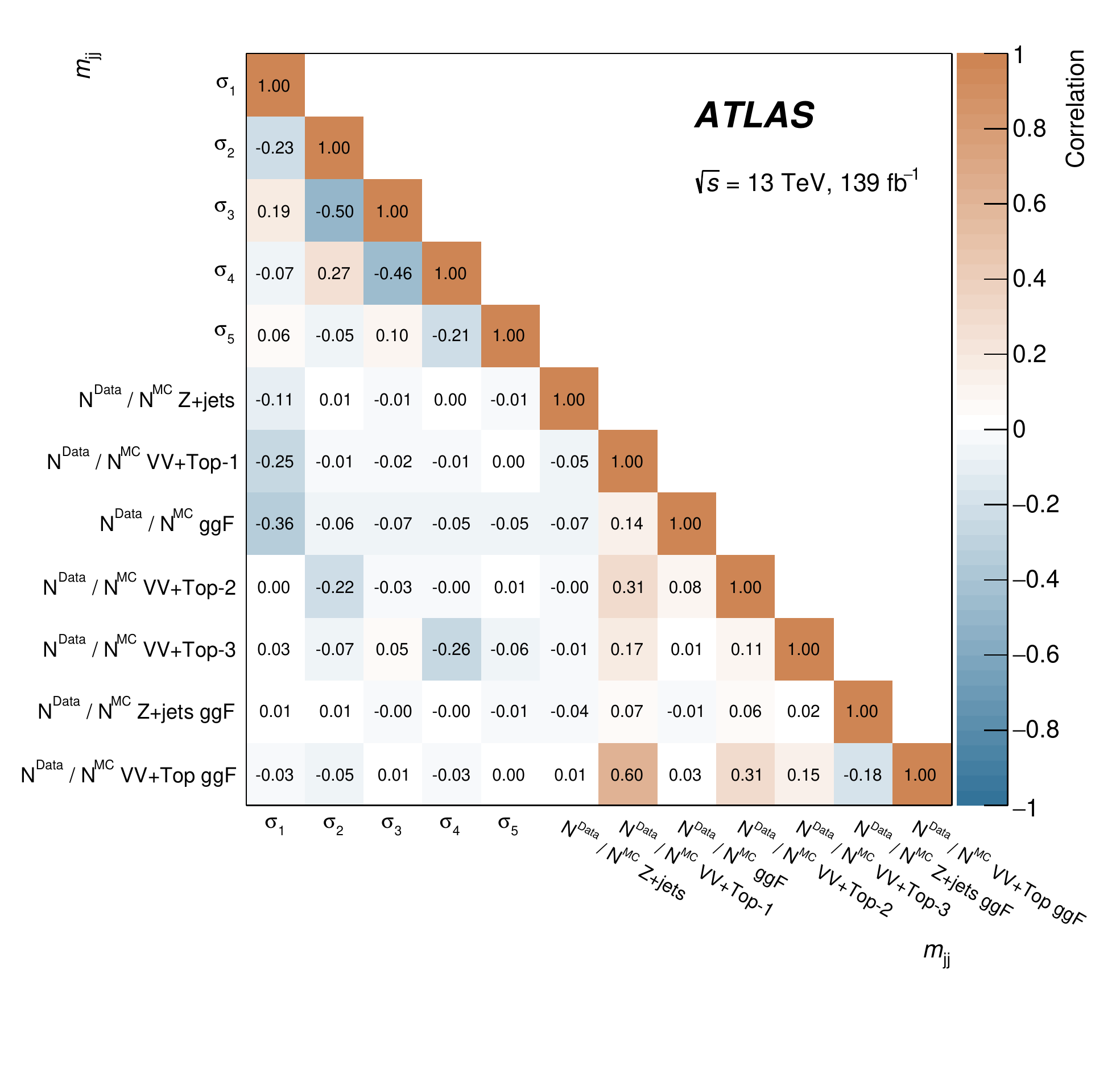}\label{fig:corMatri_mjj}}
\subfloat[]{\includegraphics[width=0.40\linewidth]{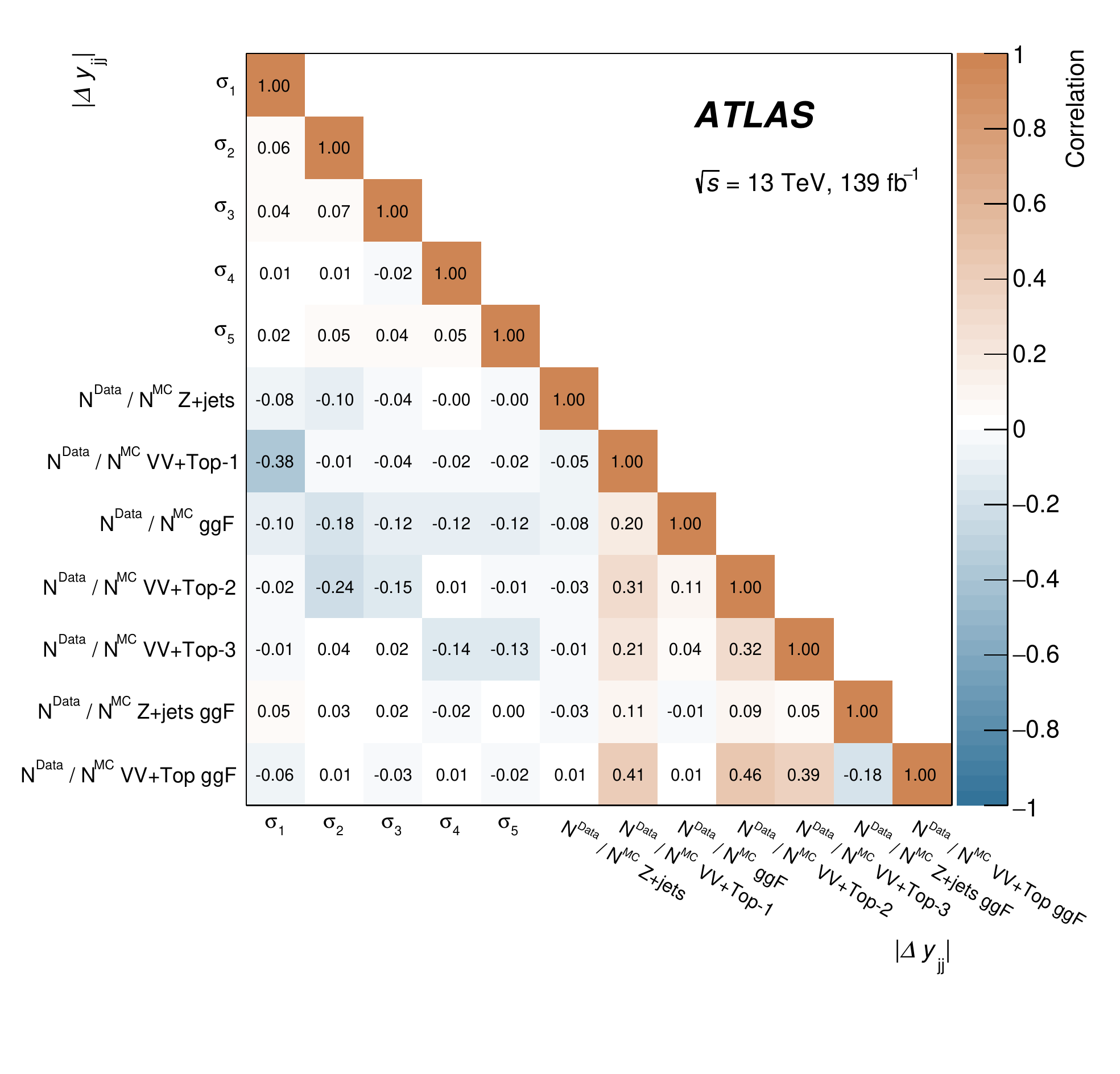}\label{fig:corMatri_dyjj}} \\
\subfloat[]{\includegraphics[width=0.40\linewidth]{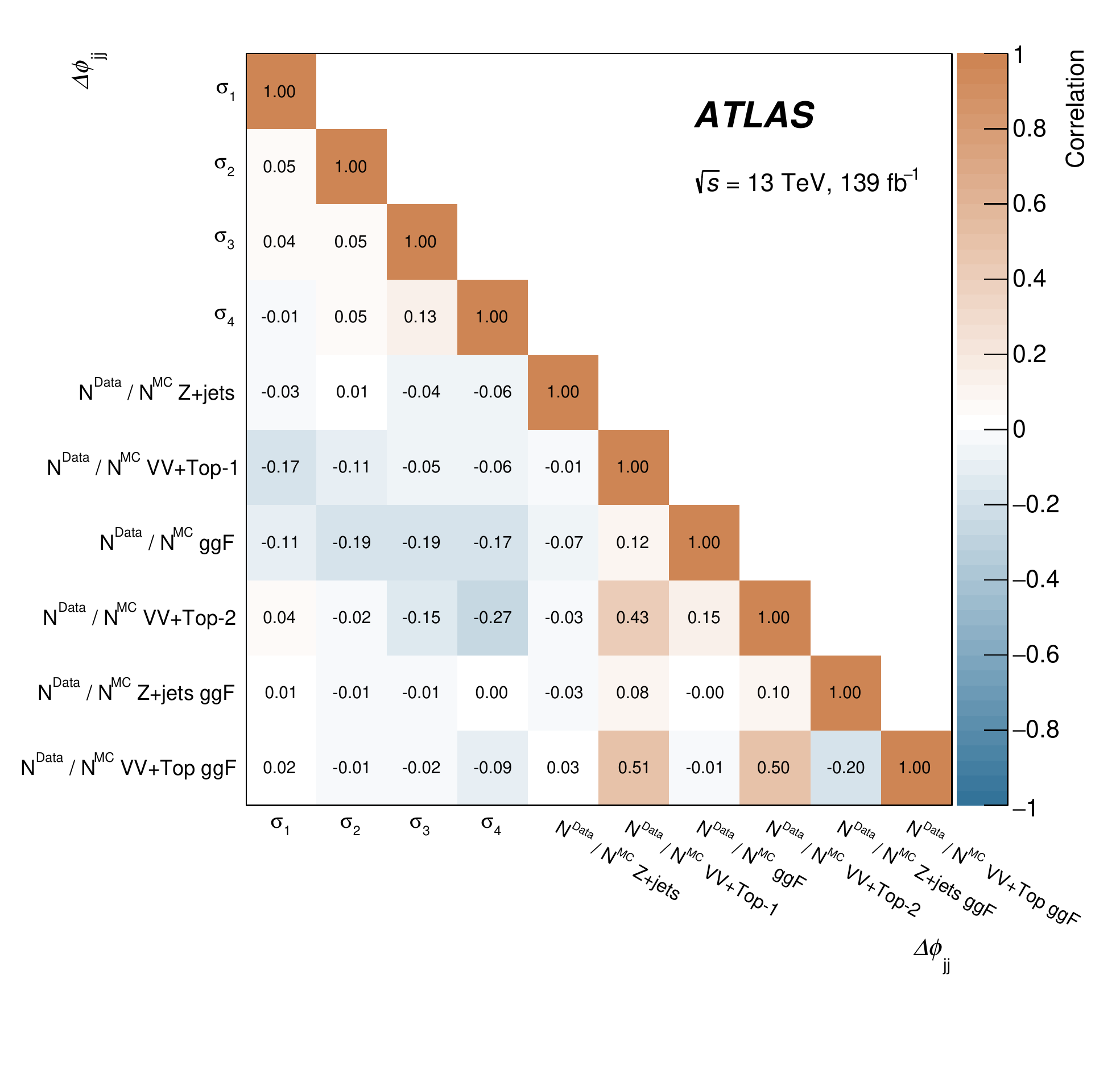}\label{fig:corMatri_dphijj}}
\caption{The correlation matrices between the background normalization factors and the measured differential cross-sections in bins of jet kinematic variables such as \protect\subref{fig:corMatri_jet0_pt} $\pTjone$, \protect\subref{fig:corMatri_jet1_pt} $\pTjtwo$, \protect\subref{fig:corMatri_mjj} $\mjj$, \protect\subref{fig:corMatri_dyjj} $|\dyjj|$, and \protect\subref{fig:corMatri_dphijj} $\dphijj$. The symbol $\sigma_{i}$ is the cross-section in bin-$i$. }
\label{fig:Result_CorrMatrixt_jets}
\end{figure}


\clearpage
 
\printbibliography

\clearpage
 
\begin{flushleft}
\hypersetup{urlcolor=black}
{\Large The ATLAS Collaboration}

\bigskip

\AtlasOrcid[0000-0002-6665-4934]{G.~Aad}$^\textrm{\scriptsize 102}$,
\AtlasOrcid[0000-0002-5888-2734]{B.~Abbott}$^\textrm{\scriptsize 120}$,
\AtlasOrcid[0000-0002-1002-1652]{K.~Abeling}$^\textrm{\scriptsize 55}$,
\AtlasOrcid[0000-0001-5763-2760]{N.J.~Abicht}$^\textrm{\scriptsize 49}$,
\AtlasOrcid[0000-0002-8496-9294]{S.H.~Abidi}$^\textrm{\scriptsize 29}$,
\AtlasOrcid[0000-0002-9987-2292]{A.~Aboulhorma}$^\textrm{\scriptsize 35e}$,
\AtlasOrcid[0000-0001-5329-6640]{H.~Abramowicz}$^\textrm{\scriptsize 151}$,
\AtlasOrcid[0000-0002-1599-2896]{H.~Abreu}$^\textrm{\scriptsize 150}$,
\AtlasOrcid[0000-0003-0403-3697]{Y.~Abulaiti}$^\textrm{\scriptsize 117}$,
\AtlasOrcid[0000-0003-0762-7204]{A.C.~Abusleme~Hoffman}$^\textrm{\scriptsize 137a}$
% ...
\bigskip
\\

% ...
$^{29}$Physics Department, Brookhaven National Laboratory, Upton NY; United States of America.\\
% ...
$^{35}$$^{(a)}$Facult\'e des Sciences Ain Chock, R\'eseau Universitaire de Physique des Hautes Energies - Universit\'e Hassan II, Casablanca;$^{(b)}$Facult\'{e} des Sciences, Universit\'{e} Ibn-Tofail, K\'{e}nitra;$^{(c)}$Facult\'e des Sciences Semlalia, Universit\'e Cadi Ayyad, LPHEA-Marrakech;$^{(d)}$LPMR, Facult\'e des Sciences, Universit\'e Mohamed Premier, Oujda;$^{(e)}$Facult\'e des sciences, Universit\'e Mohammed V, Rabat;$^{(f)}$Institute of Applied Physics, Mohammed VI Polytechnic University, Ben Guerir; Morocco.\\
% ...
$^{49}$Fakult\"{a}t Physik , Technische Universit{\"a}t Dortmund, Dortmund; Germany.\\
% ...
$^{55}$II. Physikalisches Institut, Georg-August-Universit\"{a}t G\"ottingen, G\"ottingen; Germany.\\
% ...
$^{102}$CPPM, Aix-Marseille Universit\'e, CNRS/IN2P3, Marseille; France.\\
% ...
$^{117}$Department of Physics, New York University, New York NY; United States of America.\\
% ...
$^{120}$Homer L. Dodge Department of Physics and Astronomy, University of Oklahoma, Norman OK; United States of America.\\
% ...
$^{137}$$^{(a)}$Departamento de F\'isica, Pontificia Universidad Cat\'olica de Chile, Santiago;$^{(b)}$Millennium Institute for Subatomic physics at high energy frontier (SAPHIR), Santiago;$^{(c)}$Instituto de Investigaci\'on Multidisciplinario en Ciencia y Tecnolog\'ia, y Departamento de F\'isica, Universidad de La Serena;$^{(d)}$Universidad Andres Bello, Department of Physics, Santiago;$^{(e)}$Instituto de Alta Investigaci\'on, Universidad de Tarapac\'a, Arica;$^{(f)}$Departamento de F\'isica, Universidad T\'ecnica Federico Santa Mar\'ia, Valpara\'iso; Chile.\\
% ...
$^{150}$Department of Physics, Technion, Israel Institute of Technology, Haifa; Israel.\\
$^{151}$Raymond and Beverly Sackler School of Physics and Astronomy, Tel Aviv University, Tel Aviv; Israel.\\
% ...
$^{a}$ Also Affiliated with an institute covered by a cooperation agreement with CERN.\\
$^{b}$ Also at An-Najah National University, Nablus; Palestine.\\
$^{c}$ Also at Borough of Manhattan Community College, City University of New York, New York NY; United States of America.\\
$^{d}$ Also at Center for High Energy Physics, Peking University; China.\\
$^{e}$ Also at Center for Interdisciplinary Research and Innovation (CIRI-AUTH), Thessaloniki; Greece.\\
$^{f}$ Also at Centro Studi e Ricerche Enrico Fermi; Italy.\\
% ...

\end{flushleft}


\end{document}